%% file: ds1878s.tex
\documentclass[psfig]{aa}
\usepackage{graphics}
\usepackage{psfig}

\begin{document}

\thesaurus{ 08.01.2; 08.02.1; 08.02.4; 08.03.3; 08.12.1 }

\title {Multiwavelength optical observations of chromospherically
active binary systems}
\subtitle{III. High resolution echelle spectra from 
Ca~{\sc ii} H \& K to Ca~{\sc ii} IRT 
\thanks{Based on observations made
 with the Isaac Newton Telescope (INT)
 operated on the island of La Palma by the Isaac Newton Group in
 the Spanish Observatorio del Roque de Los Muchachos of the
 Instituto de Astrof\'{\i}sica de Canarias,
and with the Nordic Optical Telescope (NOT),
 operated on the island of La Palma jointly by Denmark, Finland,
 Iceland, Norway and Sweden, in the Spanish Observatorio del 
 Roque de los Muchachos of the Instituto de Astrof\'{\i}sica de Canarias.
} }

\author{
D.~Montes\inst{1,}\inst{2}
\and M.J.~Fern\'{a}ndez-Figueroa\inst{1}
\and E.~De Castro\inst{1}
\and M.~Cornide\inst{1}
\and A.~Latorre\inst{1}
\and J.~Sanz-Forcada\inst{1,}\inst{3}
}

\offprints{ D.~Montes}
\mail{dmg@astrax.fis.ucm.es}

\institute{
Departamento de Astrof\'{\i}sica,
Facultad de Ciencias F\'{\i}sicas,
 Universidad Complutense de Madrid, E-28040 Madrid, Spain\\
E-mail: dmg@astrax.fis.ucm.es
\and Guest observer at McDonald Observatory, Texas, USA
\and Harvard-Smithsonian Center for Astrophysics,
60 Garden St., Cambridge, MA 02138, USA
}

\date{Received April 10, 2000; accepted .. .., 2000}

\maketitle

\begin{abstract}

This is the third paper of a series aimed at studying
the chromosphere of active binary systems
using the information provided for several optical spectroscopic
features.
High resolution echelle spectra 
including all the optical chromospheric activity indicators
from the Ca~{\sc ii} H \& K to Ca~{\sc ii} IRT lines
are analysed here for 16 systems.
The chromospheric contribution in these lines has been determined
using the spectral subtraction technique.
Very broad wings have been found in the subtracted H$\alpha$ profile
of the very active star HU Vir.
These profiles are well matched using a two-component 
Gaussian fit (narrow and broad) and the broad component
can be interpreted as arising from microflaring.
Red-shifted absorption features in the H$\alpha$ line 
have been detected in several systems and 
excess emission in the blue wing of FG UMa was also detected.
These features indicate that several dynamical processes, 
or a combination of them, may be involved.
Using the E$_{\rm H\alpha}$/E$_{\rm H\beta}$ ratio as a diagnostic
we have detected prominence-like extended material viewed off the limb
in many stars of the sample, and prominences viewed against the disk 
at some orbital phases in the dwarfs OU Gem and BF Lyn. 
The He~{\sc i} D$_{3}$ line has been detected
as an absorption feature in mainly all the giants of the sample.
Total filling-in of the He~{\sc i} D$_{3}$, 
probably due to microflaring activity,
is observed in HU Vir.
Self-absorption with red asymmetry
is detected in  the Ca~{\sc ii} H \& K lines of the giants
12 Cam, FG UMa and BM CVn.
All the stars analysed show clear
filled-in Ca~{\sc ii} IRT lines or even notable emission reversal.
The small values of the E$_{\rm 8542}$/E$_{\rm 8498}$ ratio we have found
indicate Ca~{\sc ii} IRT emission arises from plage-like regions.
Orbital phase modulation of the chromospheric emission 
has been detected in some systems, in the case of 
HU Vir evidence of an active longitude area has been found.

\keywords{  
   stars: activity  
-- stars: binaries: close 
-- stars: binaries: spectroscopic 
-- stars: chromospheres 
-- stars: late-type
}

\end{abstract}
 

\begin{table*}
\caption[]{Observing log
\label{tab:obslog}}
\begin{flushleft}
\scriptsize
\begin{tabular}{lccccccccccccccccccccccccc}
\noalign{\smallskip}
\hline
\noalign{\smallskip}
Name &
\multicolumn{4}{c}{NOT-SOFIN 1996/03} &\ &
\multicolumn{4}{c}{NOT-SOFIN 1998/04} &\ &
\multicolumn{4}{c}{McD-Sandiford 1998/01} &\ &
\multicolumn{4}{c}{INT-MUSICOS 1999/01} \\
\cline{2-5}\cline{7-10}\cline{12-15}\cline{17-20}
\noalign{\smallskip}
     &
\tiny Day  & \tiny UT & {$\varphi$} & \tiny S/N &\ &
\tiny Day  & \tiny UT & {$\varphi$} & \tiny S/N &\ &
\tiny Day  & \tiny UT & {$\varphi$} & \tiny S/N &\ &
\tiny Day  & \tiny UT & {$\varphi$} & \tiny S/N 
\\
     &
 &  &  & \tiny (H$\alpha$) &\ &
 &  &  & \tiny (H$\alpha$) &\ &
 &  &  & \tiny (H$\alpha$) &\ &
 &  &  & \tiny (H$\alpha$) 
\scriptsize
\\
\noalign{\smallskip}
\hline
\noalign{\smallskip}
{\bf\tiny Active Stars} \\
\noalign{\smallskip}
\hline
\noalign{\smallskip}
\object{UX Ari}    & 01 & 21:45 & 0.95 & 171  \\
\object{12 Cam}    &    &       &      &     && 09 & 20:00 & 0.64 & 228  \\
\object{V1149 Ori} &    &       &      &     && 06 & 20:07 & 0.89 &  76  \\
\object{OU Gem}    & 01 & 22:55 & 0.80 & 303 && 05 & 19:56 & 0.19 & 274 &&   &    &     &     && 07 & 01:43 & 0.70 & 137  \\
"         &    &       &      &     && 09 & 20:28 & 0.77 & 199  \\
\object{$\sigma$ Gem} &02 & 00:30 & 0.81 & 528 && 06 & 20:48 & 0.88 & 188  \\
\object{BF Lyn}    & 02 & 01:14 & 0.43 & 249 && 05 & 20:57 & 0.49 & 135 && 18 &
09:47 & 0.12 & 109 && 07 & 02:54 & 0.11 & 68  \\
"         &    &       &      &     && 07 & 21:56 & 0.02 & 176 &&  18 &
13:05 & 0.16 & 116  \\
"         &    &       &      &     && 09 & 21:17 & 0.54 & 201 &&  19 &
10:41 & 0.40 & 170  \\
"         &    &       &      &     &&    &       &       &     &&  20 &
11:46 & 0.67 & 146  \\
"         &    &       &      &     &&    &       &       &     &&  21 &
10:28 & 0.92 &  78  \\
"         &    &       &      &     &&    &       &       &     &&  22 &
10:07 & 0.18 & 181  \\
\object{IL Hya}    &    &       &      &     && 06 & 22:00 & 0.02 & 117  \\
\object{FG UMa}    &    &       &      &     && 05 & 21:55 &       & 168  \\
"         &    &       &      &     && 09 & 22:00 &       & 166  \\
\object{LR Hya}    &    &       &      &     && 06 & 22:50 & 0.58 &  73  \\
\object{HU Vir}    &    &       &      &     && 06 & 23:39 & 0.29 &  63 &&  15 &
11:33 & 0.44 &  74  \\
"         &    &       &      &     && 07 & 23:36 & 0.38 &  83 &&  16 &
11:58 & 0.54 & 120  \\
"         &    &       &      &     &&    &       &       &     &&  17 &
11:46 & 0.64 &  85  \\
"         &    &       &      &     &&    &       &       &     &&  18 &
12:05 & 0.73 &  95  \\
"         &    &       &      &     &&    &       &       &     &&  19 &
12:03 & 0.83 & 138  \\
"         &    &       &      &     &&    &       &       &     &&  20 &
12:05 & 0.93 & 113  \\
"         &    &       &      &     &&    &       &       &     &&  21 &
12:10 & 0.02 &  74  \\
"         &    &       &      &     &&    &       &       &     &&  22 &
11:50 & 0.12 & 112  \\
\object{DK Dra}    & 02 & 03:15 & 0.90 & 234 && 06 & 21:30 & 0.84 &  88  \\
\object{BQ CVn}    &    &       &      &     && 07 & 22:40 &       & 201  \\
\object{IS Vir}    &    &       &      &     && 09 & 23:05 &       & 145  \\
\object{BL CVn}    &    &       &      &     && 11 & 02:26 & 0.21 & 174  \\
\object{BM CVn}    &    &       &      &     && 09 & 22:38 & 0.26 & 191  \\
\object{MS Ser}    &    &       &      &     &  & 08 & 05:05 & 0.21 & 187  \\
"         &       &       &      &     &  & 11 & 06:04 & 0.54 &  60  \\
\noalign{\smallskip}
\hline
{\bf\tiny Ref. Stars} \\
\noalign{\smallskip}
\hline
\noalign{\smallskip}
\object{$\beta$ Gem}  & 02 & 22:05 & -  & 169 && \\
\object{$\alpha$ Tau} & 02 & 21:53 & -  & 195 &&  
    &       &    &     &&    &       &    &     && 07 & 00:42 & -  & 474 \\
\object{6 Lyn}        &    &       &    &     &&
 09 & 21:01 & -  & 196 && \\
\object{$\alpha$ Boo} &    &       &    &     && 
 06 & 01:05 & -  & 335 && 16 & 12:59 & -  & 349 && 07 & 05:13 & -  & 542 \\
"                     &    &       &    &     &&
 08 & 00:41 & -  & 305 &&    &       &    &     &&    &       &    &     \\
\object{54 Psc}       &    &       &    &     &&
    &       &    &     && 15 & 01:10 & -  & 203 &&    &       &    &     \\
\object{HR 1614}      &    &       &    &     &&
    &       &    &     && 17 & 04:35 & -  & 194 &&    &       &    &     \\
\noalign{\smallskip}
\hline
\end{tabular}

\end{flushleft}
\end{table*}

\section{Introduction}

This paper is a continuation of our ongoing project of
multiwavelength optical observations
aimed at studying
the chromosphere of active binary systems
using the information provided for several optical spectroscopic
features that are formed at different heights in the chromosphere.
In Paper I (Montes et al. \cite{M97})
we focussed our study on the analysis of the extensively
used H$\alpha$ chromospheric activity indicator together with simultaneous
observations of the less studied He~{\sc i} D$_{3}$ and
Na~{\sc i} D$_{1}$, D$_{2}$ spectral features
in a sample of 18 northern active binary systems.
In Paper II (Montes et al. \cite{M98a})
the H$\alpha$,  H$\beta$,
Na~{\sc i} D$_{1}$, D$_{2}$, He~{\sc i} D$_{3}$, Mg~{\sc i} b triplet,
Ca~{\sc ii} H \& K and Ca~{\sc ii} infrared triplet lines (Ca~{\sc ii} IRT) 
of the RS CVn system EZ Pegasi were studied at different orbital phases, 
also including a high resolution echelle spectrum.

As shown in Paper I and II, 
with the simultaneous analysis of the different optical 
chromospheric activity indicators and using the spectral subtraction technique,
it is possible to study in detail the chromosphere, discriminating between
the different structures: plages, prominences, flares and microflares
(see recent studies by Gunn \& Doyle \cite{G&D97}; Gunn et al. \cite{G97};
L\'{a}zaro \& Ar\'{e}valo \cite{L&A97}; Ar\'{e}valo \& L\'{a}zaro \cite{A&L99}; 
Hall \& Wolovitz \cite{H&W98};
Montes et al. \cite{M98b}; 
Montes \& Ramsey \cite{MR98}, \cite{MR99}; Montes et al. \cite{M99};
 Eibe et al. \cite{Eibe99}; Oliveira \& Foing 1999; Berdyugina et al. \cite{Ber99}).

In this paper we present 
high resolution echelle spectra of 16 systems 
selected from "A Catalog of Chromospherically Active Binary Stars
(second edition)" (Strassmeier et al. 1993, hereafter CABS).
The spectra were taken during four observing runs (from 1996 to 1999) 
and include all the optical chromospheric activity indicators 
from the Ca~{\sc ii} H \& K to Ca~{\sc ii} IRT lines.
Preliminary results of some of the systems included here can be found in 
Sanz-Forcada et al. (1998, 1999); 
Montes et al. (\cite{M98c}, \cite{M00b}); 
Latorre et al. (1999, 2000). 

In Sect.~2 we give the details of our observations and data reduction.
In Sect.~3 we describe the spectroscopic features analysed in this paper:
the different chromospheric activity indicators 
and the Li~{\sc i} $\lambda$6707.8 line.
Individual results about stellar parameters and the behaviour of
the chromospheric excess emission in each system is reported in Sect.~4.
Finally, in Sect.~5 the discussion and conclusions are given.

\begin{table*}
\caption[]{Stellar parameters 
\label{tab:par}}
\begin{flushleft}
\scriptsize
\begin{tabular}{l l c c c c c l l l c }
\noalign{\smallskip}
\hline
\noalign{\smallskip}
{HD} & {Name} & {T$_{\rm sp}$} & {SB} & {R} & {V--R} & {B--V} &  {T$_{\rm conj}$}
& {P$_{\rm orb}$} & {P$_{\rm rot}$} & Vsin{\it i}\\
  &       &       &    &  (R$_{\odot}$)  &        &    & (H.J.D.) & (days) &
 (days) & (km s$^{-1}$)  \\
\noalign{\smallskip}
\hline
\noalign{\smallskip}
{\bf\tiny Active Stars} \\
\noalign{\smallskip}
\hline
\noalign{\smallskip}
 21242 & UX Ari  &  G5V/K0IV & 2 & 0.93/$>$4.7 &0.54/0.70& 0.91
& 2450415.0647 & 6.437876 & $\approx$P$_{\rm orb}$ & 6/37 \\
 32357 & 12 Cam & K0III & 1 & $\geq$16 & [0.77]& 1.12 &
2448111.1   & 80.898 & 81.23 & 11.3 \\
 37824 & V1149 Ori & K1III & 1 & /$\geq$11 & 0.90  & 1.14 &
2444325.93 & 53.580 & 54.1 & 11  \\
45088  & OU Gem & K3V/K5V & 2 & 0.77/0.72 & [0.82/0.99] & [0.96/1.15] &
 2443846.2 & 6.991868 & 7.36 & 5.6/5.6 \\
62044 & $\sigma$ Gem & K1III & 1 & $>$10.5 & 0.92& 1.122 &  2447227.22
& 19.604471 & 19.601562 & 27 \\
80715  & BF Lyn & K2V/[dK] & 2 & $>$0.78/$>$0.78 & [0.74/]    & 0.99 &
2445802.472 & 3.80406 & $\approx$P$_{\rm orb}$ & 10/10 \\
81410  & IL Hya & G8V/K0III-IV & 2 & 1.0/8.1 & 0.79& 1.02 &
2447103.1723 & 12.904982 & 12.730& 26.5  \\
89546  & FG UMa & G8IV & 1 & $>$6.4 & 0.64 & 1.004 &  & 21.300 &
21.50 & 15  \\
91816  & LR Hya & K0V/K0V & 2 & $\geq$0.8/$\geq$0.8 & [0.64/0.64] & [0.81/0.81] &
2446539.89 & 6.86569 & 3.1448 & 6/6  \\
106225 & HU Vir & K0III-IV & 1 & $>$5.7 & [0.705]  & 0.974 &  2450195.909 &
10.3876 & 10.42 & 31.3 \\
106677  & DK Dra & K1III/K1III & 2 & $\geq$13/$\geq$13 & [0.81/0.81]& [1.07/1.07] &
 2443445.95 & 64.44 & 63.75 & 10/10 \\
112859 & BQ CVn  & F/K0III & 2 & /$>$6.2 & [0.77]& 0.93 &        &
18.700 & 18.50 & $<$6/17 \\
113816 & IS Vir  & K0III & 1 & $>$2.3 & 0.93& 1.15 &       &
23.700 & 23.50 & 6 \\
115781 & BL CVn & G-KIV/K0III & 2 & /14.8 & [/0.77] & 1.14 &
2445284.47 & 18.6917 & $\approx$P$_{\rm orb}$ & 7/35  \\
116204 & BM CVn & K1III & 1 & $\geq$6 & [0.81] &1.16 &
24445256.776 & 20.625 & 20.6 & 15  \\
143313 & MS Ser & G8V/K2IV & 2 & 1.0/3.5  &  [0.58/0.79]    &  0.94/1.23
&  2442616.142 & 9.01490 & 9.60 & 7/15 \\
%
\noalign{\smallskip}
\hline
\noalign{\smallskip}
{\bf\tiny Active Stars} \\
\noalign{\smallskip}
\hline
\noalign{\smallskip}
62509  & $\beta$ Gem  & K0III & \\
29139  & $\alpha$ Tau & K5III & \\
45410  & 6 Lyn        & K0III-IV \\
124897 & $\alpha$ Boo & K1III \\
3651   & 54 Psc       & K0V \\
32147  & HR 1614      & K5V \\
\hline
\noalign{\smallskip}
\end{tabular}

\end{flushleft}
\end{table*}

\section{Observations and Data Reduction}
 
The spectroscopic echelle observations of the chromospherically active binaries
analysed in this paper were obtained during four observing runs.
Two of them were carried out in 1-2 March 1996 (NOT96 hereafter) 
and in 5-10 April 1998 (NOT98 hereafter), 
using the 2.56~m Nordic Optical Telescope (NOT) located
at the Observatorio del Roque de Los Muchachos (La Palma, Spain).
The Soviet Finnish High Resolution Echelle Spectrograph
(SOFIN) was used with an echelle grating (79 grooves/mm),
camera Astromed-3200 and a 1152$\times$770
pixel EEV P88200 CCD detector. The wavelength range covers from
3765 to 9865 \AA, for the first run, and from 3632 to 10800~\AA, 
for the second run.
The reciprocal dispersion ranges from 0.07 to 0.18 \AA/pixel
and the spectral resolution, 
determined as the full width at half maximum (FWHM) 
of the arc comparison lines, ranges from 0.15 to 0.60 \AA, in both runs.

We also analyse echelle spectra
obtained during a 10-night run in 12-21 January 1998 (McD98 hereafter), 
using the 2.1~m Otto Struve telescope at McDonald Observatory (Texas, USA)
 and the Sandiford Cassegrain Echelle Spectrograph
(McCarthy et al. 1993).
This instrument is a prism cross-dispersed echelle mounted at the
Cassegrain focus and it is used with a 1200$\times$400 Reticon CCD detector.
The spectrograph setup was chosen to cover the
H$\alpha$ (6563~\AA) and Ca~{\sc ii} IRT (8498, 8542, 8662~\AA) lines.
The wavelength coverage is about 6400-8800\AA$\ $ and the
reciprocal dispersion ranges from 0.06 to 0.08 ~\AA/pixel.
The spectral resolution, determined as the FWHM of the arc
comparison lines, ranges from 0.13 to 0.20 \AA.
In one of the nights, we changed the spectrograph setup to include the
He {\sc i} D$_{3}$ (5876~\AA)
and Na~{\sc i} D$_{1}$ and D$_{2}$ (5896, 5890~\AA) lines,
with wavelength coverage of 5600-7000~\AA.

In addition, we dispose of echelle spectra taken 
in 7-8 January 1999 (INT99 hereafter), 
with the 2.5~m Isaac Newton Telescope (INT) 
at the Observatorio del Roque de Los Muchachos (La Palma, Spain)
using the ESA-MUSICOS spectrograph.
This is a fibre-fed cross-dispersed echelle spectrograph,
built at the ESA Space Science Department in ESTEC as a replica of the
first MUSICOS spectrograph built at Meudon-Paris Observatory 
(Baudrand \& B\"ohm \cite{B&B92}) and developed as part of 
MUlti-SIte COntinuous Spectroscopy (MUSICOS
\footnote{http://www.ucm.es/info/Astrof/MUSICOS.html}) project.
During this observing run, 
a 2148$\times$2148 pixel SITe1 CCD detector was used 
obtaining wavelength coverage from 3950~\AA$\ $ to 9890~\AA. The  
reciprocal dispersion ranges from 0.06 to 0.12 \AA$\ $
and the spectral resolution (FWHM of the of the arc
comparison lines) from~0.15 to 0.4 \AA. 

In Table~\ref{tab:obslog} we give the observing log.
For each observation we list date, UT, orbital phase ($\varphi$) 
and the signal to noise ratio (S/N) obtained in the H$\alpha$ line region. 
Table~\ref{tab:par} shows the HD number, name and 
stellar parameters of the active binary systems and the
non active stars used as reference stars in the spectral subtraction. 
The B--V and V--R color indexes and the radius are obtained from the 
relation with spectral type given by Landolt-B\"{o}rnstein (Schmidt-Kaler 1982) 
when individual values are not given in the literature.
Other parameters are given by 
CABS Catalog (Strassmeier et al. 1993) or taken from the references given 
in the individual results of each star.

The spectra have been extracted using the standard
reduction procedures in the 
IRAF\footnote{IRAF is distributed by the National Optical Observatory,
which is operated by the Association of Universities for Research in
Astronomy, Inc., under contract with the National Science Foundation.}
 package (bias subtraction,
flat-field division and optimal extraction of the spectra).
The wavelength calibration was obtained by taking
spectra of a Th-Ar lamp, for NOT and McDonald runs, 
and a Cu-Ar lamp, for INT run.  
Finally, the spectra have been normalized by
a polynomial fit to the observed continuum.

\begin{table*}
\caption[]{H$\alpha$ line measures in the observed and
subtracted spectra 
\label{tab:ha}}
\begin{flushleft}
\scriptsize
\begin{tabular}{lccccccccccccc}
\noalign{\smallskip}
\hline
\noalign{\smallskip}
 &   &   &  &  &
\multicolumn{4}{c}{Observed H$\alpha$ Spectrum} &\ &
\multicolumn{4}{c}{Subtracted H$\alpha$ Spectrum} \\
\cline{6-9}\cline{11-14}
\noalign{\smallskip}
 {Name} & Obs. & {$\varphi$} & {E} &  S$_{\rm H}$/S$_{\rm C}$  & H$\alpha$ &
 {W$_{\rm obs}$} & {R$_{\rm c}$ } & {EW} &  &
 {W$_{\rm sub}$} & {I} & {EW$^{*}$} & {$\log {\rm F}_{\rm S}$} \\
  & & &  &   &   & {\scriptsize (\AA) } &  & (\AA)  & &
 {\scriptsize (\AA)} &  & {\scriptsize  (\AA)} &  \\
\noalign{\smallskip}
\hline
\noalign{\smallskip}
%
 UX Ari   & NOT 96 & 0.96 & C & 0.30/0.70 & E & 1.44 & 1.27  & 0.42  & &
 2.28 & 0.65  & 2.13  & 6.89 \\
 12 Cam   & NOT 98 & 0.64 & - &     -     & A & 2.04 & 0.47  & 1.16  & &
 1.02 & 0.28  & 0.34  & 5.99 \\
 V1149 Ori& NOT 98 & 0.89 & - &     -     & A & 2.35 & 0.62  & 0.93  & &
 1.10 & 0.47  & 0.58  & 6.04 \\
 OU Gem   & NOT 96 & 0.80 & H  & 0.70 & A  & 1.17& 0.61 &
 0.49 & & 1.41& 0.27 & 0.61  & 6.29 \\
   "      &    "   &   "  & C  & 0.30 & A  & 1.16& 0.93 &
 0.08 & & 1.77& 0.17 & 1.03  & 6.31 \\
 "        & NOT 98 & 0.19 & H  & 0.70 & A & 1.23 & 0.60& 0.53 & &
 1.28& 0.24 & 0.51 & 6.21 \\
 "        &   "    &  "   & C  & 0.30 & A & 0.89 & 0.91& 0.07 & &
 1.32& 0.14 & 0.70 & 6.14 \\
 "        & NOT 98 & 0.77 & H  & 0.70& A & 1.34 & 0.60& 0.56 & &
 1.12& 0.25& 0.46 & 6.17 \\
 "        &   "    &  "   & C  & 0.30& A & 0.60 & 0.96& 0.02 & &  
 1.40& 0.21& 0.93 & 6.26 \\
 "        & INT 99 & 0.70 & H  & 0.70& A & 1.26 & 0.67&0.46 & &
 1.35& 0.29 & 0.67  & 6.33 \\
 "        &    "   &   "  & C  & 0.30& A & 0.60 & 0.92&0.05  & &
 1.48& 0.10 & 0.50  & 5.99 \\
$\sigma$ Gem& NOT 96 & 0.81& - &    -      & A & 1.85 & 0.37  & 1.24  & &
0.96 & 0.17  & 0.18  & 5.50 \\
 "          & NOT 98 & 0.88& - &    -      & A & 1.69 & 0.49  & 0.95  & &
1.13 & 0.27  & 0.34  & 5.77 \\
 BF Lyn   & NOT 98 & 0.02 & T & 0.50/0.50 & F &  -    & 0.87  &  -    & &
 1.53& 0.62  & 0.99  & 6.60 \\
 "        & INT 99 & 0.11 &H&  0.50     & A & 0.94  & 0.82  & 0.18  & &    
1.32 & 0.32  & 0.91      & 6.56 \\
 "        &    "   &  "   &C&  0.50     & F &  -    & 0.94  &  -    & &  
1.45 & 0.30  & 0.87      & 6.54 \\
 "        & McD 98 & 0.12 &H  & "         & A & 1.06 & 0.80  & 0.20  & &
1.25 & 0.29  & 0.80      & 6.51 \\
 "        &    "   &   "  &C  & "         & A & 1.07 & 0.85  & 0.16  & &
1.21 & 0.23  & 0.57      & 6.36 \\
 "        & McD 98 & 0.16 &H  & "         & A & 1.26 & 0.81  & 0.24  & &  
1.21 & 0.27  & 0.77      & 6.49 \\
 "        &   "    &   "  &C  & "         & A & 0.94 & 0.80  & 0.17  & &
1.21 & 0.20  & 0.52      & 6.32 \\
 "        & McD 98 & 0.18 &H  & "         & A & 1.31 & 0.82  & 0.24  & &    
1.23 & 0.27  & 0.77      & 6.49 \\
 "        &    "   &   "  &C  & "         & A & 0.99 & 0.83  & 0.16  & &
1.22 & 0.21  & 0.55      & 6.34 \\
 "        & McD 98 & 0.40 &H  & "         & A & 1.16 & 0.84  & 0.20  & &      
1.41 & 0.28  & 0.87      &6.54  \\
 "        &   "    &   "  &C  & "         & A & 0.94 & 0.87  & 0.12  & &
1.29 & 0.25  & 0.59      &6.37  \\
 "        & NOT 96 & 0.43 & T & "         & F & -    & 0.88  & -     & &  2.06      & 0.50  &   0.99    &  6.60\\
 "        & NOT 98 & 0.49 & T & "         & F & -    & 0.78  & -     & & 1.38       & 0.60  &   0.96    & 6.58 \\
 "        & NOT 98 & 0.54 & T & "         & F &  -   & 0.84  &  -    & & 1.84       & 0.54  &   0.98    &  6.59\\
 "        & McD 98 & 0.67 &H  & "         & A & 1.03 & 0.78  & 0.23  & & 1.29       & 0.26  & 0.76      & 6.48 \\
 "        &   "    &  "   &C  & "         & A & 1.15 & 0.89  & 0.14  & & 1.28       & 0.22  & 0.67      & 6.43 \\
 "        & McD 98 & 0.92 &H  & "         & A & 1.04 &0.81   & 0.22  & & 1.12       & 0.29  & 0.67      & 6.43 \\
 "        &    "   &   "  &C  & "         & A & 1.15 &0.89   & 0.14  & & 1.21       & 0.26  & 0.70      & 6.45 \\
 IL Hya   & NOT 98 & 0.02 & C & 0.03/0.97 & A & 1.63 & 0.63  & 0.63  & &  1.32 & 0.41 & 0.63 & 6.23 \\
 FG UMa   & NOT 98 & -    & - &     -     & A & 1.37 & 0.73  & 0.39  & &        1.29& 0.51  & 0.81  & 6.56 \\
 "        & NOT 98 & -    & - &     -     & A & 1.05 &  0.80 & 0.22  & &       
1.85& 0.60  & 1.03      & 6.66 \\
 LR Hya   & NOT 98 & 0.58 & H & 0.50 & A & 1.45 & 0.61 & 0.59 & &
 1.36 & 0.07 & 0.14 & 5.87 \\ 
    "      &   "    &  "   & C & 0.50 & A & 1.73 & 0.67 & 0.60 & &
 1.04 & 0.09 & 0.22 & 6.07 \\
 HU Vir   & NOT 98 & 0.29 & - & - & E & 2.59 & 1.43  & 1.34  & &
2.07& 1.12  & 2.47      & 6.71 \\
 "        & NOT 98 & 0.38 & - & - & E & 2.21 & 1.31  & 0.71  & &
1.80& 0.96  & 1.91      & 6.60 \\
 "        & McD 98 & 0.44 & - & - & E & 2.81 & 1.23  & 0.63  & &
2.09& 0.85  & 1.95      & 6.61 \\
 "        & McD 98 & 0.54 & - & - & E & 2.10 & 1.25  & 0.54  & &
2.02& 0.79  & 1.66      & 6.54 \\
 "        & McD 98 & 0.64 & - & - & E & 2.93 & 1.22  & 0.59  & &
2.18& 0.81  & 1.82      & 6.58 \\
 "        & McD 98 & 0.73 & - & - & E & 3.57 & 1.09  & 0.25  & &
1.90& 0.68  &  1.41     & 6.47 \\
 "        & McD 98 & 0.83 & - & - & E & 3.28 & 1.11  & 0.26  & &
1.99& 0.68  &  1.41     & 6.47 \\
 "        & McD 98 & 0.93 & - & - & E & 1.19 & 1.13  & 0.15  & &
1.91& 0.65  & 1.31      & 6.44 \\
 "        & McD 98 & 0.02 & - & - & E & 1.70 & 1.20  & 0.30  & &
2.01& 0.71  & 1.58      & 6.52 \\
 "        & McD 98 & 0.12 & - & - & E & 2.37 & 1.28  & 0.61  & &
2.21& 0.80  & 1.89      & 6.60 \\
 DK Dra   & NOT 96 & 0.90 & T & 0.50/0.50 & A & 2.07 & 0.64  & 0.74  & &
1.87& 0.32  & 0.57  & 6.16 \\
 "        & NOT 98 & 0.84 & H &   0.50    & A & 1.88 & 0.68 & 0.64& &
 1.24 & 0.18 & 0.43 & 6.04 \\
 "        &   "   &   "  & C &   0.50    & A & 1.17 & 0.81 & 0.21 & &
 1.23 & 0.25 & 0.66 & 6.22 \\
 BQ CVn   & NOT 98 & -    & C & 0.20/0.80 & A & 1.94 & 0.58  & 0.77  & &
1.39& 0.34  & 0.61      & 6.25 \\
 IS Vir   & NOT 98 & -    & - &    -      & A & 1.50 & 0.65  & 0.58  & &
1.24& 0.47  & 0.70      & 6.07 \\
 BL CVn   & NOT 98 & 0.21 & C &  0.07/0.93& A & 2.10 & 0.55  & 0.98  & &
1.03& 0.24  & 0.30      & 5.94 \\
 BM CVn   & NOT 98 & 0.26 & - &     -     & A & 1.55 & 0.76  & 0.35  & &
1.30& 0.60  & 0.89      & 6.35 \\
 MS Ser   & CAHA95 & 0.93 & P & 0.15/0.85 & F & -  & 0.95 & - & & 
1.22 & 0.63 & 1.01 & 6.44 \\ 
  "       & NOT 98 & 0.21 & P &    "      & A & 0.77 &0.90   & 0.08  & &
1.33& 0.55  & 1.02      &6.44  \\
 "        & NOT 98 & 0.54 & P &   "       & A & 1.00 & 0.85  & 0.16  & &
1.52& 0.57  & 1.13      & 6.48 \\
\noalign{\smallskip}
\hline
\noalign{\smallskip}
\end{tabular}
\end{flushleft}

{\scriptsize $^{*}$ EW  corrected for the contribution
of each component to the total continuum}
\end{table*}

\begin{table*}
\caption[]{Parameters of the broad and narrow Gaussian components
used in the fit of the H$\alpha$ subtracted spectra 
\label{tab:ha_nb}}
\begin{flushleft}
\scriptsize
\begin{tabular}{lccccccccccc}
\noalign{\smallskip}
\hline
\noalign{\smallskip}
\noalign{\smallskip}
   &     &    &
\multicolumn{4}{c}{H$\alpha$ broad component} &\ &
\multicolumn{4}{c}{H$\alpha$ narrow component} \\
\cline{4-7}\cline{9-12}
\noalign{\smallskip}
 {Name} & Obs.  & {$\varphi$} &
{I} & FWHM & EW$_{\rm B}$ & EW$_{\rm B}$/EW$_{\rm T}$ & &
{I} & FWHM & EW$_{\rm N}$ & EW$_{\rm N}$/EW$_{\rm T}$ \\
        &       &             &
    & {\scriptsize (\AA)} & {\scriptsize (\AA)} & (\%) & &
    & {\scriptsize (\AA)} & {\scriptsize (\AA)} & (\%) \\
\noalign{\smallskip}
\hline
\noalign{\smallskip}
HU Vir  & NOT 98 & 0.29 & 0.302 & 3.643 & 1.172 & 47.5 & & 
0.787 & 1.545 & 1.294 & 52.5 \\
"       & NOT 98 & 0.38 & 0.431 & 2.434 & 1.117 & 58.8 & &             
0.566 & 1.297 & 0.782 & 41.2 \\
"       & McD 98 & 0.44 & 0.275 & 3.848 & 1.128 & 57.9 & &
0.589 & 1.303 & 0.818 & 42.1 \\
"       & McD 98 & 0.54 & 0.223 & 2.296 & 0.544 & 32.2 & &
0.724 & 1.486 & 1.145 & 67.8 \\
 "      & McD 98 & 0.64 & 0.351 & 3.283 & 1.226 & 66.3 & &
0.464 & 1.257 & 0.621 & 33.7    \\
 "      & McD 98 & 0.73 & 0.267 & 2.993 & 0.849 & 59.9 & &
0.436 & 1.223 & 0.568 & 40.1    \\
 "      & McD 98 & 0.83 & 0.231 & 3.085 & 0.759 & 53.3 & &
0.459 & 1.359 & 0.664 & 46.7 \\ 
 "      & McD 98 & 0.93 & 0.223 & 3.017 & 0.717 & 54.1 & &
0.437 & 1.304 & 0.606 & 45.9 \\
 "      & McD 98 & 0.02 & 0.191 & 3.675 & 0.748 & 47.1 & &
0.536 & 1.469 & 0.839 & 52.9 \\
 "      & McD 98 & 0.12 & 0.271 & 3.760 & 1.085 & 56.3 & &
0.552 & 1.434 & 0.844 & 43.7 \\ 
\noalign{\smallskip}
\hline
\noalign{\smallskip}
\end{tabular}

\end{flushleft}
\end{table*}
 
\begin{table*}
\caption[]{H$\beta$ line measurements in the observed and subtracted spectra
\label{tab:hb} }
\begin{flushleft}
\scriptsize
\begin{tabular}{ l c c c c c c c c c c c c c c c}
\noalign{\smallskip}
\hline
\noalign{\smallskip}
   &  &  &  &  \multicolumn{3}{c}{Observed H$\beta$ Spectrum} &\ &
\multicolumn{3}{c}{Subtracted H$\beta$ Spectrum} & & \\
\cline{5-7}\cline{9-11}
\noalign{\smallskip}
 {Name} & {Obs.} & {$\varphi$} & E &
 { W$_{\rm obs}$ } & { R$_{\rm c}$ } & {EW} &  &
 { W$_{\rm sub}$ } & { I} & { EW$^{*}$ } &
$\frac{\rm EW(H\alpha)}{\rm EW(H\beta)}$ &
$\frac{\rm E_{H\alpha}}{\rm E_{H\beta}}$ \\
  &   &  &   &  {\scriptsize (\AA) } &  & {\scriptsize (\AA) } & &
 {\scriptsize (\AA)} &  & {\scriptsize  (\AA)} &  &  \\
\noalign{\smallskip}
\hline
\noalign{\smallskip}
 &       &  \\
UX Ari   & NOT 96 & 0.95 & C & 1.02 & 0.57 & 0.48 &  &
1.23 & 0.29 & 0.54 & 3.94 & 4.24 \\
12 Cam   & NOT 98 & 0.64 & - &  1.39& 0.24  & 1.15  &  &
0.42 & 0.12  & 0.06  & 5.67 & 7.90 \\
V1149 Ori& NOT 98 & 0.89 & - & 1.70 & 0.31  & 1.24  &  &
0.51 & 0.21  & 0.13  & 4.46 & 7.14 \\
OU Gem   & NOT 96 & 0.80 & H  & 1.09 & 0.48& 0.62 &  &
0.73& 0.16 & 0.20&3.05 & 3.84 \\
   "     &   "    &  "   & C  & 0.45 & 0.74 & 0.12  &  &
0.33& 0.18  & 0.20 &5.15 & 9.03 \\
"        & NOT 98 & 0.19 &H& 1.03& 0.42  & 0.66 &  &
0.62 & 0.12 & 0.13  & 3.92& 4.93 \\
"        &   "    &  "   &C& 0.44& 0.90  & 0.05 &  &
0.93 & 0.10 & 0.30  & 2.33& 4.09 \\
"        & NOT 98 & 0.77 &H& 1.07& 0.47  & 0.61 &  &
0.78 &0.13& 0.16 &2.87  & 3.62 \\
"        &   "    &   "  &C& 0.36& 0.80  & 0.08 &  &
0.57 &0.16& 0.33 &2.82  & 4.95 \\
"        & INT 99 & 0.70 &H& 0.89&0.64  & 0.34 &  &
0.75& 0.12 & 0.14 & 4.79& 6.03\\ 
"        &   "    &   "  &C& 0.34&0.90  & 0.04 &  &
0.23& 0.06 & 0.10 & 5.00& 8.78\\
$\sigma$ Gem& NOT 96 & 0.81&  - & 1.27 & 0.24  & 1.03  &  &
0.63 & 0.07  & 0.04  & 4.50 & 7.20 \\
 "          & NOT 98 & 0.88 & - & 1.33  & 0.28  & 1.00  &  &
0.59 & 0.10  & 0.07  & 4.86 & 7.78 \\
BF Lyn   & NOT 98 & 0.02 & T & -    &  0.62 &  -    &  &
1.23 & 0.31  & 0.36  & 2.75 & 3.31 \\
"        & INT 99 & 0.11 &H  & -    & 0.40  & -     &  &
1.16 & 0.13  & 0.36      & 2.53      & 3.04      \\
"        &   "    &   "  &C  & -    & 0.89  & -     &  &
0.76 & 0.22  & 0.35      & 2.48      & 2.98      \\
"        & NOT 96 & 0.43 & T &  -   &  0.69 & -     &  &
1.71 & 0.31  & 0.57      & 1.74 & 2.09 \\
"        & NOT 98 & 0.49 & T &   -  & 0.56  & -     &  &
0.77 &0.39   & 0.35 & 2.74  & 3.29 \\  
"        & NOT 98 & 0.54 & T &  -   & 0.61  & -     &  &
1.05 & 0.40  & 0.39 & 2.51  & 3.02 \\
IL Hya   & NOT 98 & 0.02 & C & 1.27 & 0.35  & 0.87  &  &
0.68 & 0.17  & 0.13  & 4.84 & 6.26 \\
FG UMa   & NOT 98 & -    & - & 1.42 & 0.35  & 0.98  &  &
0.39 & 0.15  & 0.10  & 8.10  & 9.00   \\
"        & NOT 98 & -    & - & 1.28 & 0.38  & 0.85  &  &
0.59 & 0.19  & 0.14  & 7.36 & 8.17 \\
LR Hya   & NOT 98 & 0.58 & H & 1.42 & 0.56  & 0.63 &  &
  -      &   0.00    &  0.00 &   -  &    - \\
  "      &   "    &  "   & C & 1.25 & 0.59  & 0.57 &  &
 -  &  0.00      &  0.00  &    -  &    -  \\
HU Vir   & NOT 98 & 0.29 & - & 1.27 & 0.73  & 0.32  &  &
1.19 & 0.49  & 0.60  & 4.12 & 5.58 \\  
"        & NOT 98 & 0.38 & - & 1.57 & 0.77  & 0.37  &  &
1.08 & 0.46  & 0.52  & 3.67 & 4.95 \\
DK Dra   & NOT 96 & 0.90 & T  & 1.27 & 0.45  & 0.69  &  &
1.36 & 0.16  & 0.25  & 2.29 & 3.31 \\
"        & NOT 98 & 0.84 & H& 1.69 & 0.44  & 0.85 &  &
 - &   0.00&  0.00 &   -  &   -   \\
"        &     "  &   "  & C& 0.60 & 0.60   & 0.29  &  &
0.22& 0.12 & 0.07 & 9.43& 13.0 \\
BQ CVn   & NOT 98 & -    & C & 1.32 & 0.40  & 0.83  &  &
0.96 & 0.14  & 0.18  & 3.89 & 4.55 \\
IS Vir   & NOT 98 & -    & - & 1.27 & 0.31  & 0.97  &  &
0.60 & 0.20  & 0.13  & 5.38 & 8.93 \\
BL CVn   & NOT 98 & 0.21 & C & 1.49 & 0.34  & 1.04  &  &
 -   & 0.00  &  0.00 &  -   &   -  \\
BM CVn   & NOT 98 & 0.26 & - & 1.28 & 0.39  & 0.85  &  &
0.67 & 0.27  & 0.20  &  4.45& 6.67 \\
MS Ser   & CAHA95 & 0.93 & P & 1.45 & 0.55 & 0.59 & &
0.95 & 0.29 & 0.33 & 3.06 & 4.04 \\
   "     & NOT 98 & 0.21 & P &1.22  & 0.60  & 0.50  &  &
0.80 & 0.37  & 0.40  & 2.55 & 3.36 \\
"        & NOT 98 & 0.54 & P  & 1.42 & 0.50  & 0.72  &  &
0.79 & 0.29  & 0.29  & 3.90 & 5.14 \\
\hline
\noalign{\smallskip}
\end{tabular}
\end{flushleft}

{\scriptsize $^{*}$ EW  corrected for the contribution
of each component to the total continuum}
\end{table*}

\begin{table*}
\caption[]{Ca~{\sc ii} H \& K and H$\epsilon$ lines measures in
the observed and subtracted spectra
\label{tab:hyk} }
\begin{flushleft}
\scriptsize
\begin{tabular}{l c c c c c c c c c c c c c c}
\hline
\noalign{\smallskip}
 &  &  &   & \multicolumn{3}{c}{Reconstruction} &\ &
\multicolumn{3}{c}{Spectral subtraction} &\ &
\multicolumn{3}{c}{Absolute flux} \\
\cline{5-7}\cline{9-11}\cline{13-15}
\noalign{\smallskip}
\noalign{\smallskip}
 Name & Obs.  & $\varphi$ & E &
EW & EW & EW & &
EW$^{*}$ & EW$^{*}$ & EW$^{*}$ & &
$\log{\rm F}$ & $\log{\rm F}$ & $\log{\rm F}$  \\
&  & & & (K) & (H) & (H$\epsilon$)
& & (K) & (H) & (H$\epsilon$) & &
  (K) & (H) & (H$\epsilon$) \\
\noalign{\smallskip}
\hline
\noalign{\smallskip}
UX Ari   & NOT 96 & 0.95 & C & 1.61 & 1.34  & 0.12 & & 2.76  & 2.68  & 0.44 & &
6.92 & 6.90 & 6.12 \\
12 Cam   & NOT 98 & 0.64 & - & 1.52 & 1.24 & - & & 1.98  & 1.75  &   -   & &
6.45 & 6.40 &   -  \\
V1149 Ori& NOT 98 & 0.89 & - &   1.96& 1.51  & - & & 2.45  & 2.33  &   -   & &
6.15 & 6.12 &   -  \\
OU Gem   & NOT 96 & 0.80 &H&0.48 &0.35 & 0.07 & &
0.77 & 0.54 & 0.10 & & 6.02& 5.87 & 5.13    \\
  "      &    "   &   "  &C&0.31 &0.19 & -  & &
1.07 & 0.70 & - & & 5.68& 5.50 & -    \\
"        & NOT 98 & 0.19 &H&0.63 &0.52 & -   & & 1.06 & 0.94 & - & &
6.16&6.11 &   -                            \\
"        &   "    &   "  &C&0.45 &0.44 &0.16& & 1.60 & 1.70 & 0.70 & &
5.86&5.88 & 5.50\\
"        & NOT 98 & 0.77 &H&0.65  & 0.66 &0.07& & 1.08 &1.08&  0.31& &
 6.17&6.17  & 5.63 \\
"        &   "    &  "   &C&0.49  &0.38 &  -   & & 1.83 &1.43& -  & &
 5.92&5.81  & -  \\
$\sigma$ Gem& NOT 96 & 0.81& - & 1.35& 1.14 & 0.07 & & 1.53  & 1.35  &  -    & &
5.88 & 5.82 &  -   \\
 "          & NOT 98 & 0.88& -  & 1.32& 1.20 & 0.02 & & 1.83  & 1.87  &  -    & &
5.96 & 5.97 &  -   \\
BF Lyn   & NOT 98 & 0.02 &H   & 1.06 &0.94  &0.19& &
2.38      & 2.10      & 0.56      & &
6.74 &6.68  & 6.11 \\
  "      &    "   &   "  &C   & 0.93 &0.94  &0.16& &
2.10      & 2.08      & 0.48      & &
6.68 &6.68  & 6.04 \\
"        & NOT 96 & 0.43 &H   & 1.12 & 0.96  &0.13& &
2.64      & 2.52      & 0.84      & &
6.78 & 6.76 & 6.29 \\
"        &   "    &   "  &C   & 1.04 & 0.89  &0.25 & &
2.48      & 2.30      & 0.96      & &
6.76 & 6.72 & 6.34 \\
"        & NOT 98 & 0.49 & T & 1.89 & 1.75 & 0.30 & &
3.96      & 3.72      & 0.88      & &
6.96 & 6.93 & 6.31 \\
"        & NOT 98 & 0.54 &H   & 0.95 & 0.83  &0.19 & &
2.20      & 2.06      & 0.50      & &
6.70 &6.68  & 6.06 \\
"        &    "   &   "  &C   & 0.97 & 0.79  &0.15& &
2.20      & 1.96      &   0.52    & &
6.70 &6.65  & 6.08 \\
IL Hya   & NOT 98 & 0.02 & C & 1.79& 1.60 & -  & & 1.94  & 1.91  &   -   & &
6.38 & 6.38 &   -  \\
FG UMa   & NOT 98 & -    & - & 1.55&  1.35 & - & & 1.95  & 1.77  &   -   & &
6.93 &  6.89&  -   \\
"        & NOT 98 & -    & - & 1.68& 1.35 & - & & 2.05  & 1.82  &   -   & &
6.96 & 6.90 &   -  \\
LR Hya   & NOT 98 & 0.58 & H & 0.29& 0.27& -    & &
1.06 & 1.04  &  -    & & 6.67 & 6.66 & - \\
  "      &   "    &   "  & C & 0.29& 0.24 & -  & &
1.06 & 0.88  &  -   & & 6.67 & 6.59 & - \\
HU Vir   & NOT 98 & 0.29 & - & 2.67& 2.68 &0.52 & & 2.75  & 2.96  & 0.62  & &
6.44 & 6.48 & 5.80 \\
"        & NOT 98 & 0.38 & - & 3.34& 2.90 &0.49 & & 3.78  & 3.00  & 0.96  & &
6.58 & 6.48 & 5.99 \\
DK Dra   & NOT 96 & 0.90 & T & 0.79& 0.37 & -  & & 1.41  & 1.19  &   -   & &
6.18 &  6.11&  -   \\
"        & NOT 98 & 0.84 & H& 1.08& 0.83  & - & &
2.62 & 1.92  &  -    & & 6.45  & 6.32 & - \\
"        &   "    &   "  & C& 1.04& 0.69  & - & &
2.76 & 1.80  &  -   & & 6.48 & 6.29 & - \\
BQ CVn   & NOT 98 & -    & C & 1.30& 1.14 &0.04 & & 1.69  & 1.65  &  0.16 & &
6.39 & 6.38 & 5.36 \\
IS Vir   & NOT 98 & -    & - &  1.07&1.05 & - & &  2.10 & 2.05  &  -    & &
5.99 & 5.97 &   -  \\
BL CVn   & NOT 98 & 0.21 & C & 1.28& 1.21 & - & & 1.85  & 1.77  &  -    & &
6.43 & 6.41 &  -   \\
BM CVn   & NOT 98 & 0.26 & - & 3.08& 2.45 &0.17 & & 3.25  & 2.98  &   -   & &
6.55 & 6.51 &    - \\
MS Ser   & NOT 98 & 0.21 & P & 2.05& 1.83 & - & & 2.59  & 2.74  & -     & &
 6.70& 6.66 &  -   \\
  "      &  "   &  "   &S& 0.13 & 0.33 & - & &  2.65 & 3.64 & - &   &
7.24 & 7.38 &   -  \\
"        & NOT 98 & 0.54 & T & 1.95&1.93 & 0.16  & & 2.20  & 2.31  & 0.41  & &
6.56 & 6.58 & 5.83 \\
\noalign{\smallskip}
\hline
\end{tabular}
\end{flushleft}

{\scriptsize $^{*}$ EW  corrected for the contribution
of each component to the total continuum}
\end{table*}


\begin{table*}
\caption[]{Ca~{\sc ii} IRT lines
measures in the observed and subtracted spectra
\label{tab:cairt} }
\begin{flushleft}
\scriptsize
\begin{tabular}{l c c c c c c c c c c c c c c}
\hline
\noalign{\smallskip}
 & &  &     & \multicolumn{3}{c}{Reconstruction} &\ &
\multicolumn{3}{c}{Spectral subtraction} & &
\multicolumn{3}{c}{Absolute flux} \\
\cline{5-7}\cline{9-11}\cline{13-15}
\noalign{\smallskip}
\noalign{\smallskip}
 Name & Obs.  & $\varphi$ & E &
EW & EW & EW & &
EW$^{*}$ & EW$^{*}$ & EW$^{*}$ &  
$\frac{\rm EW(\lambda8542)}{\rm EW(\lambda8498)}$ &  
$\log{\rm F}$ & $\log{\rm F}$ & $\log{\rm F}$\\
 & & & &
$\lambda$$8498$ & $\lambda$$8542$ & $\lambda$$8662$ & &
$\lambda$$8498$ & $\lambda$$8542$ & $\lambda$$8662$ & &    
$\lambda$$8498$ & $\lambda$$8542$ & $\lambda$$8662$  \\
\noalign{\smallskip}
\hline
\noalign{\smallskip}
UX Ari   & NOT 96 & 0.95 & C &   -   & - & 0.31  & &  -    &  -    &  1.33 & - &
   -  &  -   & 6.63  \\
12 Cam   & NOT 98 & 0.64 & - &   -   & - &  -    & & -     & 0.64  & 0.47  & - &
  -   &   6.28 & 6.14\\
V1149 Ori& NOT 98 & 0.89 & - &  -    & - &  -    & &  -    & 0.83  &  0.71 & - &
   -  &6.27  &  6.21 \\
OU Gem   & NOT 96 & 0.80 &H& - & - & 0.02 & & - & - & 0.56 & - &
 -    &  -   & 6.16 \\
  "     &   "    &   "  &C& - & - & 0.07 & & - & - & 0.30 & - &
 -    &  -   & 5.76 \\
"        & NOT 98 & 0.19 &H& - & 0.06 & - & & - & 0.53 & 
0.40  & - & -    & 6.14& 6.02 \\
"        &   "    &  "   &C& - & 0.14 & 0.16 & & - & 0.37 &
0.20  & - & -    & 5.85& 5.59 \\
"        & NOT 98 & 0.77 &H& - &0.05 & - & & - & 0.58 & 0.50 & - &
 -    & 6.18& 6.11 \\ 
"        &  "     &  "   &C& - &0.13 & 0.10 & & - & 0.63 & 0.27 & -  &
 -    & 6.09& 5.72 \\
"        & INT 99 & 0.70 &H& - & 0.02 & - & & 0.58& 0.96  & 1.01 & 1.66 &
6.18&6.40 & 6.42 \\
"        &   "    &  "   &C& - & 0.06 & 0.07 & & 0.27& 0.53  & 0.50 & 1.96 &
5.72&6.02 & 5.99 \\
$\sigma$ Gem& NOT 96 & 0.81& - &  -   & - &  -    & &  -    &   -   &   0.38& - &
 -    &    - &  5.93 \\
 "          & NOT 98 & 0.88& - &  -   & - &   -   & &  -    &  0.84 & 0.59  & - &
  -   & 6.26 &   6.11\\
BF Lyn   & NOT 98 & 0.02 &H   & -    &0.14&0.06       & &
 -         & 0.99      & 0.74       & - &
-     &6.47  & 6.34  \\
    "    &   "    &   "  &C   & -    &0.10&0.08       & &
 -         & 0.93      & 0.70       & - &
-     &6.45  & 6.32  \\
"        & INT 99 & 0.11 &H   & 0.16 &0.12&0.10  & &
 0.74      & 0.91      & 0.74      & 1.23 &
6.35  &6.43  & 6.34  \\
"        &   "    &   "  &C   & 0.16 &0.17&0.14  & &
 0.39      & 0.58      & 0.37      & 1.49 &
6.07  &6.24  & 6.04  \\
"        & McD 98 & 0.12 &H   & -    &0.11&0.08  & &
    0.29   & 0.42      & 0.33      & 1.45 & 
 5.94 &6.10  & 5.99  \\
"        &   "    &   "  &C   & -    &0.11&0.08  & &
    0.52   & 0.49      & 0.66      & 0.94 &
 6.19 &6.17  & 6.30  \\
"        & McD 98 & 0.16 &H   & -    &0.12&0.09  & &
    0.27   & 0.34      & 0.33      & 1.26 &
 5.91 &6.01  &5.99   \\
"        &   "    &   "  &C   & -    &0.18&0.10  & &
    0.45   & 0.45      & 0.66      & 1.00 &
 6.13 &6.13  &6.30   \\
"        & McD 98 & 0.18 &H   & -    &0.17 &0.08 & &
    0.28   & 0.38      & 0.31      & 1.36 &
 5.92 &6.06  & 5.97  \\
"        &   "    &   "  &C   & -    &0.16 &0.07 & &
    0.55   & 0.68      & 0.77      & 1.24 &
 6.21 &6.31  & 6.36  \\
"        & McD 98 & 0.40 &H   & -    &0.10& -    & &
     0.32  & 0.43      & 0.57      & 1.34 &
  5.98& 6.11 & 6.23  \\
"        &   "    &   "  &C   & -    &0.09& -    & &
     0.57  & 0.44      & 0.62      & 0.77 &
  6.23& 6.12 & 6.27  \\
"        & NOT 96 & 0.43 &H   &  -   & - & 0.10  & &
     -     & -         & 0.76      & - &
  -   &  -   &   6.36\\
"        &   "    &   "  &C   &  -   & - & 0.12  & &
     -     & -         & 0.64      & - &
  -   &  -   &   6.28\\
"        & NOT 98 & 0.49 & T  &  -   &0.26&0.22  & &
       -   & 0.82      & 0.63      & - &
 -    & 6.39 & 6.27  \\
"        & NOT 98 & 0.54 &H   & -    &0.14&0.09  & &
    -      & 0.83      & 0.72      & - &
   -  &6.39  & 6.33  \\
"        &   "    &   "  &C   & -    &0.15&0.11  & &
    -      & 0.90      & 0.73      & - &
   -  &6.43  & 6.34  \\
"        & McD 98 & 0.67 &H   & -    &0.11&0.06  & &
     0.29  & 0.38      & 0.36      & 1.31 &
  5.94& 6.06 & 6.03  \\
"        &   "    &   "  &C   & -    &0.13&0.08  & &
     0.57  & 0.70      & 0.48      & 1.23 &
  6.23& 6.32 & 6.16  \\
"        & McD 98 & 0.92 &H   &   -  &0.07&0.11  & &
      0.39 & 0.38      & 0.42      & 0.97 &
 6.07 & 6.05 & 6.10  \\
"        &   "    &   "  &C   &   -  &0.12&0.13  & &
      0.53 & 0.62      & 0.58      & 1.17 &
 6.20 & 6.27 & 6.24  \\
IL Hya   & NOT 98 & 0.02 & C & - & 0.15 & 0.05 & & - & 1.01  & 0.79  & - &
   -  & 6.44 & 6.34  \\
FG UMa   & NOT 98 & -    & - & - & 0.04  & 0.01 & & - & 0.84  & 0.55  & - &
   -  &  6.48& 6.29  \\
"        & NOT 98 & -    & - &  - & - & -  & & - & 0.64  & 0.55  & - &
  -   & 6.36 & 6.29  \\
LR Hya   & NOT 98 & 0.58 & H & - & - & - & & - & 0.60  &0.52& - &
   -   & 6.33& 6.27  \\
  "      &   "    &  "   & C & - & - & - & & - & 0.80  & 0.58& - &
   -   & 6.45 & 6.32 \\
HU Vir   & NOT 98 & 0.29 & - &   -   & 0.32 & 0.19   & & -     &  1.59 & 1.46  & - & -    & 6.58 & 6.55  \\
"        & NOT 98 & 0.38 & - &    -  & 0.25  & 0.14  & &  -    &  1.19 & 1.04  & - &  -  & 6.46 & 6.40  \\
"        & McD 98 & 0.44 & - &  0.32 & 0.30  & 0.22  & & 0.83  &  1.41 & 0.65  & 1.70 & 6.30& 6.53 & 6.19  \\
"        & McD 98 & 0.64 & - &  0.24 & 0.13  & 0.20  & & 0.66  & 1.10  &  0.88 & 1.67 & 6.20& 6.42 & 6.32  \\
"        & McD 98 & 0.73 & - &  0.38 & 0.17  & 0.32  & &  0.64 &  0.93 &  0.71 & 1.45 & 6.19& 6.35 & 6.23  \\
"        & McD 98 & 0.83 & - &  0.15 & 0.12  & 0.12  & &  0.62 & 0.89  &  0.81 & 1.43 & 6.17&  6.33& 6.29  \\
"        & McD 98 & 0.93 & - &  0.09 & 0.18  & 0.14  & &  0.63 & 0.89  &  0.57 & 1.41 & 6.18 & 6.33 & 6.14  \\
"        & McD 98 & 0.02 & - &  0.19 & 0.27  & 0.18  & &   0.63& 1.13  & 0.60  & 1.79 & 6.18 & 6.43 & 6.16  \\
"        & McD 98 & 0.12 & - &  0.35 & 0.18  & 0.09  & &0.71   & 1.10  & 0.47  & 1.55 & 6.23 & 6.42 & 6.05  \\
DK Dra   & NOT 96 & 0.90 & T & -     & - & -     & & -     &  -    & 1.12  & - &
  -   &  -   &   6.47\\
"        & NOT 98 & 0.84 & H  & - & -  &  - & & -  & 0.65 & 0.47& - &
   -  &  6.24& 6.09  \\
"        &   "    &   "  & C  & - & -  &  - & & -  & 1.59 & 0.98& - &
   -  &  6.63& 6.41  \\
BQ CVn   & NOT 98 & -    & C & - & 0.08& 0.03 & & -     & 1.11  & 0.85  & - &
   -  & 6.50 &  6.38 \\
IS Vir   & NOT 98 & -    & - &  -    & - &  -    & & -     &  0.92 & 0.70  & - &
   -  & 6.30 & 6.18  \\
BL CVn   & NOT 98 & 0.21 & C & -     & - & -     & & -     &  0.52 & 0.30  & - &
   -  &6.17  & 5.93  \\
BM CVn   & NOT 98 & 0.26 & - &  - & 0.14  & 0.04 & & - & 0.86    & 0.75      & - &
    - & 6.36 &  6.30 \\
MS Ser   & NOT 98 & 0.21 & P & -  & 0.32 & 0.26  & &  -    & 0.96  & 0.95  & - &
 -    & 6.42 &  6.42 \\
  "      &   "    &   "  & S & - & - & - & & - & 0.47 & 0.53 & - & 
 - & 6.11 & 6.16 \\
"        & NOT 98 & 0.54 & T & - & 0.30  & 0.21  & &  -    & 1.05  & 0.75  & - &
 -    & 6.46 & 6.31  \\
\noalign{\smallskip}
\hline
\end{tabular}
\end{flushleft}

{\scriptsize $^{*}$ EW  corrected for the contribution
of each component to the total continuum}
\end{table*}

\section{Spectroscopic features analysed}

\subsection{Chromospheric activity indicators}

The echelle spectra analysed in this paper allow us to study
the behaviour of the different optical
chromospheric activity indicators formed at different atmospheric heights:
Na~{\sc i} D$_{1}$, D$_{2}$ and Mg~{\sc i} b triplet
(upper photosphere and lower chromosphere),
Ca~{\sc ii} IRT lines (lower chromosphere),
H$\alpha$, H$\beta$, Ca~{\sc ii} H \& K (middle chromosphere) and
He~{\sc i} D$_{3}$ (upper chromosphere).
The chromospheric contribution in
these features has been determined  using
the spectral subtraction technique described in
detail by Montes et al. (1995a, b, c), (see Paper I and II).
The synthesized spectrum was constructed using the program STARMOD
developed at Penn State (Barden \cite{B85}).
The inactive stars used as reference stars in the spectral subtraction 
were observed during the same observing run as the active stars
or were taken from our libraries of late-type stars (see Montes 1998).
We have determined the excess emission equivalent width (EW) (measured in the 
subtracted spectra) and converted to absolute chromospheric flux at 
the stellar surface.
We have estimated the errors in the measured EW taking into account  
the typical internal precisions of STARMOD 
(0.5 - 2 km s$^{-1}$ in velocity shifts,
$\pm$5 km s$^{-1}$ in  V$\sin{\it i}$,
and 5\% in intensity weights), 
the rms obtained in the fit between observed and
synthesized spectra in the regions outside the chromospheric features 
(typically in the range 0.01-0.03)
 and the standard deviations resulting in the
EW measurements. The estimated errors are in the range 10-20\%. 
For low active stars errors are larger and 
we have considered as a clear detection of excess emission or absorption
in the chromospheric lines  only
when these features in the difference spectrum are
larger than 3~$\sigma$.

Table~\ref{tab:ha} gives the H$\alpha$ line parameters, measured in the
observed and subtracted spectra of the sample.
The column (3) gives the orbital phase ($\varphi$)
for each spectrum.     
In the column (4), H and C mean emission from the hot and
cool components, respectively, and T means that at these phases
the spectral features cannot be deblended. The
column (5) gives the relative contribution of the hot and cool components
to the total continuum (S$_{\rm H}$ and S$_{\rm C}$), respectively. The
column (6) describes the observed H$\alpha$ profile, i.e. whether the
line is in absorption (A), in emission (E) or totally filled-in by
emission (F). The
columns (7), (8) and (9) give the following parameters
measured in the observed spectrum:
the full width at half maximum (W$_{\rm obs}$), 
the residual intensity (R$_{\rm c}$ =
 $\frac{\rm F_{c}}{\rm F_{cont}}$)
 and the equivalent width (EW).
The last four columns give the following parameters
measured in the subtracted spectrum:
the full width at half maximum
(W$_{\rm sub}$), the peak emission intensity (I),
 the excess H$\alpha$ emission equivalent width (EW(H$\alpha$)),
corrected for the contribution of the components to the total continuum and
the logarithm of the absolute flux at the stellar surface
(logF$_{\rm S}$(H$\alpha$)) obtained with the calibration 
of Hall (\cite{Hall96}) as a function of (V--R).

In Table~\ref{tab:ha_nb} we list the parameters (I, FWHM, EW)
of the broad and narrow components used in the two Gaussian-component fit  
to the H$\alpha$ subtracted emission profile. We have performed this fit in the
stars that show broad wings. 
 See the comments for each individual star in
Sect.~4 and the interpretation of these components given in Sect.~5.

Table~\ref{tab:hb}
gives the  H$\beta$ line parameters, measured in the
observed and subtracted spectra, as in the case of the H$\alpha$ line.
In this table we also give the ratio of
excess emission EW in the H$\alpha$ and H$\beta$ lines, 
$\frac{\rm EW(H\alpha)}{\rm EW(H\beta)}$, and the ratio
of excess emission $\frac{\rm E_{H\alpha}}{\rm E_{H\beta}}$
with the correction:

\[ \frac{\rm E_{H\alpha}}{\rm E_{H\beta}} =
\frac{\rm EW(H\alpha)}{\rm EW(H\beta)}*0.2444*2.512^{(B-R)}\]
given by Hall \& Ramsey (\cite{H&R92}) that takes into account the absolute flux
density in these lines and the color difference in the components.
We have used this ratio as a diagnostic for
discriminating between the presence of plage-like 
and prominence-like material at the stellar
surface, following the theoretical modelling by Buzasi (\cite{Buz89}) 
who found that low E$_{\rm H\alpha}$/E$_{\rm H\beta}$  ($\approx$~1-2)
can be achieved both in plages and prominences viewed against the disk, but
that high ratios ($\approx$~3-15) can only be achieved in extended regions
viewed off the limb.
The study of chromospherically active binaries 
by Hall \& Ramsey (\cite{H&R92})
has demonstrated the presence of large amounts of extended, prominence-like 
material in these stars.

We also analyse the possible filling-in of the core of
the Na~{\sc i} D$_{1}$ and D$_{2}$ lines as other chromospheric 
activity indicator as well as the behaviour of the He~{\sc i} D$_{3}$ line,
which can be in absorption, filled-in due to
frequent low-level flaring or in emission due to flares
(see Paper I and II; Saar et al. 1997; Montes et al. \cite{M96b}, \cite{M99}).

Table~\ref{tab:hyk} gives
the Ca~{\sc ii} H \& K and H$\epsilon$ lines parameters, measured in the
observed and subtracted spectra.
In columns (5) and (6) we list the EW for the K and H lines,
obtained by reconstruction of the absorption line profile
(described by Fern\'andez-Figueroa et al. \cite{FFMCC94}, hereafter FFMCC).
In columns (7), (8) and (9) we give the EW for
the K, H and H$\epsilon$ lines using the
spectral subtraction technique (explained by Montes et al. 1995c, 1996a)
and corrected for the contribution of the components to the total continuum. 
In columns (10), (11) and (12) we list the corresponding logarithm of
the surface flux
obtained by means of
the linear relationship between the absolute surface
flux at 3950~\AA$\ $ (in erg cm$^{-2}$ s$^{-1}$ \AA$^{-1}$)
and the colour index (V--R) by Pasquini et al. (1988).

\input{ds1878_figs1s.tex}

Table~\ref{tab:cairt} gives
the Ca~{\sc ii} IRT lines parameters, measured in the
observed spectra by reconstruction of the absorption line profile
and using the spectral subtraction. The columns of this table
have the same meaning as in Table~\ref{tab:hyk} 
for the Ca~{\sc ii} H \& K lines.
The absolute fluxes at the stellar surface have been obtained using
the calibration of Hall (\cite{Hall96}) as a function of (V--R).
For the observing runs in which the  $\lambda$8542 and $\lambda$8498 lines
are included we also give the ratio of excess emission EW, 
E$_{\rm 8542}$/E$_{\rm 8498}$, which is also an indicator of the 
type of chromospheric structure that produces the observed emission.
In solar plages, values of  
E$_{\rm 8542}$/E$_{\rm 8498}$ $\approx$~1.5-3 are measured, 
while in solar prominences the values are $\approx$~9, the limit of an optically
thin emitting plasma (Chester \cite{Ch91}).
However, the observations of active stars 
(Chester et al. \cite{Ch94}; Ar\'{e}valo \& L\'{a}zaro \cite{A&L99})
indicate that these lines exhibit markedly different behaviour.
The E$_{\rm 8542}$/E$_{\rm 8498}$ ratios found in these stars are smaller
(closer to the optically thick value of one) than solar plages.
These values indicate that the Ca~{\sc ii} IRT emission arises predominantly 
in chromospheric plages.

\subsection{The Li~{\sc i} $\lambda$6707.8 line}

The resonance doublet of  Li~{\sc i} at $\lambda$6708 \AA\
is an important diagnostic of age in late-type stars
since it is destroyed easily by thermonuclear reactions in the
stellar interior.
It is well-known that a large number of
chromospherically active binaries 
shows Li~{\sc i} abundances higher than
other stars of the same mass and evolutionary
stage (Barrado et al. \cite{Ba97}, \cite{Ba98}; 
Paper I, Montes \& Ramsey \cite{MR98}).
This line is only included in 
our echelle spectra in the McD98 and INT99 observing runs.
In Fig.~\ref{fig:li} we have plotted representative spectra of 
OU Gem, BF Lyn and HU Vir in this spectral region. 
A K1III reference star with
some photospheric lines identified has been also plotted in order to better 
identify the expected position of the Li~{\sc i} line.
It was only possible to measure the equivalent width of the Li~{\sc i} absorption 
line in the SB1 system HU Vir.
In the case of the SB2 systems OU Gem and BF Lyn 
the possible small absorption Li~{\sc i} of one or both components 
is blended with photospheric lines of the other component.

\section{Individual Results}

In this section we describe the behaviour of the above mentioned 
chromospheric activity indicators for each star of the sample.
The profiles of H$\alpha$, H$\beta$, Ca~{\sc ii} H \& K  
and Ca~{\sc ii} IRT are displayed from Fig.~1 to 23.
For each system we have plotted the observed spectrum (solid-line) and the
synthesized spectrum (dashed-line) in the left panel 
and the subtracted spectrum (dotted line) in the right panel.
The name of the star, the observing run (NOT96, NOT98, McD98, INT99) 
and the orbital phase ($\varphi$) of each spectrum 
 are given in every figure.
The He~{\sc i} D$_{3}$ line, for selected stars of the sample, is displayed in 
Fig.~\ref{fig:heid3}.

\subsection{UX Ari (HD 21242)}

This double-lined spectroscopic binary (G5V/K0IV)
is a well known RS CVn system and
extensively studied in the literature (Carlos \& Popper \cite{C&P71};
 Bopp \& Talcott \cite{BoT78}; Huenemoerder et al. 1989; 
Raveendran \& Mohin 1995).
Recently, Duemmler \& Aarum (\cite{D&A00}) have given a new orbit
 determination, which we have adopted in Table~2.
Our previous H$\alpha$ observations (Montes et al. 1995a, b) showed clear
H$\alpha$ emission above the continuum from the cool component. This
emission was superimposed to the weak absorption of the hot component.
Our spectrum in the Ca~{\sc ii} H \& K region (Montes et al. 1995c)
showed strong emission from the cool component
and a weak H$\epsilon$ emission line.
We also detected a flare in this system through
simultaneous H$\alpha$, Na~{\sc i} D$_{1}$, D$_{2}$
and He~{\sc i} D$_{3}$ observations (Montes et al. 1996b; Paper~I).

In the new observation (NOT96) we observe intense emission in the 
Ca~{\sc ii} H \& K, H$\epsilon$, H$\alpha$ and Ca~{\sc ii} $\lambda$8662 
lines and a filled-in absorption H$\beta$ line
 from the cool component (see Fig.~\ref{fig:uxari_all}). 
The excess emission measured in all these activity indicators is larger 
than in our previous observations of this system in quiescent state 
in 1992 and 1995.
The small emission we observe in He~{\sc i} D$_{3}$ 
(Fig.~\ref{fig:heid3}) confirms the high level of activity 
of UX Ari in this observation.
The He~{\sc i} D$_{3}$ line has been observed as clear emission in this system 
in other occasions associated to flare-like events 
(Montes et al. 1996b and references therein).

\subsection{12 Cam (BM Cam, HD 32357)}

This single-lined spectroscopy binary was classified 
by Bidelman (\cite{Bi64}) as a K0 giant. 
He also noticed Ca~{\sc ii} H \& K emission.
Later, Hall et al. (\cite{Hall95}) revised the system spectroscopically and
photometrically and
obtained new values of orbital parameters, given in Table~2.
Eker et al. (\cite{Ek95}) observed that H$\alpha$ profiles showed an asymmetric
shape with a round shoulder in the red wing and a steeper blue wing. They
also confirmed the variable H$\alpha$ filling, which was       
suspected by Strassmeier et al. (1990) too.
In our previous observations of this system (FFMCC)
we found the H$\alpha$ line filled-in and
strong Ca~{\sc ii} H \& K emission.

In the new spectrum (Fig.~\ref{fig:12cam_all}) 
we find strong emission in the Ca~{\sc ii} H \& K lines.
Thanks to the higher resolution of this spectrum it is possible to see 
now that both lines exhibit self-absorption with blue asymmetry. 
Small H$\epsilon$ emission is observed, 
but it was impossible to deblend it from the Ca~{\sc ii} H line.
Both H$\alpha$ and H$\beta$
appear in absorption with a slight filling-in, the H$\alpha$ filling-in
is smaller than the one corresponding to other epochs.
The H$\alpha$ line shows excess absorption in the red wing similar to the
asymmetric shape of H$\alpha$ observed by Eker et al. (\cite{Ek95}).
The application of the spectral subtraction technique
 reveals that the He {\sc i} D$_{3}$
line appears as an absorption feature (Fig.~\ref{fig:heid3}). 
This fact is more frequent in giants than in dwarfs (Paper I). 
Finally, the Ca~{\sc ii} IRT ($\lambda$$\lambda$ 8542, 8662 \AA) 
absorption lines are clearly filled-in.

\subsection{V1149 Ori (HD 37824)}

This single-lined spectroscopic binary, classified as K1III~+ F
by Bidelman \& MacConnell (\cite{BiMcC73}),
was listed as a G5IV star by Hirshfeld \& Sinnott (1982).
Our previous observations
revealed clear excess H$\alpha$ emission (Montes et al. 1995a, b; Paper I),
strong Ca~{\sc ii} H \& K and H$\epsilon$ emission 
(Montes et al. 1995c) and clear absorption in the He~{\sc i} D$_{3}$ line
in the subtracted spectrum (Paper I).

In the new observation (NOT98, Fig.~\ref{fig:v1149ori_all}), 
the H$\alpha$ and H$\beta$ lines show a filled-in absorption profile, 
and clear absorption is observed in the blue wing of both lines.
This excess absorption in the blue wing, not observed in our previous 
observations of this system, could be indicative of variable mass motion. 
To confirm this behaviour we took 
a new spectrum with the ESA-MUSICOS spectrograph in 
January 2000 (forthcoming paper) in 
which the blue wing of the H$\alpha$ line is in emission, 
confirming the high variability of the H$\alpha$ line profile in this system.
Strong emission is observed in the Ca~{\sc ii} H \& K lines 
but the H$\epsilon$ line is not detected.
The Ca~{\sc ii} IRT lines ($\lambda$$\lambda$ 8542, 8662 \AA) 
show a strong filling-in.
The He~{\sc i} D$_{3}$ line appears in absorption (Fig.~\ref{fig:heid3}). 

\subsection {OU Gem (HD 45088)}

OU Gem is a bright (V= 6.79, Strassmeier et al. 1990) 
and nearby (d= 14.7 pc, ESA \cite{ESA97}) BY Dra-type
SB2 system (K3V/ K5V) with an orbital period of 6.99 days 
and a noticeable eccentricity (Griffin \& Emerson \cite{Gr&E75}).
Both components show Ca~{\sc ii} H~\&~K
emission, though the primary shows slightly stronger emission
than the secondary. The H$\alpha$ line is in 
absorption for the primary and filled-in for the secondary
(Bopp \cite{Bo80}; Bopp et al. \cite{Bo&81a}, b; Strassmeier et al. 1990; 
Montes et al. 1995a, b, 1996). 
Dempsey et al. (\cite{De93a}) observed that the Ca~{\sc ii} IRT lines
were filled-in.
This binary was detected by the WFC on board the ROSAT
satellite during the all-sky survey (Pounds et al. 1993; 
Pye et al. 1995). OU Gem has 1.7$\times$10$^{29}$ ergs$^{-1}$
X-ray luminosity, typical value of the BY Dra systems 
(Dempsey et al. \cite{De93b}, \cite{De97}).
The photometric
variability was discovered by Bopp et al. (\cite{Bo&81a}) and they also 
computed a 7.36-day photometric period.
It is interesting that the orbital and rotational
periods differ in 5\% due to the appreciable orbital
eccentricity (e= 0.15), according to Bopp (\cite{Bo80}).
Although BY Dra systems are main-sequence stars, their
evolutionary stage is not clear. 
OU Gem has been listed by Soderblom et al. (1990) 
and Montes~et~al.~(\cite{M00a}, \cite{M00c}) 
as a possible member of the UMa moving group (300~Myr),
indicating that it may be a young star.

{\it The H$\alpha$ line}:
In the observed spectra, we see an absorption line for the 
primary star and a nearly complete filling-in for the secondary star. 
After applying the spectral subtraction technique, clear excess H$\alpha$ 
emission is obtained for the two components, being stronger for the hot one
(see Fig.~\ref{fig:ougem_ha_hb} upper panel).
 The excess H$\alpha$ emission EW is measured
in the subtracted spectrum and corrected for the contribution of the 
components to the total continuum. 
We took one spectrum in this region in Dec-92 (Montes et al. 1995b). 
At the orbital phase of this observation ($\varphi$=~0.48) we could  
not separate the emission from both components and we measured the
 total excess H$\alpha$ emission EW relative to the combined continuum. 
We obtained a similar value to Mar-96, 
Apr-98 and Jan-99 values obtained adding up the excess emission EW
 from the two components.
 
{\it The H$\beta$ line}:
Looking at the observed spectra, we only see the H$\beta$ 
line for the primary, in absorption. After applying the spectral subtraction 
technique small excess H$\beta$ emission is obtained for the two components
(see Fig.~\ref{fig:ougem_ha_hb} lower panel). 
We have obtained, in general,  $\frac{ E_{H\alpha} } { E_{H\beta} }$ 
values larger than three for the two components, 
so the emission can come from prominences.

{\it The Ca~{\sc ii} H $\&$ K and H$\epsilon$ lines}:
We observe that both components of this binary have
 the Ca~{\sc ii} H \& K and H$\epsilon$ lines in emission. 
We can also see that the excess Ca~{\sc ii} H \& K
emission of the hot star is larger than the one of the cool star
(Fig.~\ref{fig:ougem_hyk_cairt} upper panel). 
The measured excess Ca~{\sc ii} H \& K emission of both components is larger
in the two spectra of the NOT98 observing run than in the NOT96 spectrum.  
Overlapping between the H$\epsilon$ line of one star and the 
Ca~{\sc ii}~H line of
the other only allows to see the H$\epsilon$ line of the cool star 
at orbital phase 0.19, and H$\epsilon$ of the hot star otherwise.
%

{\it The Ca~{\sc ii} IRT lines}:
In the observed spectra, we can see that both components of 
OU Gem show the Ca~{\sc ii} IRT lines 
 in emission superimposed to the corresponding absorption. After applying 
the spectral subtraction technique, clear excess emission appears for the two 
components, being clearly stronger for the hot one
(see Fig.~\ref{fig:ougem_hyk_cairt} lower panel).

\subsection{$\sigma$ Gem (HD 62044)}

This single-lined spectroscopic binary belongs to the long-period
group of RS~CVn binary systems.
It was classified as K1III, 
but the radius obtained by Duemmler et al. (\cite{Du97})
seems to be too small for a giant star.
In Table~2 we have adopted the orbital and physical parameters
updated by Duemmler et al. (\cite{Du97}).
Strong and variable Ca~{\sc ii} H \& K emission always centered 
at the absorption
line has been reported by Bopp (\cite{Bo83}), Strassmeier et al. (1990),
FFMCC, Montes \cite{M95}, Montes et al. \cite{M96b}.
Our previous observation in the H$\alpha$ line region 
(Montes et al. 1995a, b) revealed  
 small excess emission, similar to that reported by 
Strassmeier et al. (1990) and Frasca \& Catalano (\cite{F&C94}). 
Variable excess H$\alpha$  emission anti-correlated with spot regions 
has been found by Zhang \& Zhang (1999).

The new observations (NOT96, NOT98) show strong Ca~{\sc ii} H \& K emission 
lines and small emission in the H$\epsilon$ line.
After applying the spectral subtraction, a small 
 filling-in is observed in the H$\alpha$, H$\beta$ and  Ca~{\sc ii} IRT lines
(Fig.~\ref{fig:sgem_all}). 
We observe in this system notable
He~{\sc i} D$_{3}$ absorption (Fig.~\ref{fig:heid3}). 
All the activity indicators show an increase in the
 1998 observation ($\varphi$= 0.88) with respect to 1996 one ($\varphi$= 0.81). 
The emission in the Ca~{\sc ii} H \& K lines in these two observations is 
noticeably larger than in our previous observations
at different epochs and orbital phases.

\subsection{BF Lyn (HD 80715)}

This double-lined spectroscopic binary with spectral types K2V/[dK] 
shows variable H$\alpha$ emission and strong
Ca~{\sc ii} H \& K, H$\epsilon$ and Ca~{\sc ii} IRT emission 
from both components
(Barden \& Nations \cite{BN85}; Strassmeier et al. 1989b). 
In our previous observation (Montes et al. 1995c) we found strong 
emission in the Ca~{\sc ii} H \& K
lines from both components with very similar intensity 
and the H$\epsilon$ line in emission.
The orbital period is 3.80406 days (Barden \& Nations \cite{BN85}), 
and Strassmeier et al. (1989b), from photometric
observations, found that BF Lyn is a synchronized binary with a circular
orbit.

In the four runs analysed in this paper we have obtained 11 spectra 
of BF Lyn at different orbital phases.
We have used the original heliocentric Julian date on conjunction 
(T$_{\rm conj}$) given by Barden \& Nations (\cite{BN85}) to calculate 
the orbital phases since we discovered a mistake in the 
Strassmeier et al. (1993) catalog
where the orbital data from Barden \& Nations (\cite{BN85})
are compiled. In the original paper the epoch was determined using Modified
Julian Date (MJD), that is why 0.5 days must be added to the
Strassmeier et al. (1993) date, who used the
2440000.0 Julian day as a reference date.

{\it The H$\alpha$ line}: 
We took several spectra of BF Lyn in the H$\alpha$ line region in
four different epochs and at different orbital phases. In all the
spectra (Fig.~\ref{fig:bflyn_ha}) we can see the H$\alpha$ line in 
absorption from both components. The
spectral subtraction reveals that both stars have excess H$\alpha$ emission.
At some orbital phases, near to the conjunction, it is      
impossible to separate the contribution of both components.
The excess H$\alpha$ emission of BF Lyn shows
variations with the orbital phase for both components, but the hot star is
the most active in H$\alpha$. 
In Fig~\ref{fig:bflyn_ew} we have plotted for the McD~98
observing run the excess H$\alpha$ emission EW versus the orbital phase
 for the hot and cool components.
 The highest excess H$\alpha$ emission EW for the hot component has
been reached at about 0.4 orbital phase and the lowest value is placed at
about 0.9 orbital phase, whereas the cool component shows the highest excess
H$\alpha$ emission 
EW at near 0.9 orbital phase and the lowest value between 0.2 and 0.4
orbital phases. 
The variations of the excess H$\alpha$ emission EW for both components are
anti-correlated, which indicates that the chromospheric active regions
are concentrated on faced hemispheres of both components, but at about
0.4 and 0.9 orbital phases for the hot and cool component, respectively.
The same behaviour is also found in Ca~{\sc ii} IRT.
  The excess H$\alpha$ emission EW also shows seasonal variations, for
instance, the values of the INT99 observing run are very different,
specially for the cool component, from McD 98 values at very similar
orbital phase.

{\it The H$\beta$ line}:
Five spectra in the H$\beta$ line region are available.
In all of them the H$\beta$ line
appears in absorption from both components. The application of the
spectral subtraction technique reveals clear excess H$\beta$ emission from
both stars (see Fig.~\ref{fig:bflyn_hb}).
The ($ \frac{ E_{H\alpha} } { E_{H\beta} } $) values that
we have found for this star allow us to say that the emission comes
from extended regions viewed off the limb.

{\it The Ca~{\sc ii} H $\&$ K and H$\epsilon$ lines}:
We took four spectra in the Ca~{\sc ii} H \& K region during the
NOT (96 \& 98) observing runs (Fig.~\ref{fig:bflyn_cahyk}).
 Another spectrum was taken in
1993 with the 2.2~m telescope at the German Spanish Astronomical
Observatory (CAHA) (Montes et al., 1995c). 
These spectra exhibit
clear and strong Ca~{\sc ii} H \& K and H$\epsilon$ emission lines.
At 0.02, 0.43, and 0.54 orbital phases the emission from both
components can be deblended using 
a two-Gaussian fit (see Fig.~\ref{fig:bflyn_cahyk}).
In the case of CAHA 93 run, the H$\epsilon$ emission line from
the hot component is overlapped with the Ca~{\sc ii} H emission of the
cool component.
The excess Ca~{\sc ii} H $\&$ K and H$\epsilon$ emission changes
with the orbital phase during the NOT 98 run in the same way as the
corresponding excess Ca~{\sc ii} $\lambda$8542 and H$\alpha$ emission.
The excess Ca~{\sc ii} H $\&$ K
emission EW also shows seasonal variations, for instance, the values
of CAHA 93 observing run are lower than NOT 96 \& 98 values.

{\it The Ca~{\sc ii} IRT lines}:
In all the spectra we can see
the Ca~{\sc ii} IRT lines in emission from both components 
(Fig.~\ref{fig:bflyn_cairt}).
As in the case of H$\alpha$, the
Ca~{\sc ii} IRT emission shows variations with the orbital phase for
 both components.
In Fig.~\ref{fig:bflyn_ew}  we have plotted, for the McD 98  observing run,
the excess Ca~{\sc ii} $\lambda$8542 emission
EW versus the orbital phase for the hot and cool components.
The variations of the excess Ca~{\sc ii} emission EW for both components are
anti-correlated and they show the same behaviour as the excess H$\alpha$
emission EW.

\subsection{IL Hya (HD 81410)}

IL Hydrae is a typical RS CVn star with
very strong Ca~{\sc ii} H \& K emission
(Bidelman \& MacConnell \cite{BiMcC73}). 
The 12.86833-day photometric period derived by Raveendran et
al. (1982) is very close to the orbital period.
It was found to be an X-ray source and a microwave emitter (Mitrou
et al. 1996). From a photometric analysis, 
Cutispoto (\cite{Cut95}) estimated the
secondary to be a G8V star. Later, Donati et al. (\cite{Do97}) 
detected the secondary component in the optical range and 
they calculated a 1.0~R$_{\odot}$ radius for it.
Weber \& Strassmeier (1998) gave a K0III-IV type for the primary and     
computed the first double-lined orbit of IL Hya. 
Later, Raveendran \& Mekkaden (1998) gave a new orbital solution and 
just recently, Fekel et al. (\cite{F99}) have presented updated SB2
orbital elements which we have adopted
and they are given in Table~2 
(we have corrected the heliocentric Julian date 
on conjunction (T$_{\rm conj}$) so that the primary is in front). 
The multiwavelength Doppler images presented by
Weber \& Strassmeier (1998) revealed a cool polar spot and several 
features at low latitudes. 
These authors also found that the H$\alpha$ EW showed
sinusoidal variation which was in phase with the photospheric light
curve. 

We have taken one spectrum of IL Hya (NOT98) with 0.02                          
orbital phase which is very close to the conjunction, 
so the very weak
lines of the secondary can not be detected in any wavelength. 
In the observed spectrum (Fig.~\ref{fig:ilhya_all}), 
the H$\alpha$ line can be seen as a filled-in absorption line with a
superimposed 1.04~\AA$\ $ (48 km~s$^{-1}$) red-shifted absorption feature,
as obtained from a two-Gaussian fit.
After applying the spectral subtraction, clear excess emission is observed.
The excess emission shows an asymmetric profile due to the 
red-shifted absorption feature.
Similar H$\alpha$ profiles were observed by Weber \& Strassmeier (1998)
 in this system, but the red-shifts measured in their spectra were larger 
(1.24~\AA) and remained constant during a rotational cycle.
This behaviour could be due to mass motions that change 
from one epoch to another, but a combination of several 
dynamical processes may be involved.
%
A filling-in is also found in the H$\beta$ line. 
According to the value obtained for the
corrected ratio of the excess emission EW of both lines, the
emission may be ascribed to an extended region viewed off the limb.
The He {\sc i} D$_{3}$ line appears in absorption (Fig.~\ref{fig:heid3}),       
but no filling-in is detected in the Na {\sc i} D$_{1}$, D$_{2}$ lines. 
The Ca~{\sc ii} H $\&$ K lines are observed in emission.
Furthermore, the Ca {\sc ii} IRT lines 
show clear central emission reversal.

\subsection{FG UMa (HD 89546)}

FG UMa is the least studied star of our sample.
Bidelman (\cite{Bi81}) included it in
his Catalogue of stars with Ca~{\sc ii} H \& K emission.
This star is identified
as a single-lined binary by CABS.
A 21.50-day photometric period has been obtained from the automated
 monitoring that Henry et al. (1995a) carried out. From spectroscopic
measurements, they confirmed V$\sin{\it i}$= (15$\pm$2) kms$^{-1}$ and
 a G8IV spectral type.
These authors also mentioned
that, according to an unpublished orbital analysis, the system is 
synchronized and circularized.
We have also taken from them the orbital period and the radius.
Some indication of possible eclipses is noted in  
 the Hipparcos Catalogue (ESA \cite{ESA97}).
Fluxes of the Ca~{\sc ii} H \& K emission lines have been calculated by
Strassmeier (1994b) and a filled-in and variable H$\alpha$ line has been 
reported by  Henry et al. (1995a).

We have not got enough data to be able to compute the orbital phases 
corresponding to the two spectra (NOT98) that we present here. 
However, there is a change of 0.2 in
the photometric phase between both observations. We have not found any
evidence
of the secondary star through the whole spectral range. Moreover,  
according to the appearance of some Ti {\sc i}
 and Fe {\sc i} lines (Paper II) we suggest that the observed spectra
correspond to a luminosity class more evolved than subgiant
(in agreement with the radius calculated by Henry et al. (1995a) who suggested
a luminosity class III-IV).
The presence of a notable He {\sc i} D$_{3}$
absorption line (Fig.~\ref{fig:heid3}) encourages this conclusion.
In the observed spectra of the H$\alpha$ and H$\beta$ lines
(Fig.~\ref{fig:fguma_all}),
 both absorption lines show a clear filling-in. The spectral subtraction allows
us to compare the two observations. As it can be read in Table~5 the
ratio of the excess emission EW is typical of extended
regions viewed off limb, and a significant variation, mainly due to H$\beta$,
is obtained for the two different nights. Furthermore, excess emission 
is detected in the blue wing of the H$\alpha$ line. Similar behaviour was
mentioned by Henry et al. (1995a).
Strong filling-in is observed in the Ca~{\sc ii} IRT lines. 
The Ca~{\sc ii} H \& K lines present strong emission with
clear self-absorption with blue-shifted asymmetry in both observations.

\subsection{LR Hya (HD 91816)} 

It is a double-lined spectroscopic binary, classified as a BY Dra-type
system, that contains two almost identical K-type dwarf 
components (Bopp et al. \cite{Bo&84}).
 The orbital parameters were determined by Fekel et al. (\cite{F88}) who  
suggested a K0V spectral type for both components. The photometric period
of 3.1448 days, given in CABS, was reported by Bopp et al. (\cite{Bo&84}), 
but following observation campaigns could not confirm that
value. In fact, the
results obtained are not consistent (Strassmeier 1989;
Cutispoto \cite{Cut91}, \cite{Cut93}) and point out
low-amplitude rotational modulation due to the development and decline
of small active regions at different longitudes of both components.

We only have got one spectrum (NOT98) of this system at 0.58
orbital phase, so that the chromospheric activity indicators
from both components can be easily analyzed.
The H$\alpha$ absorption line (Fig.~\ref{fig:lrhya_all}) 
shows a weak filling-in for both components,
as it has been previously mentioned by Strassmeier et al. (1990).
No filling-in is detected in the  H$\beta$ line.
Although the S/N ratio in the Ca {\sc ii} H \& K region is very low 
in this observation and the synthetic and observed spectra are not well 
matched,  we can clearly see moderate emission in the Ca {\sc ii} H \& K lines 
from both components.
The Ca {\sc ii} IRT lines of the two stars exhibit a clear filling-in.
The measured excess emission in the different lines are very similar 
in both components, although a bit larger in the red-shifted one.

\input{ds1878_figs2s.tex}

\subsection{HU Vir (HD 106225)}

HU Vir is a late-type star (K0III-IV) with strong 
Ca {\sc ii} H $\&$ K emission noted by Bidelman (\cite{Bi81}) 
for the first time.
Recently, Fekel et al. (\cite{F99}) have discovered that HU Vir is a triple 
system with a long period of about 6.3 years 
and we have taken from them the spectral type, T$_{conj}$ and the 
rotational period. The B--V colour index has been taken from 
Hipparcos Catalogue (ESA \cite{ESA97})
and V$\sin{\it i}$ from Fekel (\cite{F97}).
Fekel et al. (1986) observed the H$\alpha$ line in emission 
and Strassmeier \& Fekel (1990) found the H$\epsilon$ line in emission. 
Such emission lines are seen in the most active 
RS CVn type systems. 
Strassmeier (1994a) found a big, cool polar spot from Doppler imaging 
and two hot plages 180$^{o}$ apart from the H$\alpha$ and Ca {\sc ii} 
line-profile analysis. The chromospheric plages seemed to be spatially 
related to two large appendages of the polar spot. 
Broadening in the H$\alpha$ profile suggested mass flow in a coronal 
loop connecting the two plage regions.
Hatzes (\cite{Hat98}) used the Doppler imaging technique 
to derive the cool spot 
distribution. He found a large spot at latitude 45$^{o}$ and a weak polar 
spot with an appendage. The polar spot was considerably smaller than similar 
features found on other RS CVn stars. From an analysis of the H$\alpha$ 
variations he also found evidence for a plage located at high 
($\approx$ 70$^{o}$) latitude, near the polar extension.

{\it The H$\alpha$ line}:
Strassmeier (1994a) identified three distinctive features in the H$\alpha$ 
line: blue-shifted emission, 
central sharp absorption 
and red-shifted broad absorption. 
Hatzes (1998) found similar behaviour in this line.
In our spectra (Fig.~\ref{fig:huvir_ha}), 
the H$\alpha$ line always appears in emission, 
although its intensity is variable. Moreover, the emission  
is blue-shifted at some orbital phases. 
Hall \& Ramsey (\cite{H&R92}, \cite{H&R94}) 
explained the blue-shifted emission as prominence-like structures. 
We can see red-shifted broad absorption at the 0.54 orbital 
phase but we never observe central sharp absorption. 
Walter \& Byrne (1998) said that there was growing evidence for continuous 
low-level mass in-fall, seen as red-shifted absorption in H$\alpha$ line 
profiles. 
We can also notice that the subtracted H$\alpha$ profiles have 
broad and variable extended wings which are not well matched using 
a single-Gaussian fit.
These profiles have been fitted using two Gaussian components.
The parameters of the broad and narrow components used in the two-Gaussian fit
are given in Table~4 and the corresponding profiles are
plotted in the left panel of Fig.~\ref{fig:huvir_ha}.
These broad wings are observed at different orbital phases 
and in different epochs.
In some cases the blue wing is noticeable stronger than the red wing and the
fit is better matched when the broad component is blue-shifted with respect
to the narrow component.
 We have interpreted these broad components as 
microflaring activity that occurs in the chromosphere of this very 
active star (Paper I, II; Montes et al. \cite{M98b}).
The contribution of the broad component to the total EW of the line
ranges from 32\% to 66\% which is in the range observed in the stars 
analysed in Paper I \& II. 
Strassmeier (1994a), in 1991, and Hatzes (1998), in 1995, observed strong 
emission between the 0.27 and 0.51 orbital phases. We have obtained     
the strongest  emission at the 0.44 orbital phase, 
in January 1998, and at 0.29, in April 1998 (see Table 3). 
Thus, our 0.29-0.44 orbital phase interval is similar to Strassmeier         
and Hatzes's orbital phase interval, so we can conclude that HU Vir has an 
active longitude area (which corresponds to that orbital phase interval) 
since 1991. Similar active longitudes have been found by other authors in 
several chromospherically active binaries (Ol\'{a}h et al. 1991; 
Henry et al. 1995b; Jetsu 1996; Berdyugina \& Tuominen \cite{BerT98}).
 
{\it The H$\beta$ line}:
A nearly complete filling-in is observed 
(Fig.~\ref{fig:huvir_hb_hyk} upper panel). 
After applying the spectral subtraction technique, clear excess 
H$\beta$ emission is obtained.
We have obtained $ \frac{ E_{H\alpha} } { E_{H\beta} } $ values larger 
than three (see Table 5), so the emission can come from prominences.
 
{\it The Ca {\sc ii} H $\&$ K and H$\epsilon$ lines}:
We can observe very strong  
Ca {\sc ii} H $\&$ K emission and an important H$\epsilon$ emission 
line superimposed to the wide Ca~{\sc  ii} H line 
(Fig.~\ref{fig:huvir_hb_hyk} lower panel).
The H$\epsilon$ line in emission indicates that HU Vir is a very active system. 
Similar strong emission was found in our previous observation
obtained in March 1993 at 0.71 orbital phase (Montes et al. 1995c).
We can also notice that the largest values of the excess 
Ca {\sc ii} H $\&$ K emission EW appear at the 0.29 and 0.38 orbital phases
(see Table 6). 
It is in agreement with the H$\alpha$ line behaviour.
 
{\it The Ca {\sc ii} IRT lines}:
HU Vir shows the Ca {\sc ii} IRT  
lines in emission above the continuum (Fig.~\ref{fig:huvir_cairt}).  
We observe that the emission 
is centered at its corresponding absorption. 
We also notice that the subtracted profiles have broad wings due to 
microflares according to Montes et al. (1997). 
The excess Ca~{\sc ii}~$\lambda$8542 emission EW (see Table 7)
behaves like the excess 
H$\alpha$ emission EW.
 
{\it The He {\sc i} D$_{3}$ line}:
We have not detected
any significant absorption or emission for He~{\sc i} D$_{3}$
(Fig.~\ref{fig:heid3}), 
contrary to the absorption observed in other giants.
This total filling-in of the He~{\sc i} D$_{3}$ line could be explained
(Saar et al., 1997; Paper II)
 as a filling-in due to the low-level flaring (microflares) 
that takes place in this very active star according to the
H$\alpha$ broad component that we have found. 

{\it The Li~{\sc i} $\lambda$6707.8 line}:
The Li~{\sc i} absorption line is clearly observed 
in the eight spectra of the McD98 observing run (see Fig.\ref{fig:li}).
The mean EW obtained is 56~$\pm$11 (m\AA).
At this spectral resolution the Li~{\sc i} line is blended with
the nearby Fe~{\sc i} $\lambda$6707.41~\AA\ line.
We have corrected the total EW measured, EW(Li~{\sc i}+Fe~{\sc i}),
by subtracting the EW of Fe~{\sc i} calculated from the empirical
relationship with (B--V) given by Soderblom et al. (1990)
(EW(Fe~{\sc i})=24 (m\AA)).
Finally, the corrected  EW(Li~{\sc i})$_{\rm corr}$=32 (m\AA) was converted
into abundances by means of the curves of growth computed by Pallavicini et al.
(1987), obtaining $\log{\sc N(Li~{\sc i})}$ =1.2 
(on a scale where $\log{\sc N(H)}$ =12.0)
with an accuracy of the $\approx$~0.30~dex.
This value is larger than the lower limit reported 
by Barrado et al. (\cite{Ba98}) for this star.

\subsection{DK Dra (HD 106677)}

This is a double-lined spectroscopic binary with almost identical
components of spectral types K1III
 and  Ca~{\sc ii} H \& K emission from both components
 (Bopp et al. \cite{Bo&79}; Fekel et al. \cite{F86}; Strassmeier 1994b).
Eker et al. (\cite{Ek95}) reported variable nature of H$\alpha$ and,
using a subtracted spectrum, found emission of similar
intensity from both components.
In our previous observations (Montes et al. 1995a, b; FFMCC) we found a
broad excess H$\alpha$ emission line and the Ca~{\sc ii} H \& K lines in
emission, but all of them were taken at orbital phases near to the conjunction,
so it was impossible to distinguish the contribution of each component.

The spectra analysed in this paper were taken at 0.90 (NOT98) 
and 0.84 (NOT98) orbital phases (see Fig.~\ref{fig:dkdra_all}). 
In both spectra we observe the H$\alpha$, H$\beta$ and Ca~{\sc ii}
$\lambda$8662 lines filled-in, and the Ca~{\sc ii} H \& K lines in emission.
At 0.90, we cannot separate the contribution
of each component in H$\alpha$, H$\beta$ and  Ca~{\sc ii} H \& K lines, 
but due to the large wavelength of the Ca~{\sc ii} $\lambda$8662 line, 
a double peak is clearly observed in the subtracted spectrum of this line. 
In the 0.84 spectrum, we can see that the H$\alpha$ line
is filled-in for both stars. Moreover, we can observe the excess H$\alpha$
emission of the blue-shifted component is bigger than that of the red-shifted one.
The excess H$\beta$ emission seems to come only from the blue-shifted  
component.
Although the S/N ratio in the Ca~{\sc ii} H \& K lines region is low, 
we can clearly see broad and two-piked emission in the H \& K lines.
In the subtracted spectrum of the  Ca~{\sc ii} IRT 
$\lambda$$\lambda$ 8542 and 8662 lines the emission from both components 
is clearly separated, being the emission of the blue-shifted one 
slightly larger. 
All of this indicates that the blue-shifted component is slightly more active 
than the other component.
Very small absorption is observed in the He~{\sc i} D$_{3}$ line in the
expected position for both components
(see Fig.~\ref{fig:heid3}), contrary to the notable absorption observed
 in other 
giants of the sample. 
This is probably due to the blend with other photospheric lines 
of both stars in this SB2 system.

\subsection{BQ CVn (HD 112859)}

A quarter of century ago Schild (1973) classified this star as a peculiar
G8III-IV. Henry et al. (1995a) detected its Ca {\sc ii} H \& K emission
and noticed that
at red wavelengths the spectrum was double-lined although the intensity
of the lines of both components was very different. They suggested
 a K0III primary and a late-F spectral type for the secondary.
We have taken from these authors both the orbital and 
rotational period and the radius.
 On the other hand,
Strassmeier (1994b) confirmed the strong Ca {\sc ii} H \& K emission.

Orbital parameters have not been published, so we cannot
calculate the orbital phase of our spectrum (NOT98). 
However, some conclusions can
be obtained looking at the observed spectrum (Fig.~\ref{fig:bqcvn_all}):
at infrared and red
wavelengths, the secondary lines are not clear. A very weak blue-shifted
absorption
feature can be ascribed to the secondary in the H$\alpha$ region.
At shorter wavelengths, as in the  
Na {\sc i} D$_{1}$, D$_{2}$ lines region,
 the spectral lines of the secondary are more conspicuous.
In the Ca {\sc ii} H \& K lines region
the contribution of
the F star is evident in the broad Ca {\sc ii} absorption lines,
where a clear red-shift of the emission can be seen.
Taking into account what has been said above, 
we have calculated the synthesized
spectrum using a F8V star as a template one for the secondary.
In the observed spectrum, the H$\alpha$ line shows a filling-in for
the primary star.            
The H$\beta$ line is slightly filled-in.
The presence of the He {\sc i} D$_{3}$ absorption line is detected
 (Fig.~\ref{fig:heid3}).
In our spectrum we observe strong Ca {\sc ii} H \& K emission
and a weak H$\epsilon$ emission line from the cool star.
Finally, the Ca {\sc ii} IRT lines exhibit reversal emission.

\subsection{IS Vir (HD 113816)}

IS Vir is a single-lined spectroscopic binary classified as
K0III by Henry et al. (1995a). 
We have taken from them the orbital and rotational periods
and the radius given in Table 2.
Strong Ca~{\sc  ii} H \& K emission was observed
(Buckley et al. \cite{Bu87}; Strassmeier 1994b; Montes et al. 1995c).
In our previous observation of this system in the 
Ca~{\sc ii} H \& K lines
region in Mar-93 at 0.68 orbital phase (Montes et al. 1995c)
we found 
strong emission in the H \& K lines with intensity
above the continuum at 3950 \AA, but lower than reported by
Strassmeier (1994b). 

In the new spectrum (NOT 98) (Fig.~\ref{fig:isvir_all}), 
the H$\alpha$ line and the Ca~{\sc ii} IRT lines show intense 
filled-in absorption, whereas the H$\beta$ line only shows a
slight filling-in.
In the Ca {\sc ii} H \& K lines region the S/N ratio is low 
and the synthetic and observed spectra are not well
matched, but a strong emission in the H \& K lines well above the continuum 
is observed.
The He~{\sc i} D$_{3}$ line appears in
absorption (Fig.~\ref{fig:heid3}).

\subsection{BL CVn (HD 115781)}

This double-lined spectroscopy binary was classified by 
Hall (\cite{Hall90}) as K1III~+~FIV, and later as K1III~+~G5IV by
Stawikowski \& Glebocki (1994).
Lines et al. (1985) found this system to have a photometric period of
9.31$\pm$0.06 days and an amplitude of 0.16~mag. They
attributed the light variability to the ellipticity
effect because the orbital period was twice the photometric
period and times of maximum brightness occurred at times
of maximum positive radial velocity.
It is a nearly-synchronous
binary: its orbital period is 18.6917$\pm$0.0011 days
(Griffin \& Fekel \cite{Gr&F88}) and its rotational period is
18.70 days (Stawikowski \& Glebocki 1994). The orbital
eccentricity cannot be far from zero (Griffin \& Fekel \cite{Gr&F88}).
Fekel et al. (\cite{F86}) found V$\sin{\it i}$ values
of (35$\pm$2) km$^{-1}$ and (7$\pm$2) km$^{-1}$ for the 
primary and secondary, respectively. The great line broadening
of BL CVn and the ellipsoidal light variations might
suggest that the K giant star is close to filling its
Roche lobe (Griffin \& Fekel \cite{Gr&F88}).
Moderate H$\alpha$ absorption is found by Fekel et al. (\cite{F86}).
 Strassmeier et al. (1990) found strong Ca {\sc ii} H \& K emission.

In our present observations (NOT98, Fig.~\ref{fig:blcvn_all}) 
the H$\alpha$ and H$\beta$ lines appears in absorption 
in the observed spectrum.
After applying the spectral subtraction technique, we only obtain
small excess H$\alpha$ emission.
Broad emission is observed in the  Ca {\sc ii} H \& K lines.
We observe small absorption in the He {\sc i} D$_{3}$ line
(Fig.~\ref{fig:heid3}). 
A small filling-in is obtained in the
Ca {\sc ii} $\lambda$8542 and $\lambda$8662 lines. 

\subsection{BM CVn (HD 116204)}

This single-lined spectroscopy binary was classified by Keenan 
(1940) as K1III. It is a nearly-synchronous binary: its orbital
period is (20.6252$\pm$0.0018) days (Griffin \& Fekel \cite{Gr&F88}) 
and its rotational period is (20.66$\pm$0.03)
days (Strassmeier et al. 1989a). The orbit is judged to be 
circular (Griffin \& Fekel \cite{Gr&F88}). It is also a relatively fast 
rotator, Fekel et al. (\cite{F86}) found rotationally broadened
lines with V$\sin{\it i}$= (15$\pm$2) km$^{-1}$.
This system shows strong Ca {\sc ii} H \& K emission (Bidelman \cite{Bi83}),
together with an emission-line spectrum typical of RS CVn stars in
the IUE ultraviolet region, but H$\alpha$ is an absorption line
(Griffin \& Fekel \cite{Gr&F88}).

In our present spectrum (NOT98, Fig.~\ref{fig:bmcvn_all})
 we observe the H$\alpha$ line as nearly total filled-in absorption.
After applying the spectral subtraction technique, 
strong excess H$\alpha$ emission is obtained.
The H$\beta$ line appears as an absorption line in the observed spectrum 
and clear excess emission is observed in the subtracted spectrum.
The $ \frac{ E_{H\alpha} } { E_{H\beta} } $ ratio obtained is larger
 than 3, which indicates 
that the emission would arise from extended regions viewed off the limb. 
The Ca {\sc ii} H \& K lines show strong 
emission, with a blue-shifted self-absorption feature, 
and small emission is also detected in the H$\epsilon$ line. 
The He {\sc i} D$_{3}$ line appears in 
absorption (Fig.~\ref{fig:heid3}). 
A clear emission reversal is observed in the Ca {\sc ii} IRT 
$\lambda$8542 and $\lambda$8662 lines.

\subsection{MS Ser (HD 143313)}

Griffin (\cite{Gr78}) first observed the binary nature of MS Ser              
calculating the orbital elements for this system
(see Table~2).
Griffin gave its T$_{0}$ in MJD, and this yielded Strassmeier et al.
(1993) to a bad calculation of the T$_{0}$ in HJD.
Griffin also proposed K2V/K6V as spectral types of the components, based on
photometric arguments for the secondary star.
Bopp et al. (\cite{Bo&81b}) observed a variable filling-in 
of the H$\alpha$ line and
calculated a photometric period of 9.60 days, slightly different
from the orbital period. Miller \& Osborn (1996) confirmed the value of
the photometric period and Strassmeier et al. (1990)
observed strong emission in the Ca~{\sc ii} H \& K
composite spectrum.
Dempsey et al. (\cite{De93a}) noted some filling-in in the Ca~IRT lines for
MS Ser, but not reverse emission.
Alekseev (\cite{A99})  made a photometric and polarimetric study of MS Ser,
calculating a spot area of 15\% of the total stellar surface,
and  observed some seasonal variations.
Finally, Osten \& Saar (1998) revised the
stellar parameters for MS Ser and suggested
K2IV/G8V as a better classification. 
We have also found that the primary component may have a luminosity 
class IV or higher based on
our analysis of some metallic lines, like the
Ti~{\sc i} lines, the Hipparcos data and the Wilson Bappu effect
(see Sanz-Forcada et al. 1999).


Two spectra of MS Ser were taken in the NOT98 observing run. Moreover,   
we analyse here another spectrum taken on 12th June 1995 with the 2.2 m
telescope at the German Spanish Observatory (CAHA) in Calar Alto
(Almer\'{\i}a, Spain), using a Coud\'{e} spectrograph with the
f/3 camera, CCD RCA \#11
covering two ranges: H$\alpha$ (from 6510 to 6638~\AA) and H$\beta$
(from 4807 to 4926~\AA), with a resolution of $\Delta\lambda$ 0.26
in both cases.

{\it The H$\alpha$ and H$\beta$ lines}:
%
In both lines, we observe a nearly total filling-in in the 1995 spectra
and small absorption in the spectra 
taken in the NOT98 observing run. 
After applying the spectral subtraction, a
clear filling-in in the H$\alpha$ and H$\beta$ lines is observed in
the three spectra (see Fig.~\ref{fig:msser_all}).
The H$\alpha$ line of this system is highly variable.
We have found, in the present spectra, a variable filling-in whereas
Bopp et al. (\cite{Bo&81b}) found H$\alpha$ was a weak emission line. 
In the three new spectra we have taken of this system with the 
FOCES echelle spectrograph in July 1999 (forthcoming paper) we 
observe variable H$\alpha$ emission well above the continuum. 
%

{\it The Ca~{\sc ii} H $\&$ K and H$\epsilon$ lines}:
%
Strong emission in these lines and
the H$\epsilon$ line in emission arising from the primary component
was observed in our previous observation of this system in 
March 1993 at 0.16 orbital phase (Montes et al. 1995c).
In the present observations (Fig.~\ref{fig:msser_all}) 
we have deblended the emission
arising from both components in the spectrum taken  
 near to the quadratures ($\varphi$ =0.21).
The strongest emission, centered at the absorption line, arises from the K2IV
component, which is the component with larger contribution to the
continuum.
The red-shifted and less intense emission corresponds to the G8V secondary
component.
In the $\varphi$ =0.55 observation we
cannot separate the contribution from both stars.
The H$\epsilon$ line appears in emission in both spectra.
The emission intensity observed
in our 1993 and 1998 spectra is larger
than the emission intensity observed in the 1988 spectrum presented
by Strassmeier et al. (1990).

{\it The Ca~{\sc ii} IRT lines}:
%
A clear emission reversal is observed in the core of the
Ca~{\sc ii} IRT absorption lines $\lambda$8542 and $\lambda$8662 
in both spectra (Fig.~\ref{fig:msser_all}).
After applying the spectral subtraction technique, we can see
small excess emission arising from the secondary component,
in the 0.21 spectrum,
as in the case of the Ca~{\sc ii} H \& K lines.
This emission reversal observed here clearly contrasts with the
only filling-in in these lines reported by Dempsey et al. (\cite{De93a}).

\begin{figure}
{\psfig{figure=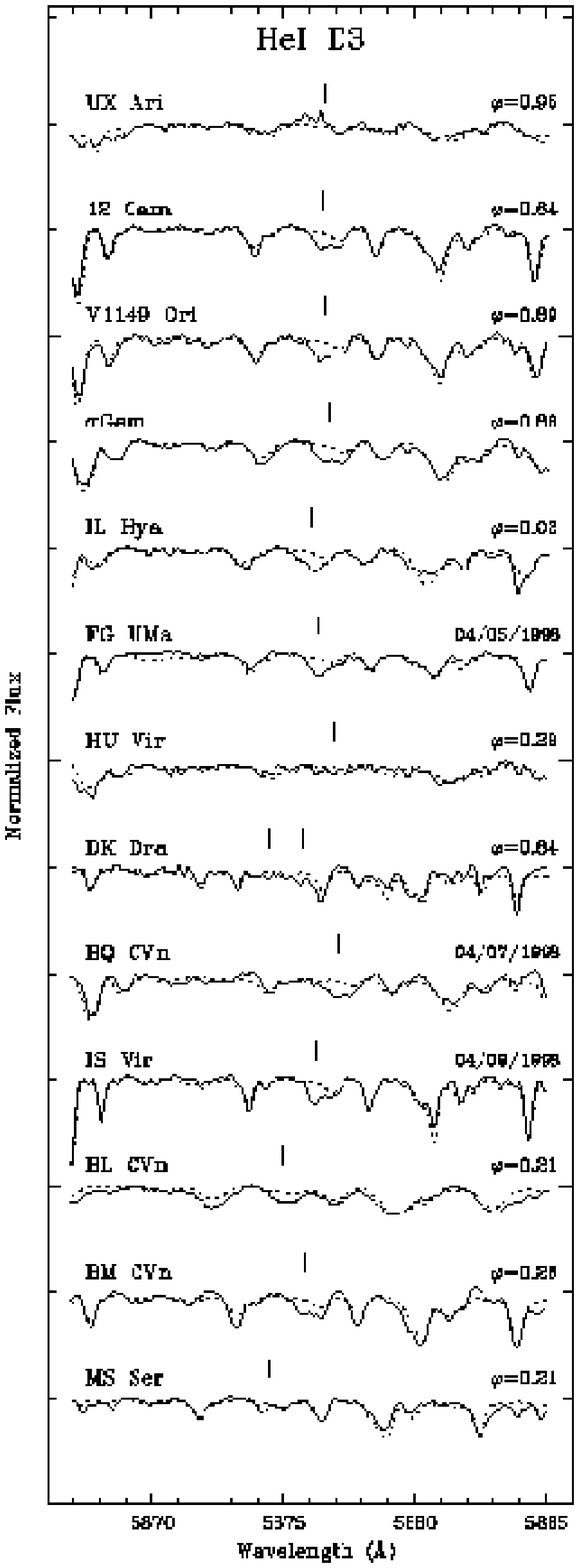,bbllx=36pt,bblly=36pt,bburx=302pt,bbury=760pt,height=21.5cm,width=8.7cm,clip=}}
\caption[ ]{He~{\sc i} D$_{3}$ line for selected stars of the sample.
We plot observed spectrum (solid-line) and the
synthesized spectrum (dashed-line).
The expected position of the He~{\sc i} D$_{3}$ line for each star
is marked with a short vertical line.
\label{fig:heid3} }
\end{figure}

\begin{figure}
{\psfig{figure=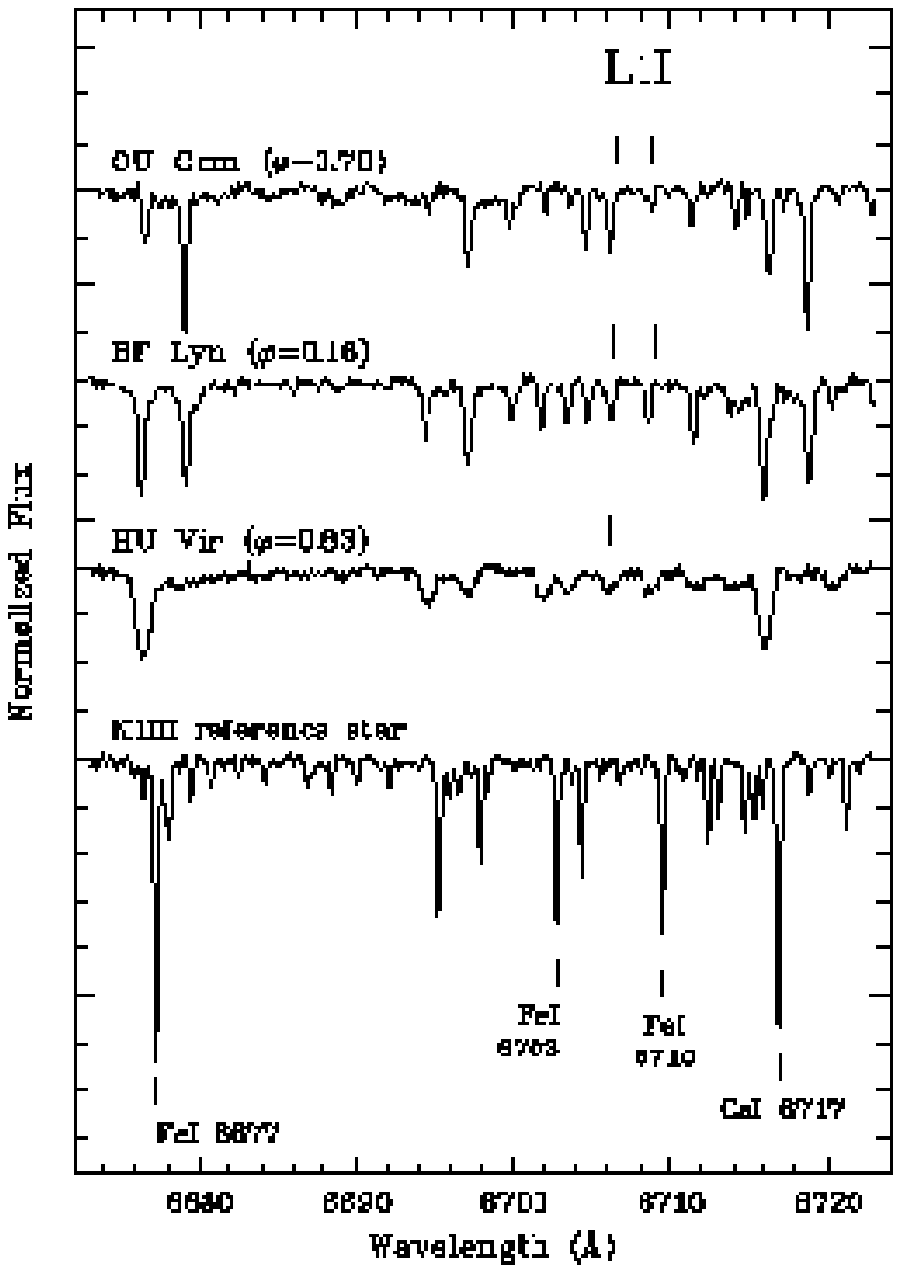,bbllx=36pt,bblly=36pt,bburx=302pt,bbury=413pt,height=10.0cm,width=8.7cm,clip=}}
\caption[ ]{Spectra of  OU Gem, BF Lyn and HU Vir in the
Li~{\sc i} $\lambda$6707.8 line region.
The expected position of the Li~{\sc i} line for each star
is marked with a short vertical line.
We have also plotted a K1III reference star with
some photospheric lines identified.
\label{fig:li} }
\end{figure}

{\it The He~{\sc i} D$_{3}$ line}:
%
We can distinguish (Fig.~\ref{fig:heid3}) 
the He~{\sc i} D$_{3}$ line as a very small absorption from the primary star. 
A slight variation is observed between the spectra at 0.21 and 0.55 orbital
phases.
The luminosity class of this star (IV-III) and the SB2 nature of the system 
could be the reason of this small absorption in the  He~{\sc i} D$_{3}$ line
in relation with that observed in other giants of the sample.

\section{Discussion and conclusions}

In this paper we have analysed, using the spectral subtraction technique,
high resolution echelle spectra of 16 chromospherically active binary systems.
These spectra include all the optical chromospheric activity indicators
from the Ca~{\sc ii} H \& K to Ca~{\sc ii} IRT lines, and in some observing
 runs the Li~{\sc i} $\lambda$6707.8 line too.

H$\alpha$ emission above the continuum is observed in UX Ari and HU Vir,
in the rest of the systems excess H$\alpha$ emission is clearly detected in
the subtracted spectra.
Filled-in absorption H$\beta$ line profiles have been found 
in all the stars of the sample except LR Hya and BL CVn, 
which also have a lower level of activity in the
other chromospheric activity indicators.

Very broad and variable extended wings have been found in the subtracted
H$\alpha$ profile of the very active star HU Vir.
These line profiles are not well matched using a single-Gaussian fit,
and  have been fitted using two Gaussian components (narrow and broad).
Similar behaviour has been found in other very active stars
(Hatzes 1995; Jones et al. 1996; Stauffer et al. 1997;
Paper I, II, Montes et al. \cite{M98b}).
We have interpreted this broad component emission
as arising from microflaring (high turbulence emitting plasma)
activity that takes place in the chromosphere by similarity with the
broad components found in the chromospheric Mg~{\sc ii} h \& k lines
(Wood et al. 1996; Bus\`{a} et al. \cite{Busa99})
and in several transition region lines of active stars observed with GHRS-HST
(Linsky \& Wood 1994; Linsky et al. 1995;
Wood et al. 1996, 1997; Dempsey et al. \cite{De96a}, b;
Robinson et al. 1996) and recently confirmed with STIS-HST observations
(Pagano et al. 2000).
In some cases the H$\alpha$ line is asymmetric and the
fit is better matched when the broad component is blue-shifted
or red-shifted with respect to the narrow component.
These asymmetries are also observed during the impulsive and gradual phases
of solar and stellar flares (Montes et al. \cite{M96b}, \cite{M99};
Montes \& Ramsey \cite{MR99}),
 and favour the interpretation of the broad
component as due to upward and downward motions produced by microflaring
in the chromosphere.

Excess absorption and emission is observed in the wings of the H$\alpha$ and
H$\beta$ lines of several systems.
Absorption features are detected by us in the blue wing of V1149 Ori
and in the red wing of
12 Cam (also observed by Eker et al. (\cite{Ek95})),
IL Hya (also detected by Weber \& Strassmeier (1998)) and
HU Vir (also seen by Strassmeier (1994a) and Hatzes (1998)).
We detected excess emission in the blue wing of FG UMa too.
Similar red-shifted absorption features were already seen by other authors
in the single star OP And (Fekel et al. \cite{F86})
and in the binaries VY Ari (Bopp et al. \cite{Bo&89}); 
XX Tri (Bopp et al. \cite{Bo&93}) and IN Vir (Strassmeier 1997).
Blue-shifted emission was also detected in the single giant 
YY Men (Vilhu et al. 1991).
Several dynamical processes, or a combination of them, could be
the origin of these blue-shifted or red-shifted absorption features:
plage- and prominence-like structures 
(Hall \& Ramsey \cite{H&R92}, \cite{H&R94}; Neff 1996; Eibe \cite{Eibe98});
continuous low-level mass infall (Walter \& Byrne 1998; Walter 1999; Eibe et al. \cite{Eibe99});
local velocity fields and mass motions due to magnetic field inhomogeneities
possibly coupled with a loop-like geometry (Strassmeier 1994a);
fluctuations of both the column density and temperature gradient within
the chromosphere (Smith \& Dupree 1988);
stellar winds and anti-winds (Linsky et al. 1995).

Prominence-like extended material viewed off the limb has been detected
in many stars of the sample according to the high ratios
E$_{\rm H\alpha}$/E$_{\rm H\beta}$ obtained.
Prominences viewed against the disk seem to be present
at some orbital phases in the dwarfs OU Gem and BF Lyn.

The application of the spectral subtraction
reveals that the He~{\sc i} D$_{3}$ line appears
as an absorption feature (Fig.~\ref{fig:heid3})
in mainly all the giants of the sample
as in the case of the stars analysed in Paper~I.
Total filling-in of He~{\sc i} D$_{3}$
is observed in the very active star HU Vir,
similar to the behaviour observed in EZ Peg (Paper II).
These results are in agreement with the behaviour reported 
by Saar et al. (1997).
These authors found that while for few active stars the He~{\sc i} D$_{3}$  line
behaves "normally", increasing in absorption with increasing rotation,
and showing consistent correlations with other activity indicators,
the behaviour clearly diverges (large filling-in) when stars become very active,
suggesting that the line could be filled-in due to
frequent low-level flaring.
In the most evolved stars the behaviour is different as a consequence of
the lower chromospheric densities of these stars.

Ca~{\sc ii} H \& K emission is observed
in all the stars of the sample.
Small emission in the close H$\epsilon$ line is also detected
in some of the more active stars.
In some systems like OU Gem, BF Lyn, LR Hya, DK Dra and MS Ser 
the emission from both components is clearly deblended in these lines.
Self-reversed absorption core with red asymmetry (I(K$_{2V}$) $<$ I(K$_{2R}$)
is detected in  the Ca~{\sc ii} H \& K lines of the giants
12 Cam, FG UMa and BM CVn.
The self-reversed feature is a consequence of the line formation process 
in the chromosphere (depth variation of the line source function in an 
atmosphere having a chromospheric temperature rise).
Asymmetries in these profiles provide information on velocity fields 
in the line formation regions. 
In these three giants we observed a small 
red asymmetry (indicative of outward mass flux, wind),
 contrary to the blue asymmetry 
(indicative of upward propagating waves, but not large wind) 
normally observed in  giant stars hotter than spectral type K3
(Stencel 1978) and also observed by us (FFMCC) 
in the giants (V1817 Cyg and V1764 Cyg).

The Ca~{\sc ii} IRT lines result to be a very useful
chromospheric activity indicator too.
We have found that all the stars analysed here show a clear
filled-in absorption line profile or even notable
emission reversal (UX Ari, OU Gem, BF Lyn, IL Hya, HU Vir, BQ CVn, BM CVn, MS Ser).
Thanks to the higher resolution of our spectra, we were able to detect
emission reversal in the Ca~{\sc ii} IRT lines in some systems in which
previous studies (Linsky et al. 1979; Dempsey et al. \cite{De93a})
only reported a filling-in.
An increase in the level of activity of these stars could also be the cause
of these detections.
When both components of the binary system are active
the excess emission detected in the Ca~{\sc ii} IRT lines
is much more  easily deblended thanks to the large wavelength of these lines
(see e.g. DK Dra, Fig.~\ref{fig:dkdra_all}).
We have found E$_{\rm 8542}$/E$_{\rm 8498}$ ratios in the range $\simeq$~1-2
which is indicative of optically thick emission in plage-like regions,
in contrast with the prominence-like material inferred by 
the E$_{\rm H\alpha}$/E$_{\rm H\beta}$ ratios.
These small E$_{\rm 8542}$/E$_{\rm 8498}$ ratios, also found by 
Chester et al. (\cite{Ch94}) and  Ar\'{e}valo \& L\'{a}zaro (\cite{A&L99}) 
in other active binaries,
indicate the existence of distinct sources of Balmer and
 Ca~{\sc ii} IRT emission and suggest that the activity 
of these very active stars is not simply a scaled-up version 
of solar activity.

Some systems have been observed at different orbital phases and 
different epochs, covering all the orbital cycle and it was possible 
to study the variability of the chromospheric emission.
The excess H$\alpha$ and Ca~{\sc ii} IRT emission of BF Lyn 
in the McD98 observing run shows 
anti-correlated variations with the orbital phase (see Fig~\ref{fig:bflyn_ew}).
This anti-correlation could indicate that the chromospheric active regions
are concentrated on faced hemispheres of both components but at about
0.4 and 0.9 orbital phases for the hot and cool component, respectively.
Evidence of an active longitude area (0.29-0.44 orbital phase interval) 
has been found in HU Vir when we compare with the higher level 
of activity in this 
phase interval also reported by Strassmeier (1994a), in 1991, 
and Hatzes (1998), in 1995.

\begin{acknowledgements}

We would like to thank Dr. L.W. Ramsey for collaborating in the 
McDonald observing run 
and the staff of McDonald observatory for their allocation of
observing time and their assistance with our observations.
We would like to thank Dr. B.H. Foing for allow us to use the
ESA-MUSICOS spectrograph at Isaac Newton Telescope.
We would also like to thank the referee S. Catalano
for suggesting several improvements and clarifications.
This work has been supported by the Universidad Complutense de Madrid
and the Spanish Direcci\'{o}n General de  Ense\~{n}anza Superior e 
Investigaci\'{o}n Cient\'{\i}fica (DGESIC) under grant PB97-0259.

\end{acknowledgements}


\listofobjects



\end{document}

%% file: ds1878_figs1s.tex




\begin{figure*}
{\psfig{figure=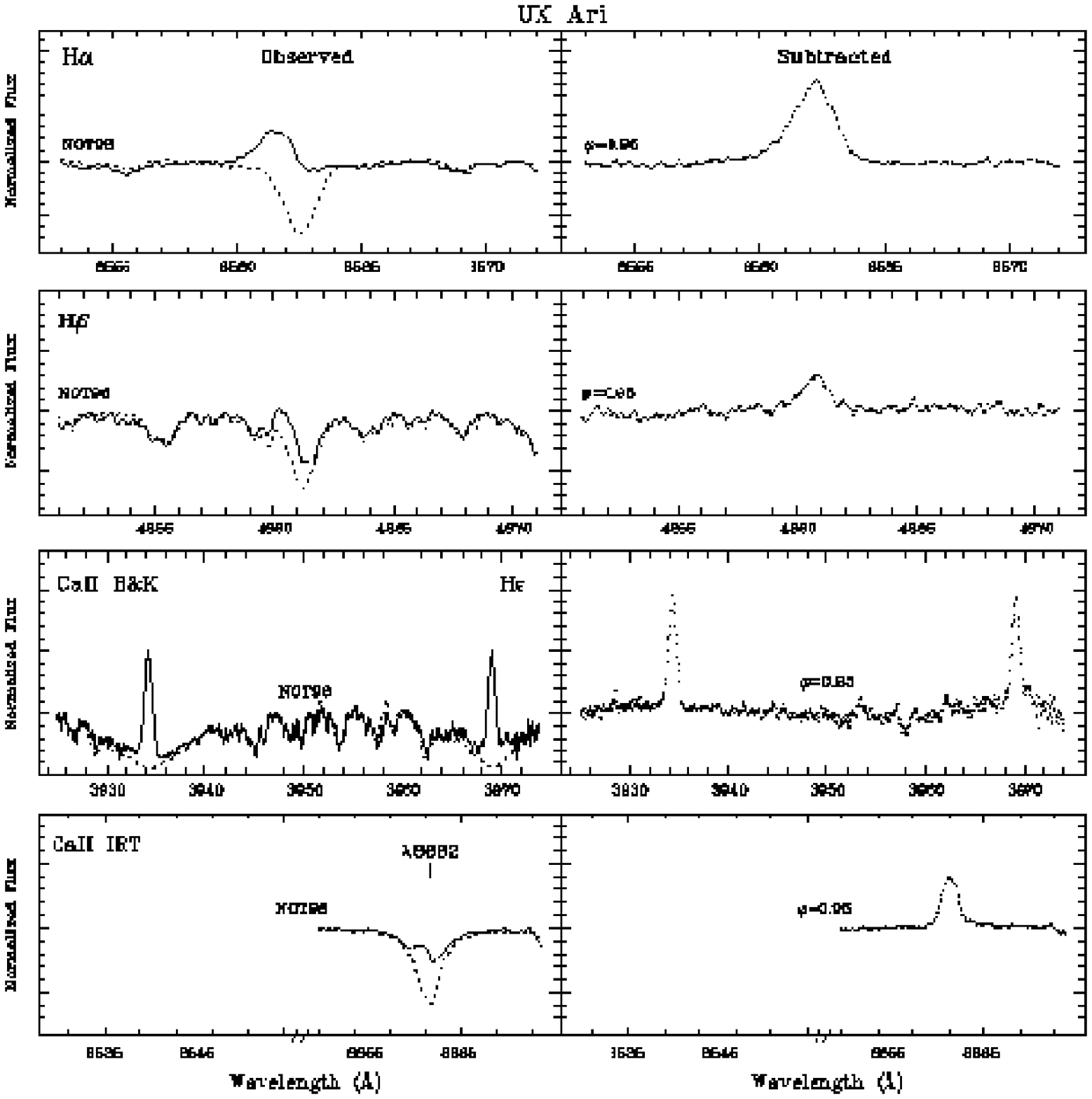,bbllx=36pt,bblly=36pt,bburx=545pt,bbury=544pt,height=15.0cm,width=17.8cm,clip=}}
\caption[ ]{H$\alpha$, H$\beta$, Ca~{\sc ii} H \& K, and Ca~{\sc ii} IRT
spectra of UX Ari. 
Observed and synthetic spectra in the left panel and subtracted spectra 
in the right panel.
\label{fig:uxari_all} }
\end{figure*}

\begin{figure*}
{\psfig{figure=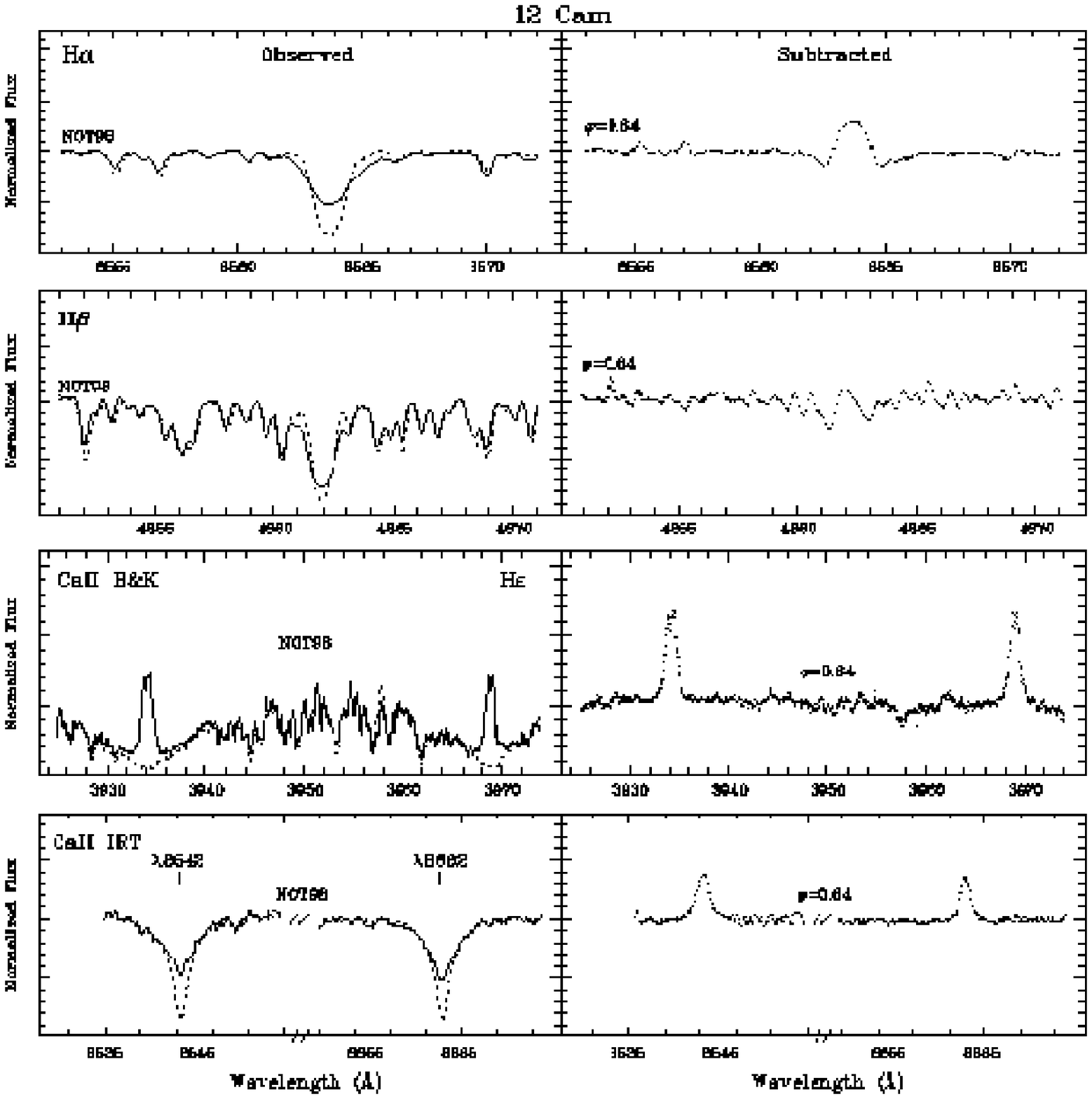,bbllx=36pt,bblly=36pt,bburx=545pt,bbury=544pt,height=15.0cm,width=17.8cm,clip=}}
\caption[ ]{H$\alpha$, H$\beta$, Ca~{\sc ii} H \& K, and Ca~{\sc ii} IRT
spectra of 12 Cam.
Observed and synthetic spectra in the left panel and subtracted spectra
in the right panel.
\label{fig:12cam_all} }
\end{figure*}

\begin{figure*}
{\psfig{figure=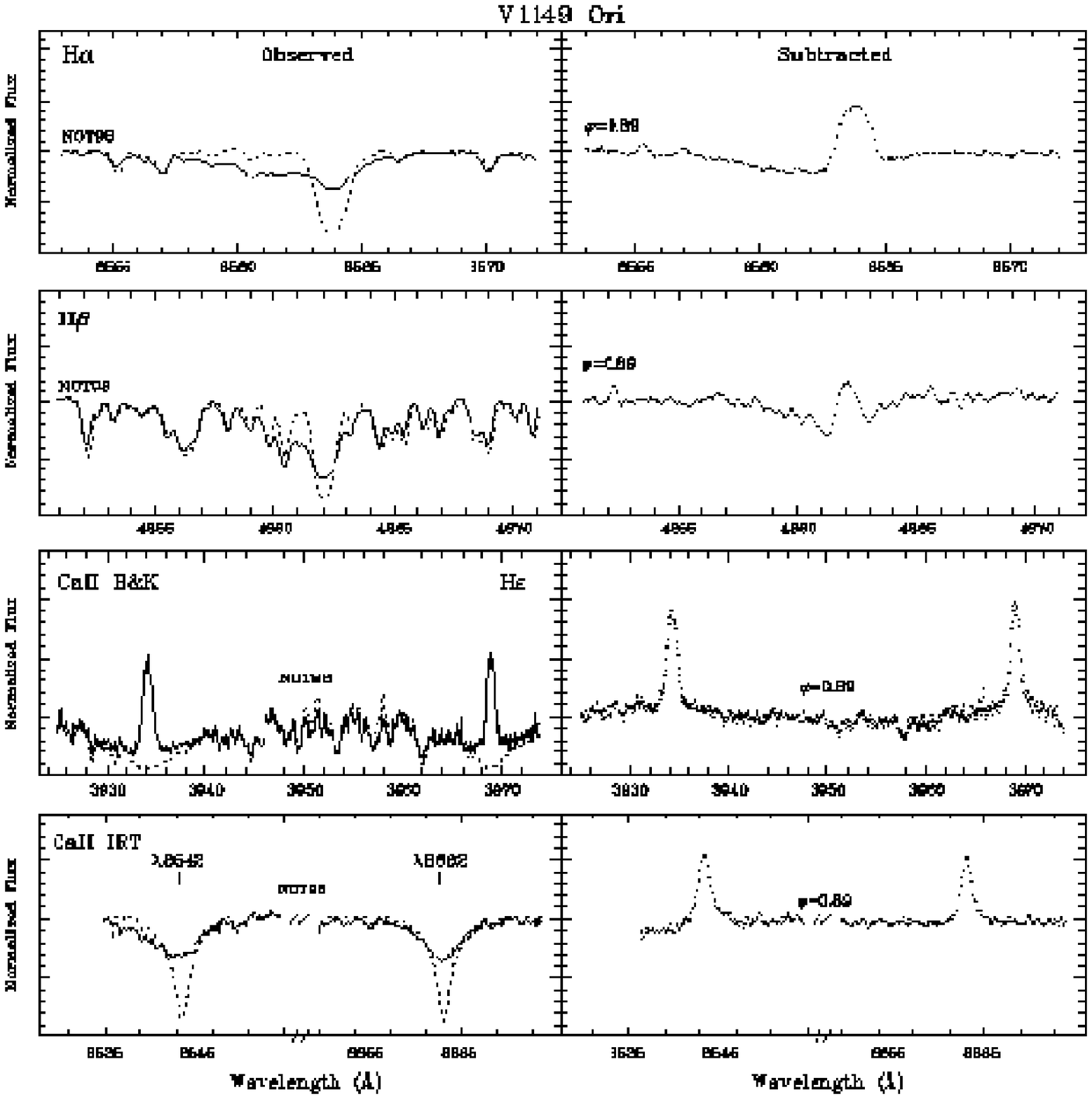,bbllx=36pt,bblly=36pt,bburx=545pt,bbury=544pt,height=15.0cm,width=17.8cm,clip=}}
\caption[ ]{H$\alpha$, H$\beta$, Ca~{\sc ii} H \& K, and Ca~{\sc ii} IRT
spectra of V1149 Ori.
Observed and synthetic spectra in the left panel and subtracted spectra
in the right panel.
\label{fig:v1149ori_all} }
\end{figure*}


\begin{figure*}
{\psfig{figure=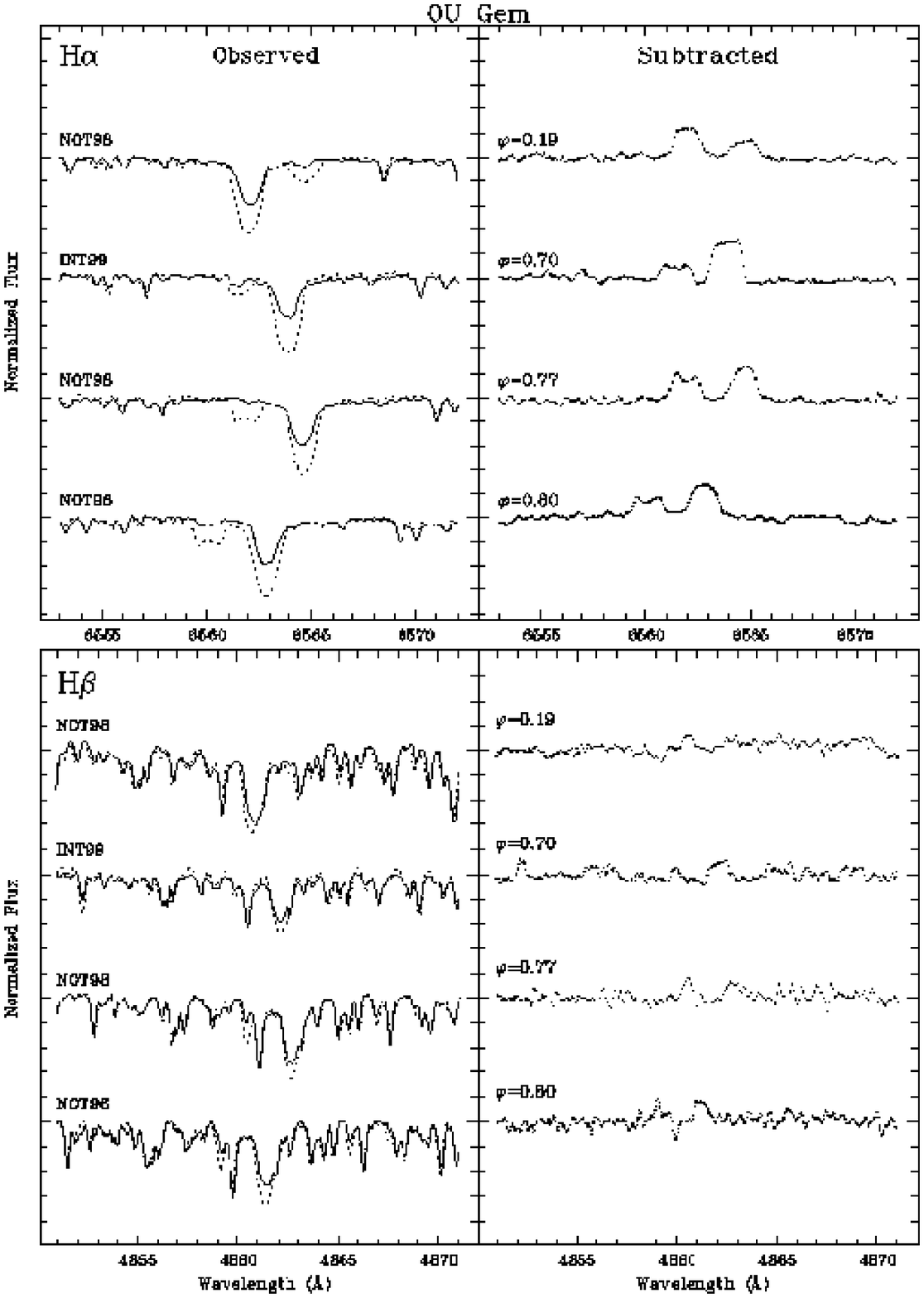,bbllx=36pt,bblly=36pt,bburx=545pt,bbury=774pt,height=22.5cm,width=17.8cm,clip=}}
\caption[ ]{H$\alpha$ and H$\beta$ spectra of OU Gem.
Observed and synthetic spectra in the left panel and subtracted spectra
in the right panel
\label{fig:ougem_ha_hb} }
\end{figure*}

\begin{figure*}
{\psfig{figure=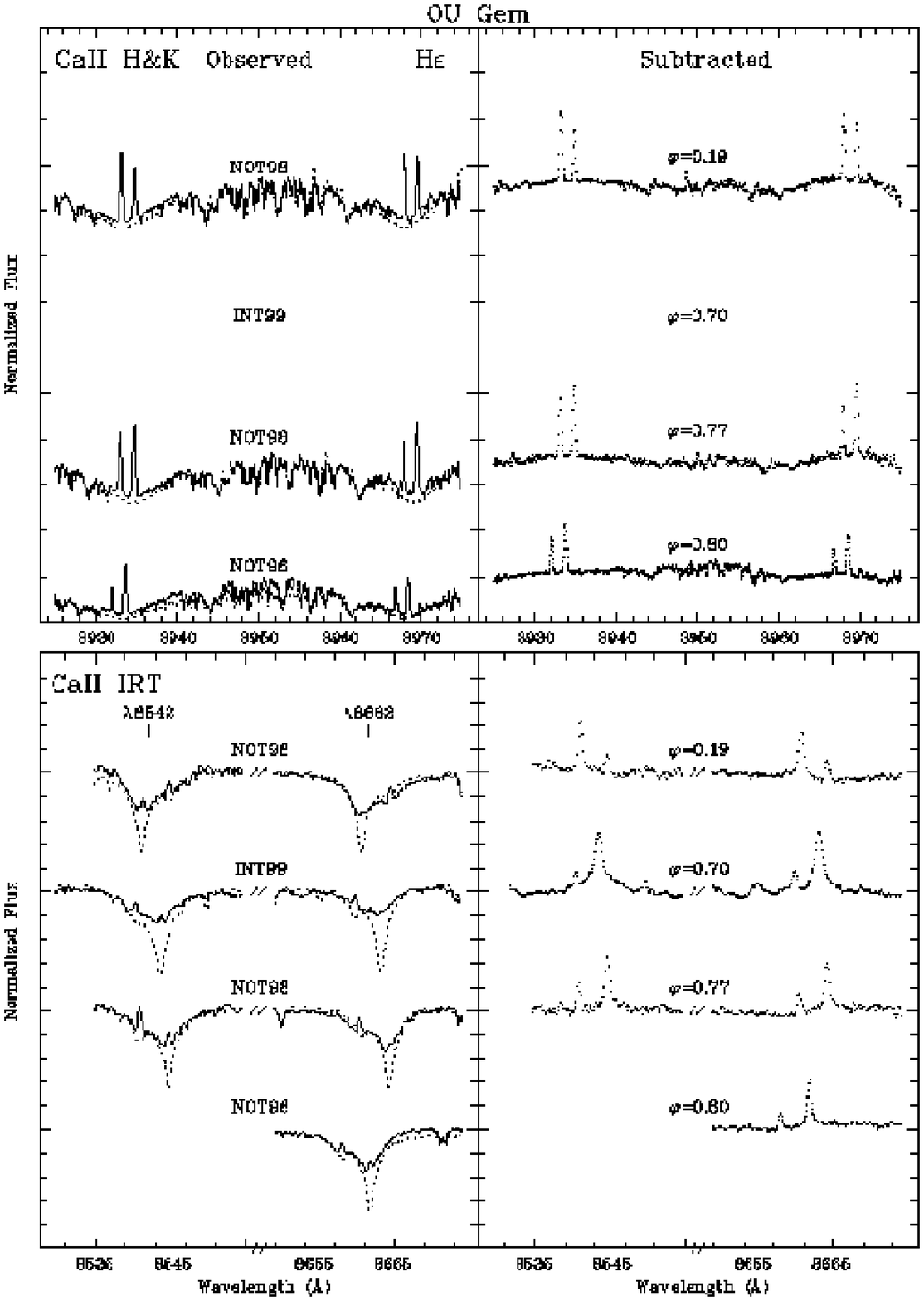,bbllx=36pt,bblly=36pt,bburx=545pt,bbury=774pt,height=22.5cm,width=17.8cm,clip=}}
\caption[ ]{Ca~{\sc ii} H \& K and Ca~{\sc ii} IRT spectra of OU Gem.
Observed and synthetic spectra in the left panel and subtracted spectra
in the right panel
\label{fig:ougem_hyk_cairt} }
\end{figure*}


\begin{figure*}
{\psfig{figure=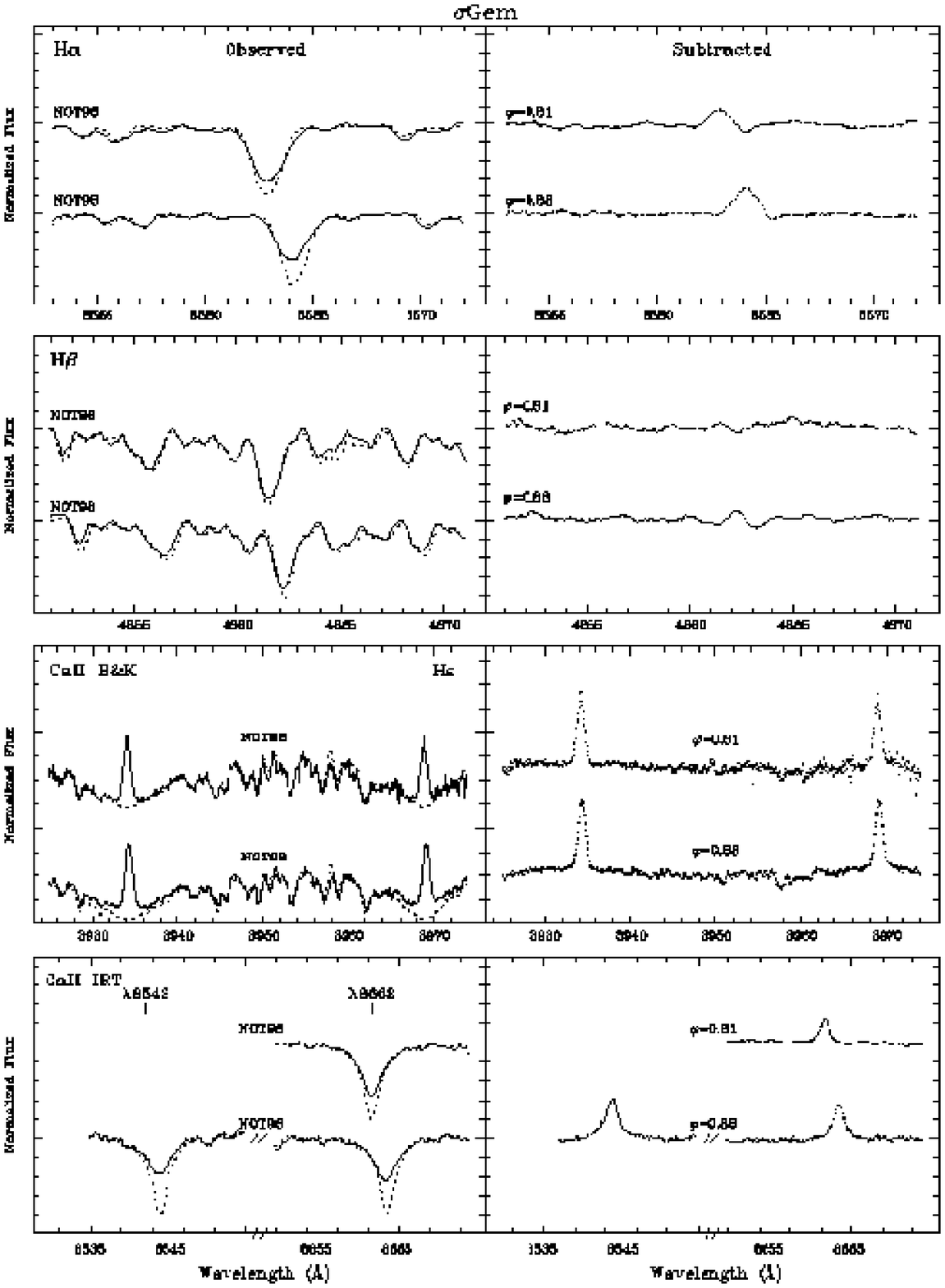,bbllx=36pt,bblly=36pt,bburx=545pt,bbury=725pt,height=22.5cm,width=17.8cm,clip=}}
\caption[ ]{H$\alpha$, H$\beta$, Ca~{\sc ii} H \& K, and Ca~{\sc ii} IRT
spectra of $\sigma$ Gem.
Observed and synthetic spectra in the left panel and subtracted spectra
in the right panel.
\label{fig:sgem_all} }
\end{figure*}


\begin{figure*}
{\psfig{figure=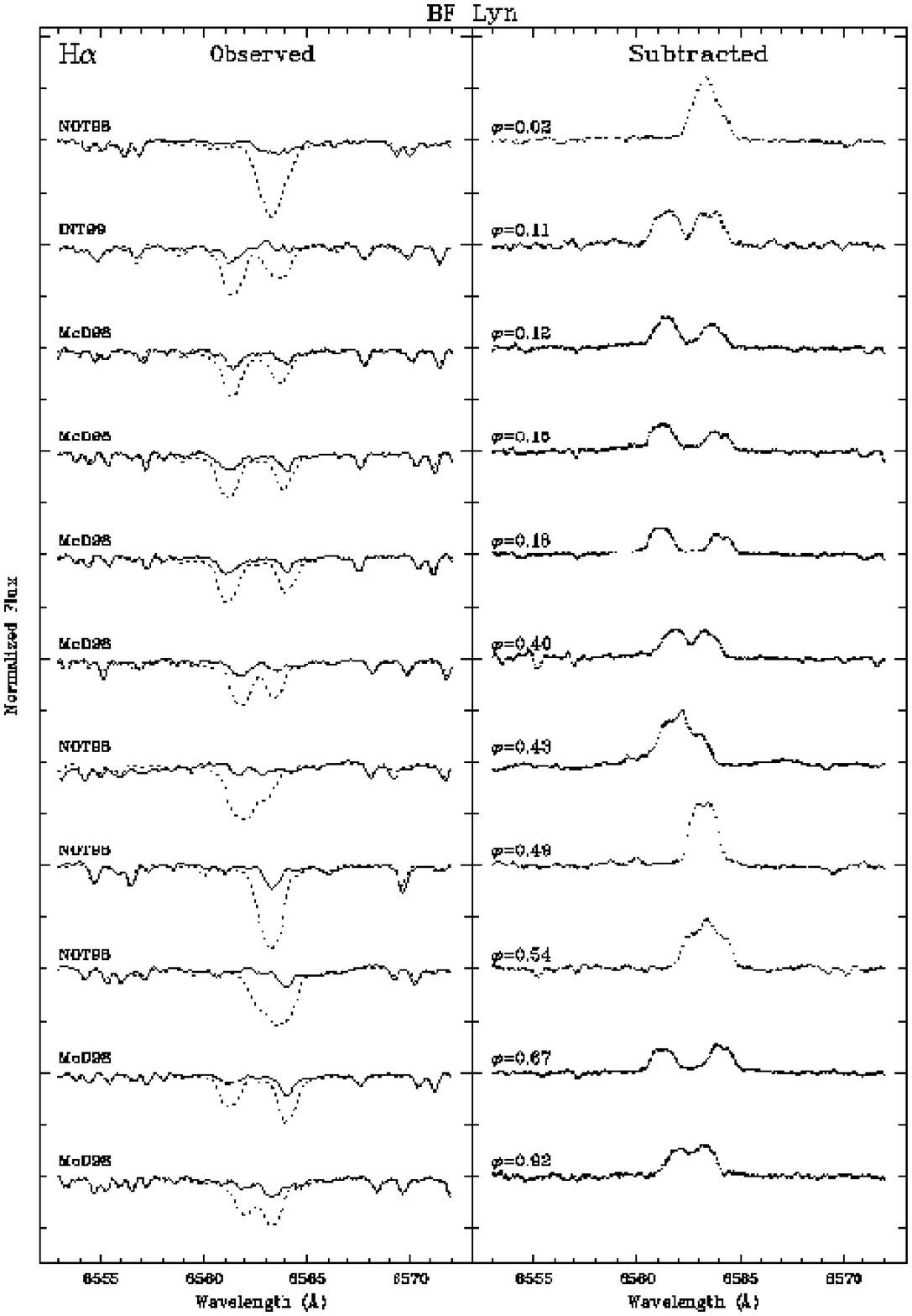,bbllx=36pt,bblly=36pt,bburx=545pt,bbury=774pt,height=22.5cm,width=17.8cm,clip=}}
\caption[ ]{H$\alpha$ observed (left panel) and subtracted (right panel) 
spectra of BF Lyn
\label{fig:bflyn_ha} }
\end{figure*}

\begin{figure*}
\resizebox{17.5cm}{7.5cm}{\includegraphics{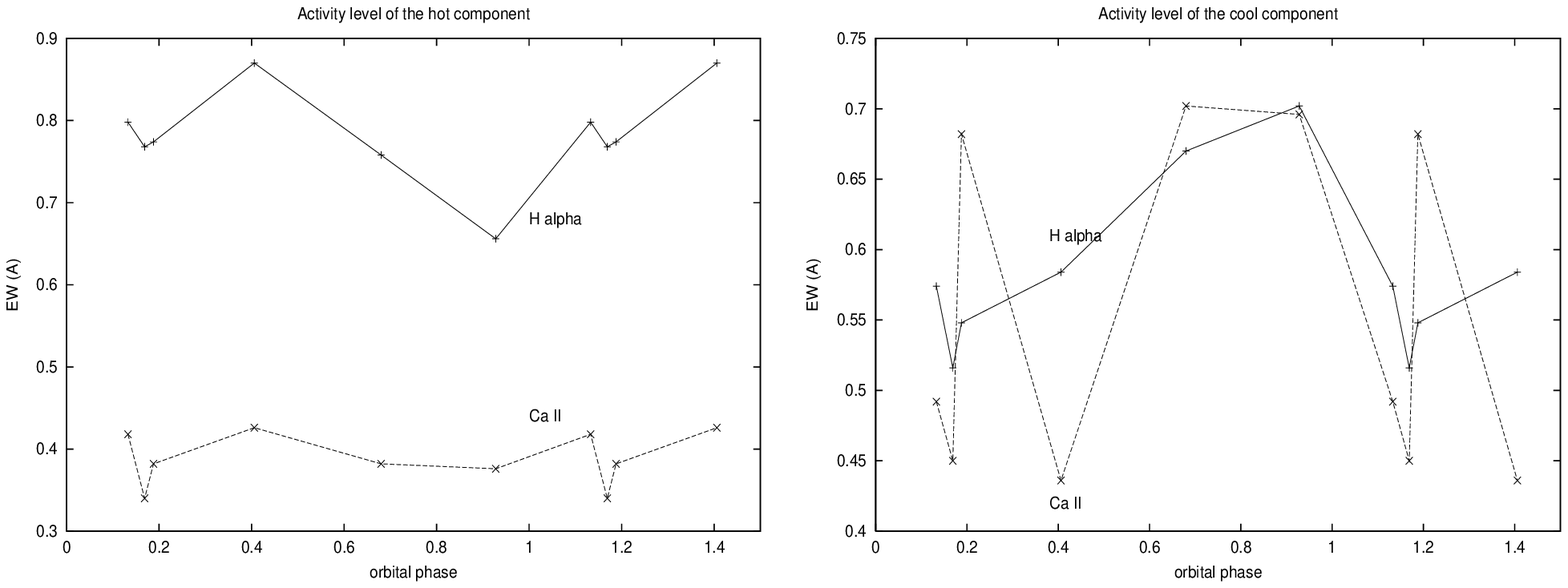}}
\caption{H$\alpha$ and Ca~{\sc II} IRT $\lambda$8542 EW of the hot (left panel)
and cool (right panel) components of BF Lyn for the McD98 run 
versus the orbital phase
\label{fig:bflyn_ew} }
\end{figure*}

%
\begin{figure*}
{\psfig{figure=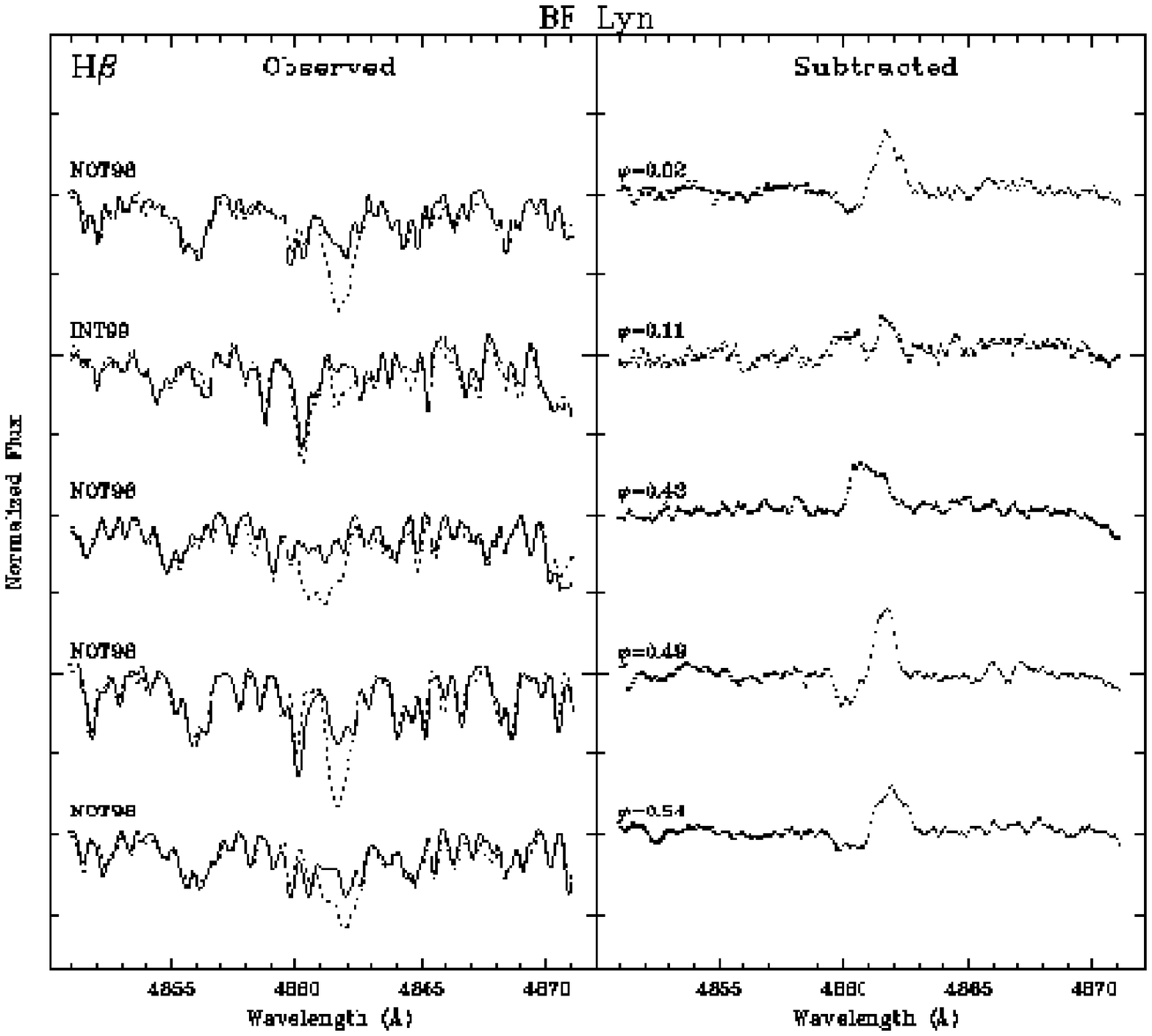,bbllx=36pt,bblly=36pt,bburx=545pt,bbury=494pt,height=14.0cm,width=17.8cm,clip=}}
\caption[ ]{H$\beta$ observed (left panel) and subtracted (right panel)
spectra of BF Lyn
\label{fig:bflyn_hb} }
\end{figure*}

\begin{figure*}
{\psfig{figure=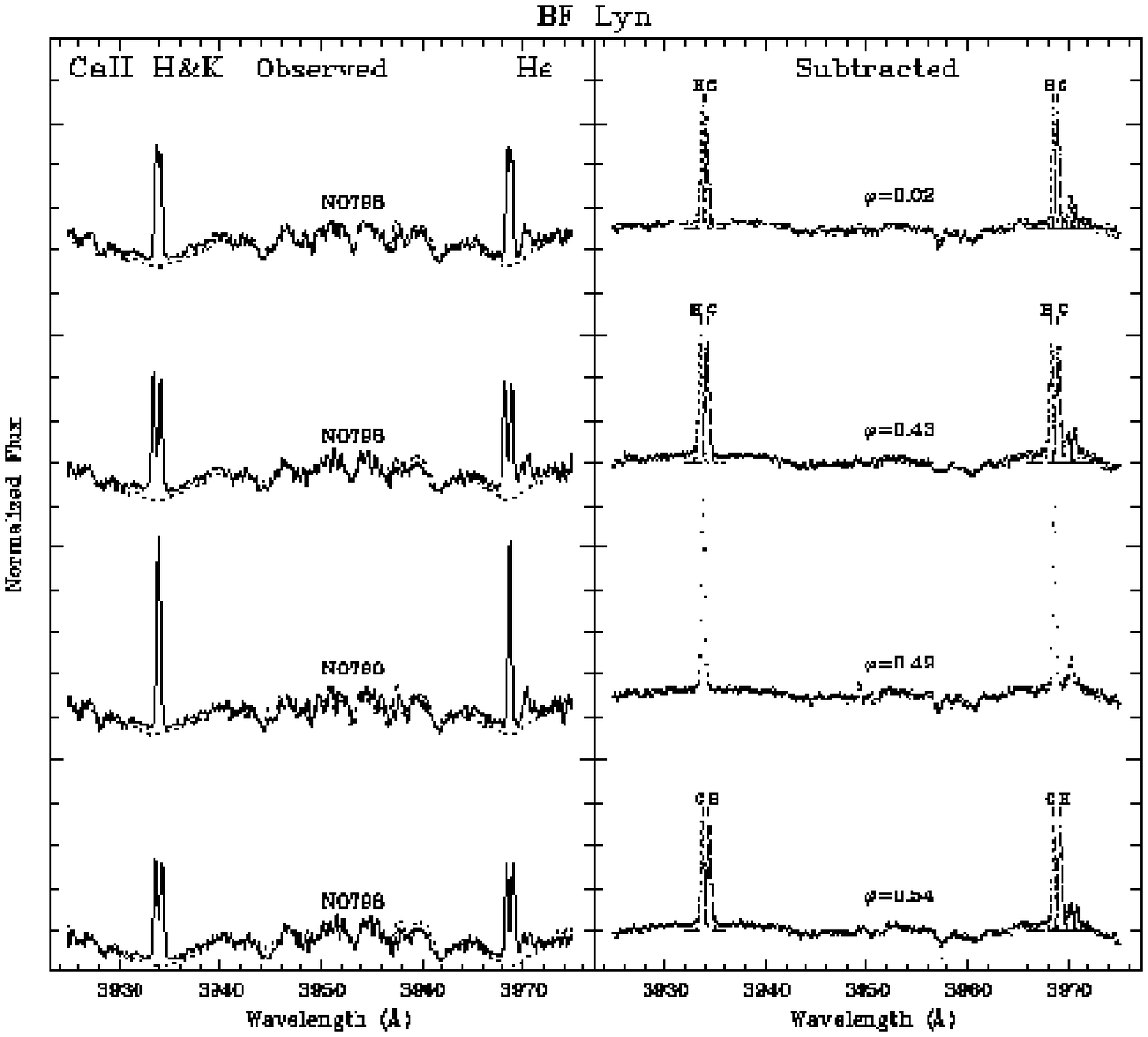,bbllx=36pt,bblly=36pt,bburx=545pt,bbury=496pt,height=12.0cm,width=17.8cm,clip=}}
\caption[ ]{Ca~{\sc ii} H \& K 
observed (left panel) and subtracted (right panel)
spectra of BF Lyn.
We have also plotted the  Gaussian fit to the subtracted spectrum 
used to deblend the contribution
of the hot (H) and cool (C) components (short- and long-dashed lines)
\label{fig:bflyn_cahyk} }
\end{figure*}

%
\begin{figure*}
{\psfig{figure=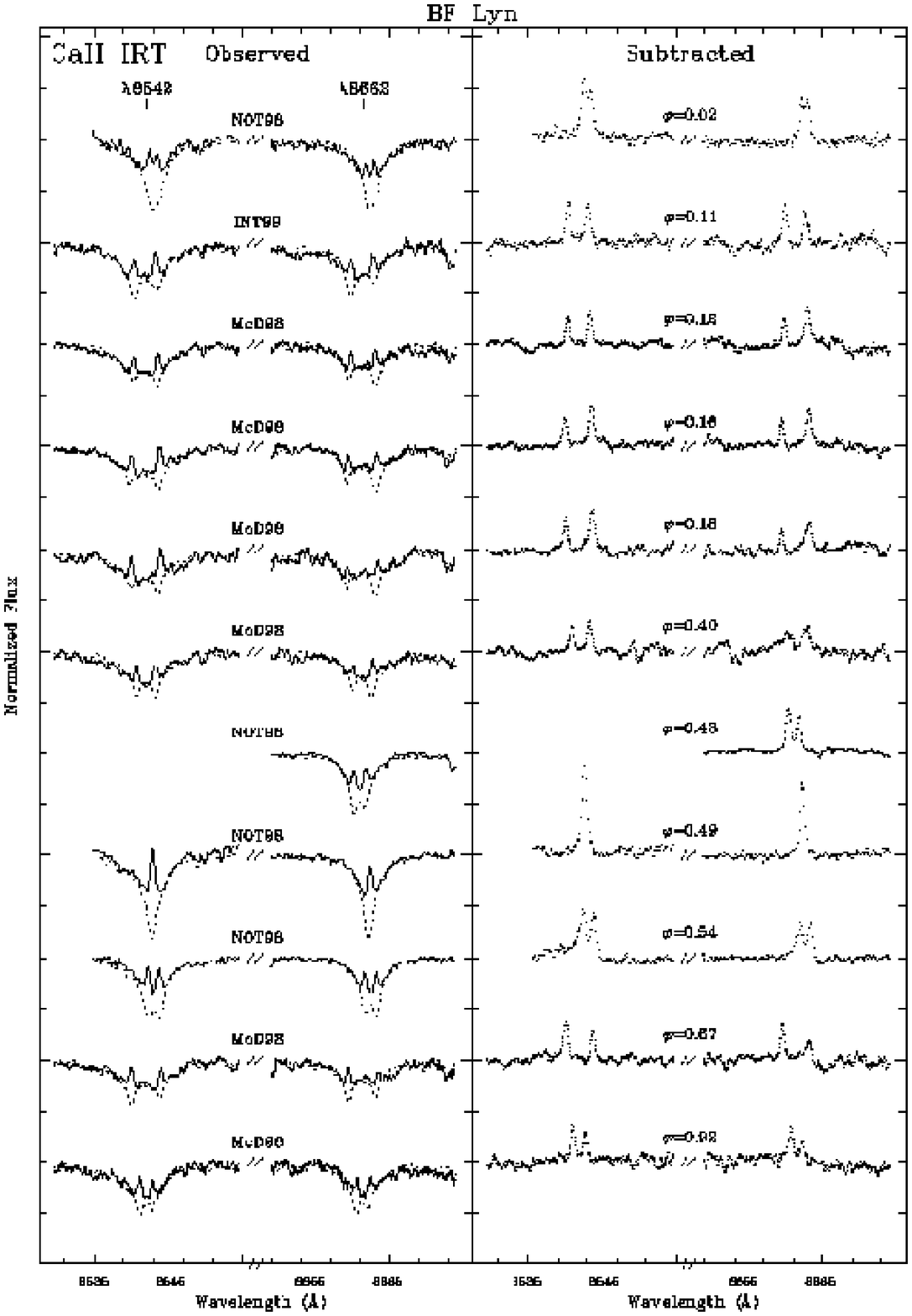,bbllx=36pt,bblly=36pt,bburx=545pt,bbury=774pt,height=22.5cm,width=17.8cm,clip=}}
\caption[ ]{Ca~{\sc II} IRT $\lambda$8542 and  $\lambda$8662
observed (left panel) and subtracted (right panel)
spectra of BF Lyn
\label{fig:bflyn_cairt} }
\end{figure*}


\begin{figure*}
{\psfig{figure=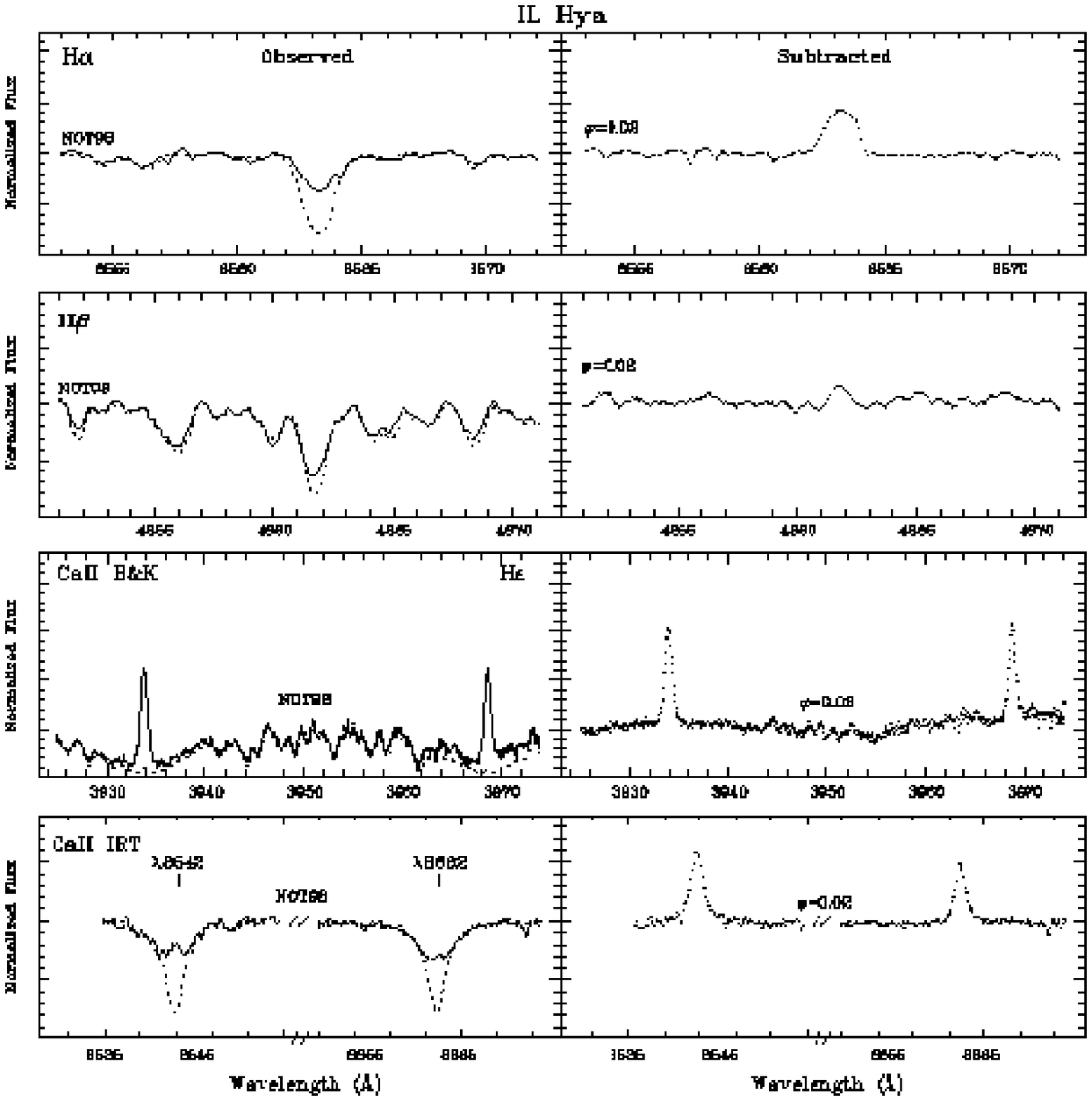,bbllx=36pt,bblly=36pt,bburx=545pt,bbury=546pt,height=15.0cm,width=17.8cm,clip=}}
\caption[ ]{H$\alpha$, H$\beta$, Ca~{\sc ii} H \& K, and Ca~{\sc ii} IRT
spectra of IL Hya.
Observed and synthetic spectra in the left panel and subtracted spectra
in the right panel.
\label{fig:ilhya_all} }
\end{figure*}


\begin{figure*}
{\psfig{figure=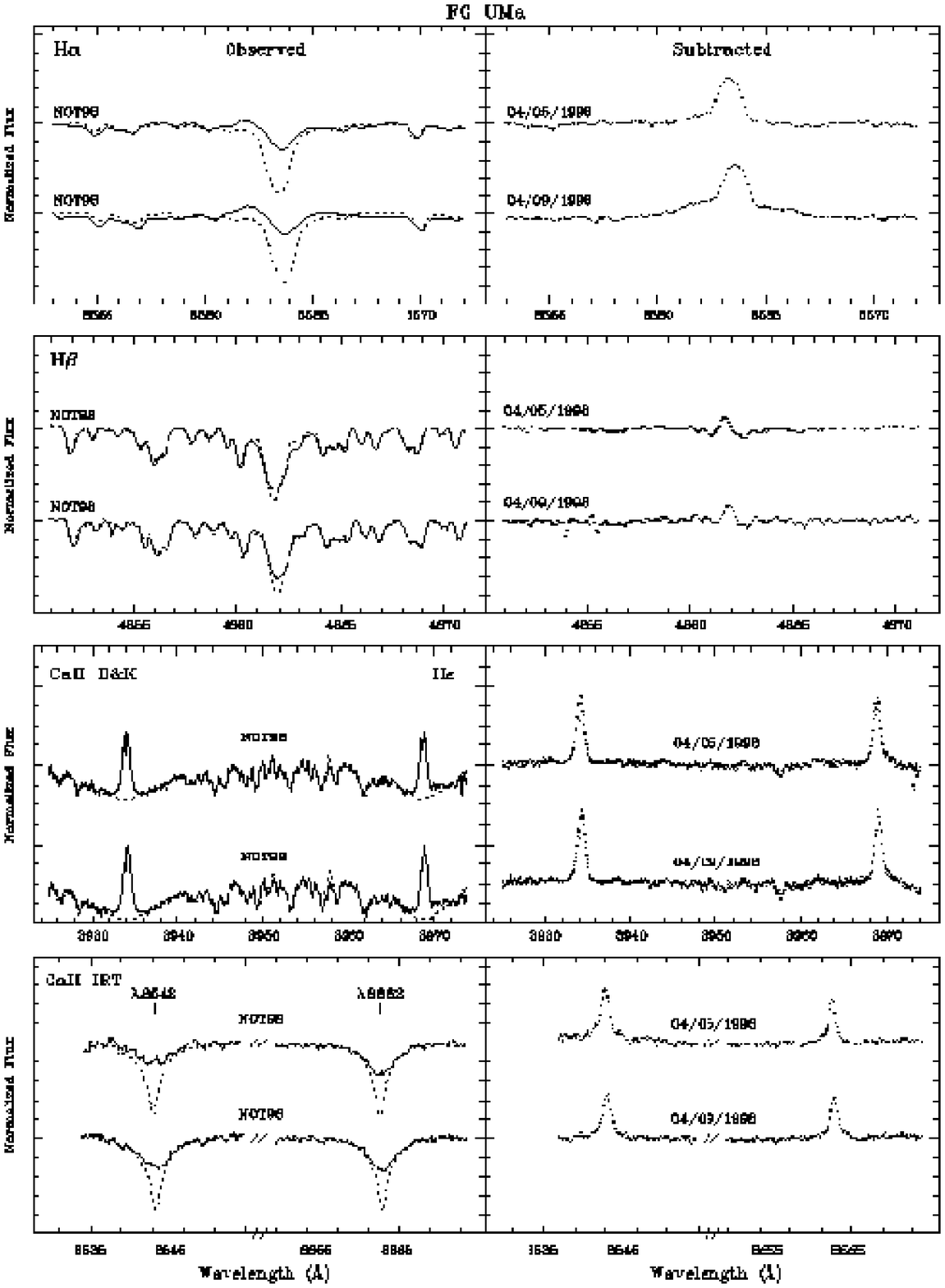,bbllx=36pt,bblly=36pt,bburx=545pt,bbury=725pt,height=22.5cm,width=17.8cm,clip=}}
\caption[ ]{H$\alpha$, H$\beta$, Ca~{\sc ii} H \& K, and Ca~{\sc ii} IRT
spectra of FG UMa.
Observed and synthetic spectra in the left panel and subtracted spectra
in the right panel.
\label{fig:fguma_all} }
\end{figure*}


\begin{figure*}
{\psfig{figure=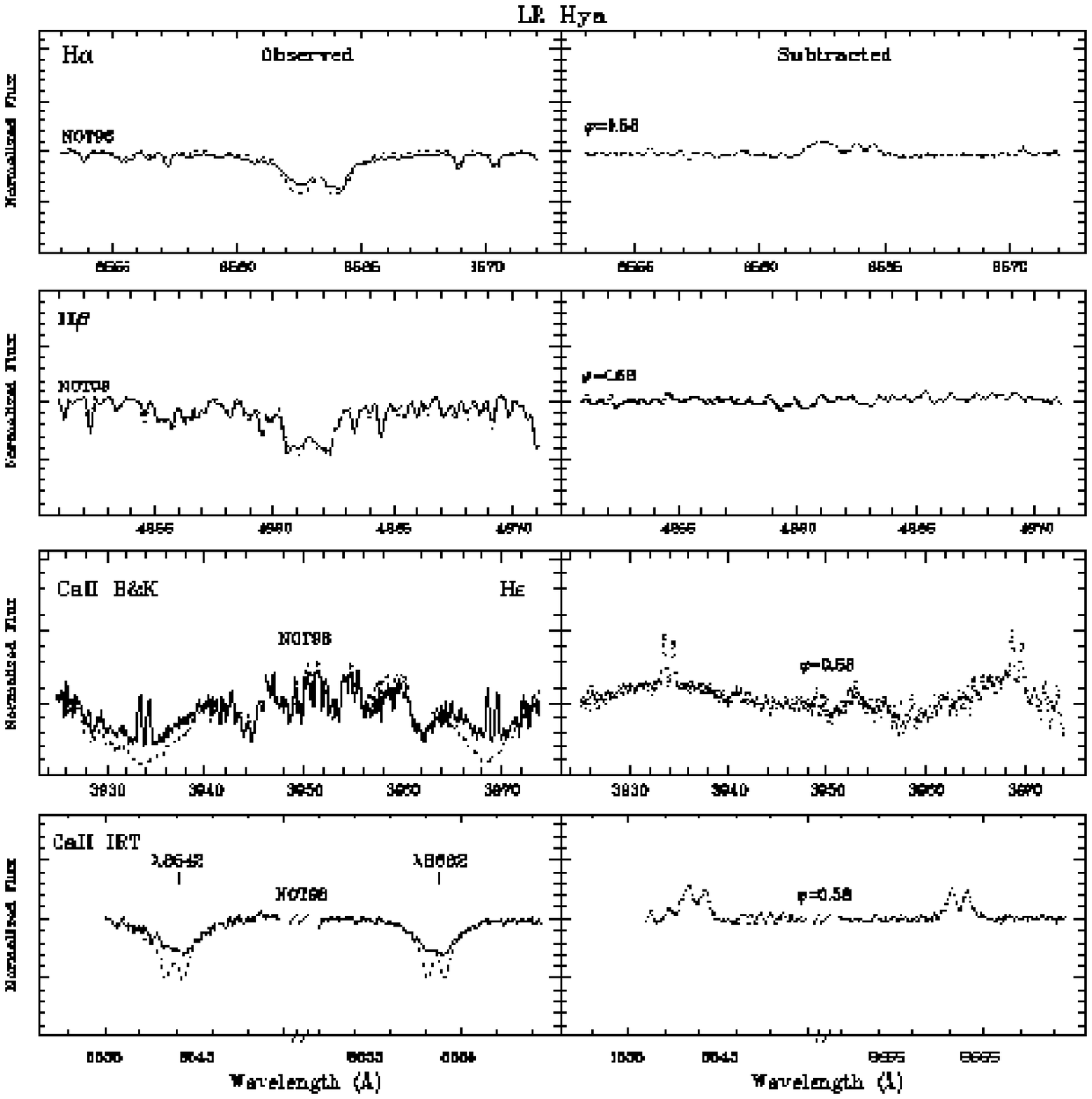,bbllx=36pt,bblly=36pt,bburx=545pt,bbury=544pt,height=15.0cm,width=17.8cm,clip=}}
\caption[ ]{H$\alpha$, H$\beta$, Ca~{\sc ii} H \& K, and Ca~{\sc ii} IRT
spectra of LR Hya.
Observed and synthetic spectra in the left panel and subtracted spectra
in the right panel.
\label{fig:lrhya_all} }
\end{figure*}



%% file: ds1878_figs2s.tex





\begin{figure*}
{\psfig{figure=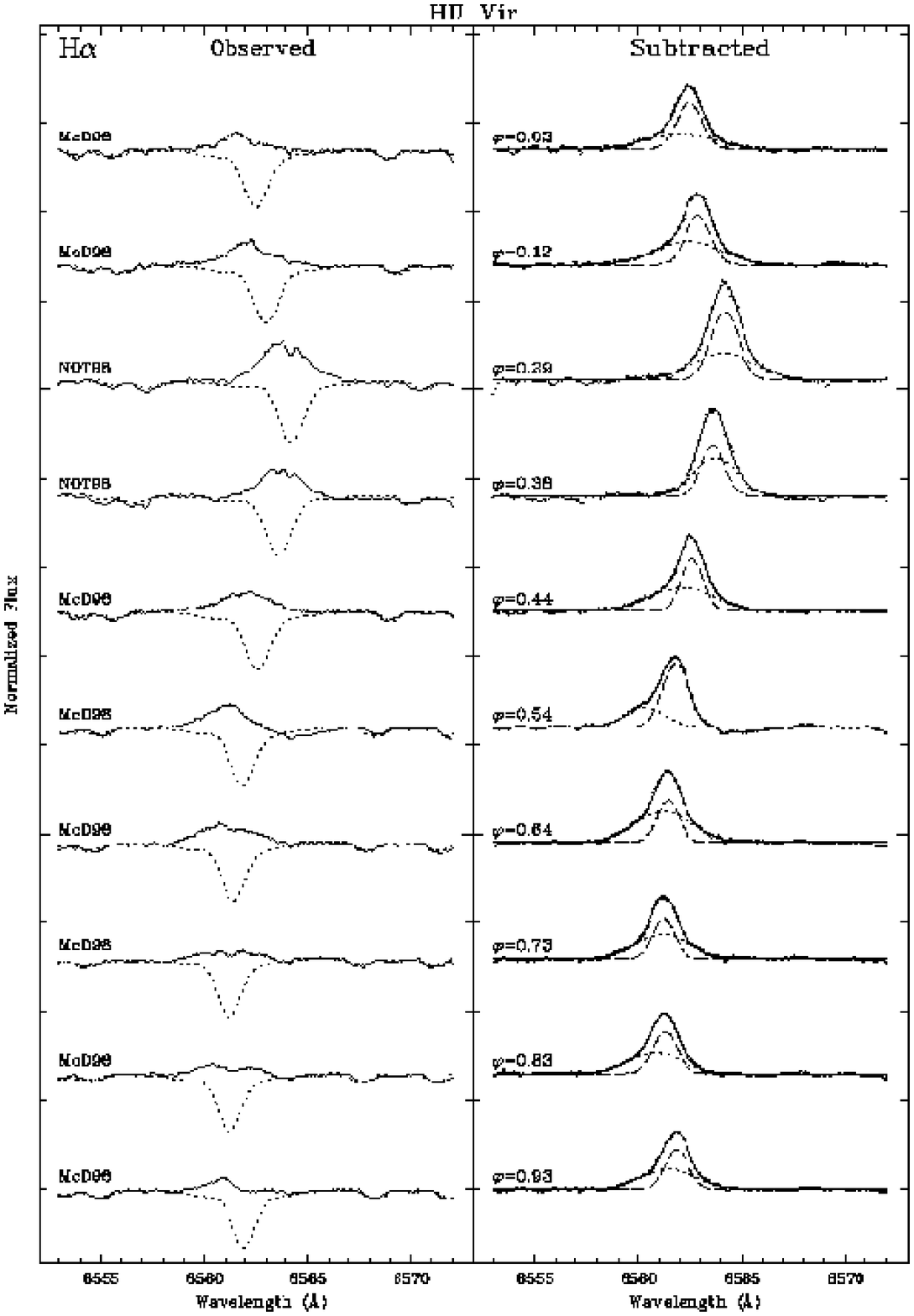,bbllx=36pt,bblly=36pt,bburx=545pt,bbury=774pt,height=22.5cm,width=17.8cm,clip=}}
\caption[ ]{H$\alpha$ observed (left panel) and subtracted (right panel)
spectra of HU Vir.
We have superposed the two Gaussian components fit (solid-line).
The sort-dashed-line represents the broad component
and the large-dashed-line the narrow one.
\label{fig:huvir_ha} }
\end{figure*}

\begin{figure*}
{\psfig{figure=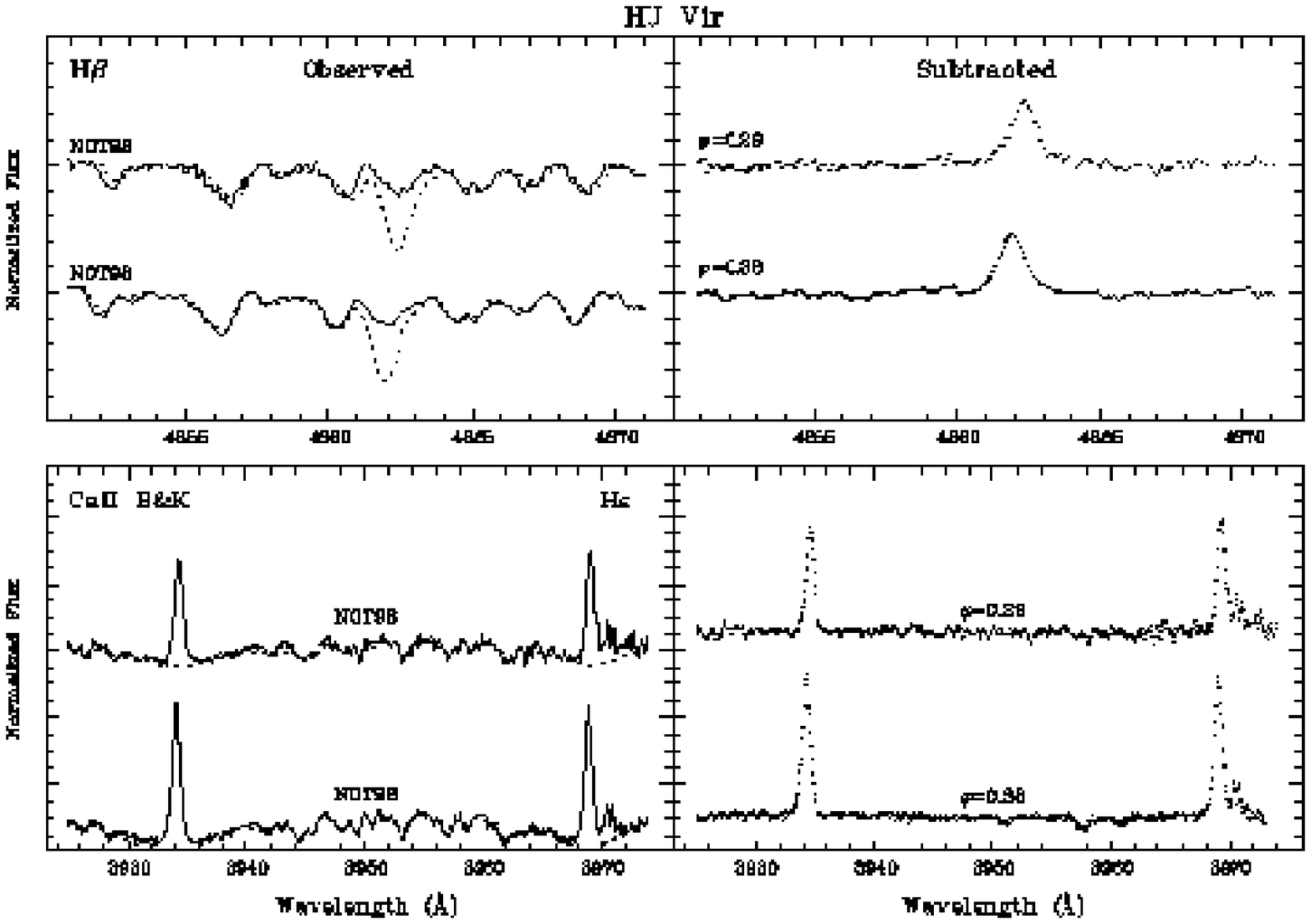,bbllx=36pt,bblly=202pt,bburx=545pt,bbury=560pt,height=11.5cm,width=17.8cm,clip=}}
\caption[ ]{H$\beta$, and Ca~{\sc ii} H \& K observed (left panel) and subtracted (right panel) spectra of HU Vir
\label{fig:huvir_hb_hyk} }
\end{figure*}

\begin{figure*}
{\psfig{figure=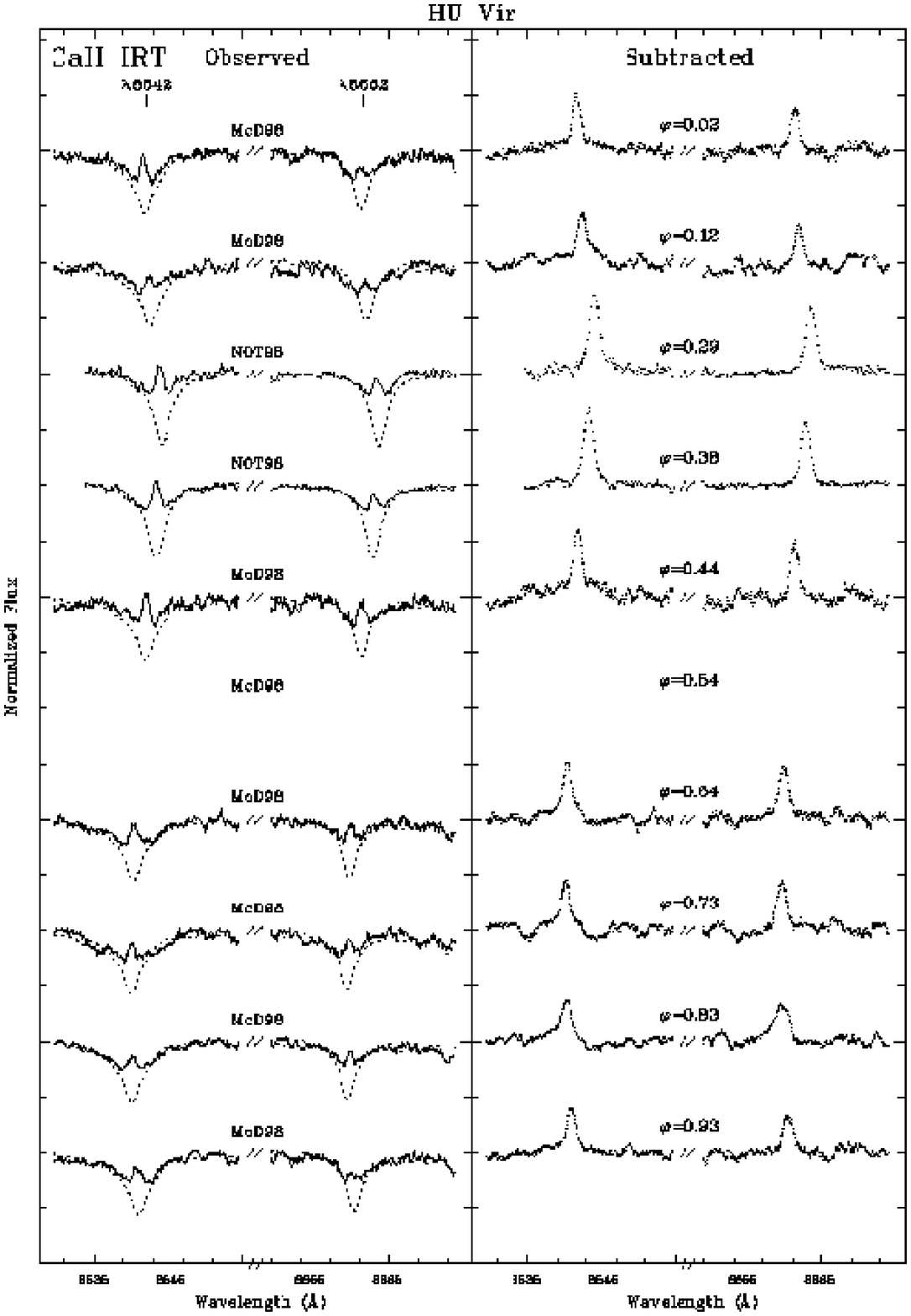,bbllx=36pt,bblly=36pt,bburx=545pt,bbury=774pt,height=22.5cm,width=17.8cm,clip=}}
\caption[ ]{Ca~{\sc II} IRT $\lambda$8542 and  $\lambda$8662
observed (left panel) and subtracted (right panel)
spectra of HU Vir
\label{fig:huvir_cairt} }
\end{figure*}


\begin{figure*}
{\psfig{figure=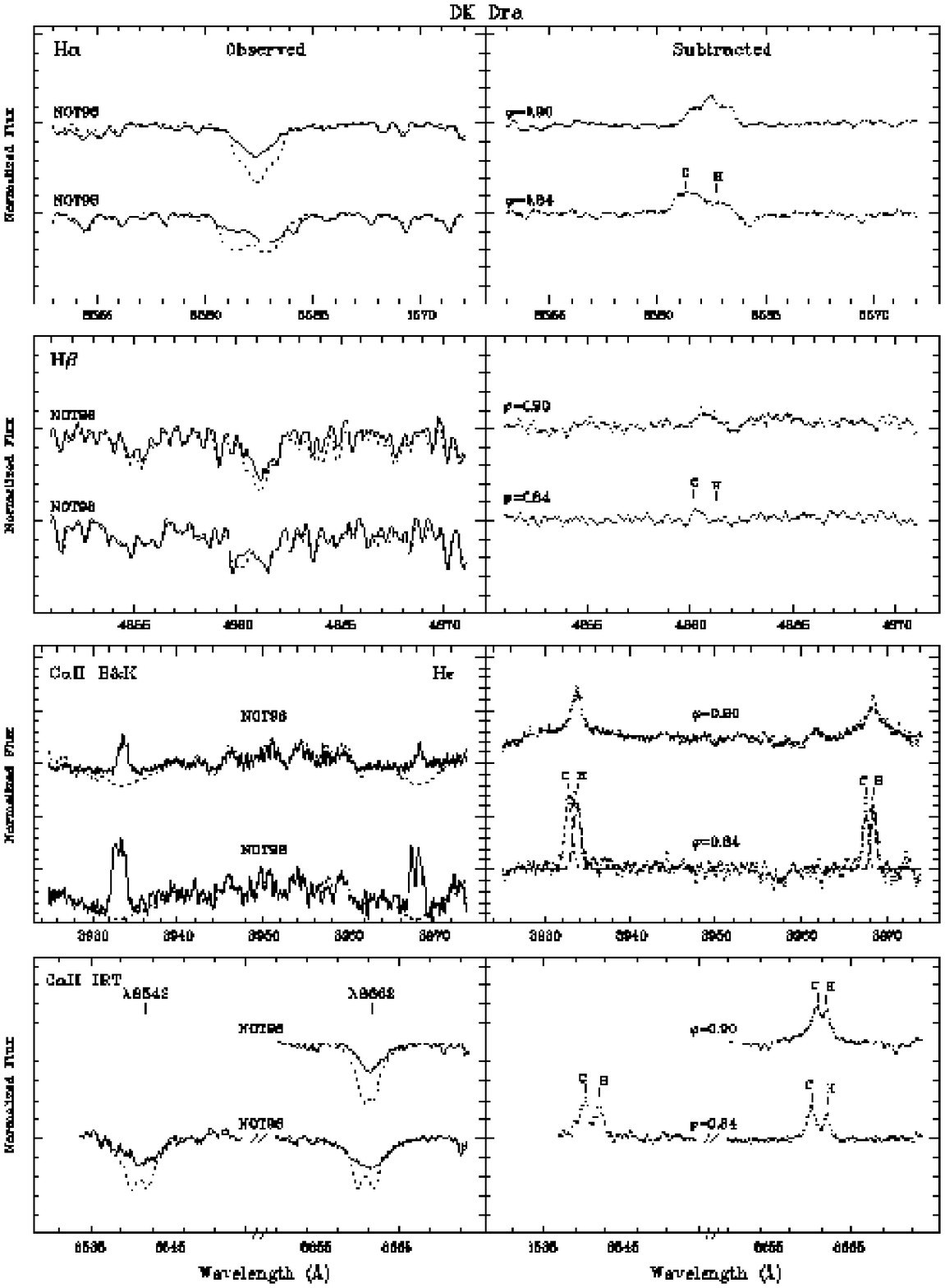,bbllx=36pt,bblly=36pt,bburx=545pt,bbury=725pt,height=22.5cm,width=17.5cm,clip=}}
\caption[ ]{H$\alpha$, H$\beta$, Ca~{\sc ii} H \& K, and Ca~{\sc ii} IRT
spectra of DK Dra.
Observed and synthetic spectra in the left panel and subtracted spectra
in the right panel.
We have also plotted the  Gaussian fit to the subtracted spectrum 
used to deblend the contribution
of the hot (H) and cool (C) components in the Ca~{\sc ii} H \& K lines 
at 0.84 orbital phase (short- and long-dashed lines).
\label{fig:dkdra_all} }
\end{figure*}

\begin{figure*}
{\psfig{figure=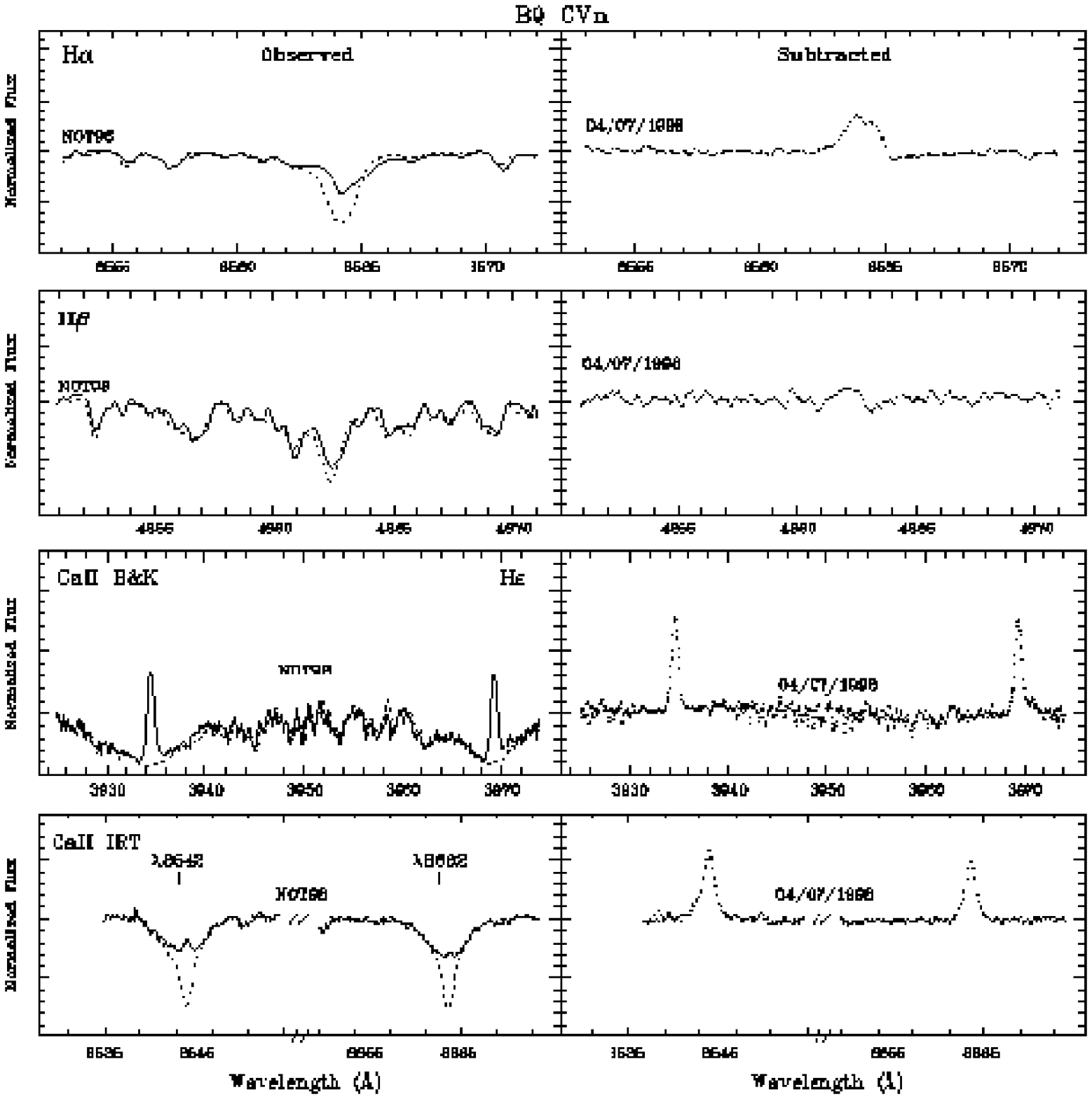,bbllx=36pt,bblly=36pt,bburx=545pt,bbury=544pt,height=15.0cm,width=17.8cm,clip=}}
\caption[ ]{H$\alpha$, H$\beta$, Ca~{\sc ii} H \& K, and Ca~{\sc ii} IRT
spectra of BQ CVn.
Observed and synthetic spectra in the left panel and subtracted spectra
in the right panel.
\label{fig:bqcvn_all} }
\end{figure*}

\begin{figure*}
{\psfig{figure=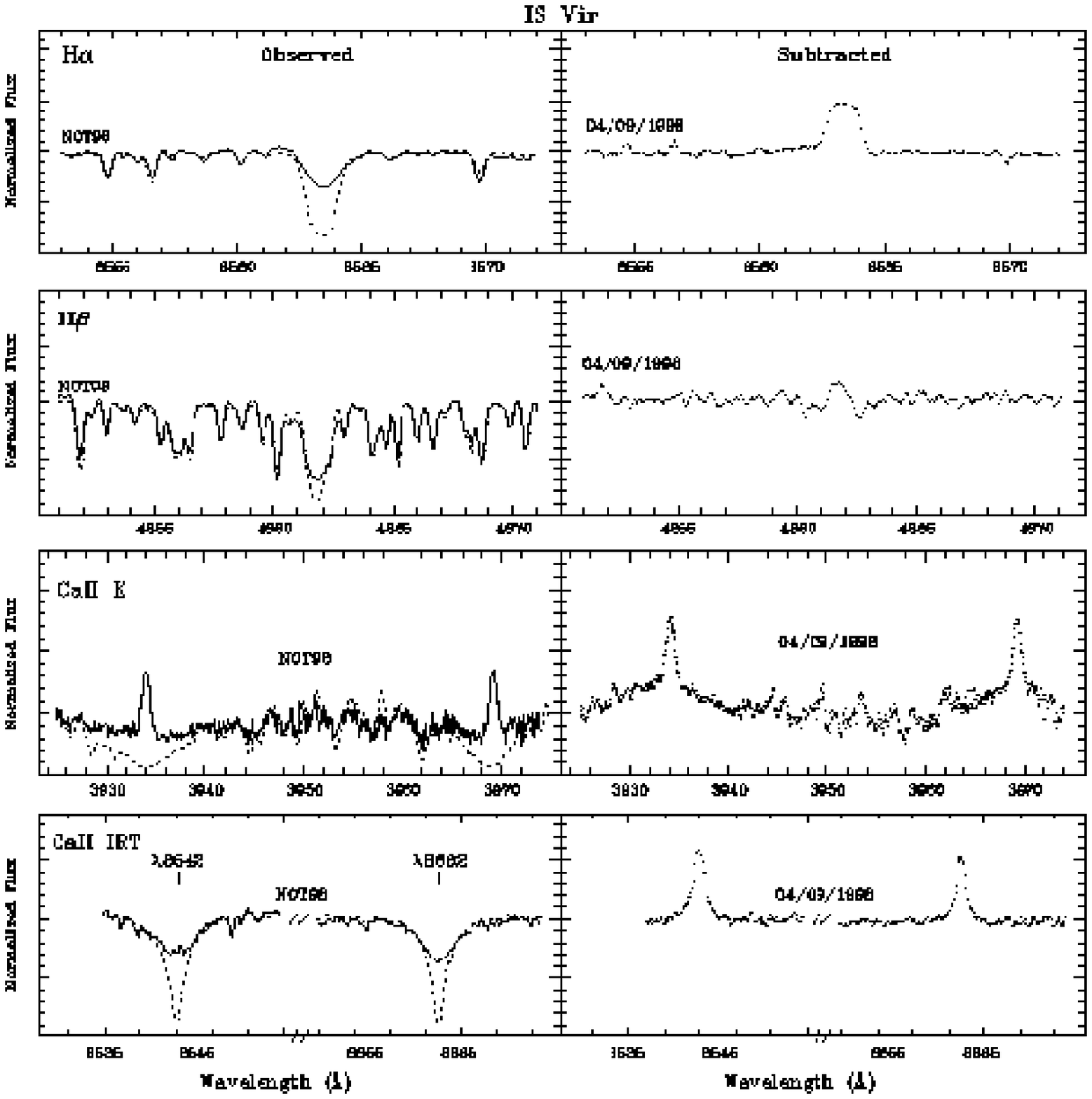,bbllx=36pt,bblly=36pt,bburx=545pt,bbury=544pt,height=15.0cm,width=17.8cm,clip=}}
\caption[ ]{H$\alpha$, H$\beta$, Ca~{\sc ii} H \& K, and Ca~{\sc ii} IRT
spectra of IS Vir.
Observed and synthetic spectra in the left panel and subtracted spectra
in the right panel.
\label{fig:isvir_all} }
\end{figure*}

\begin{figure*}
{\psfig{figure=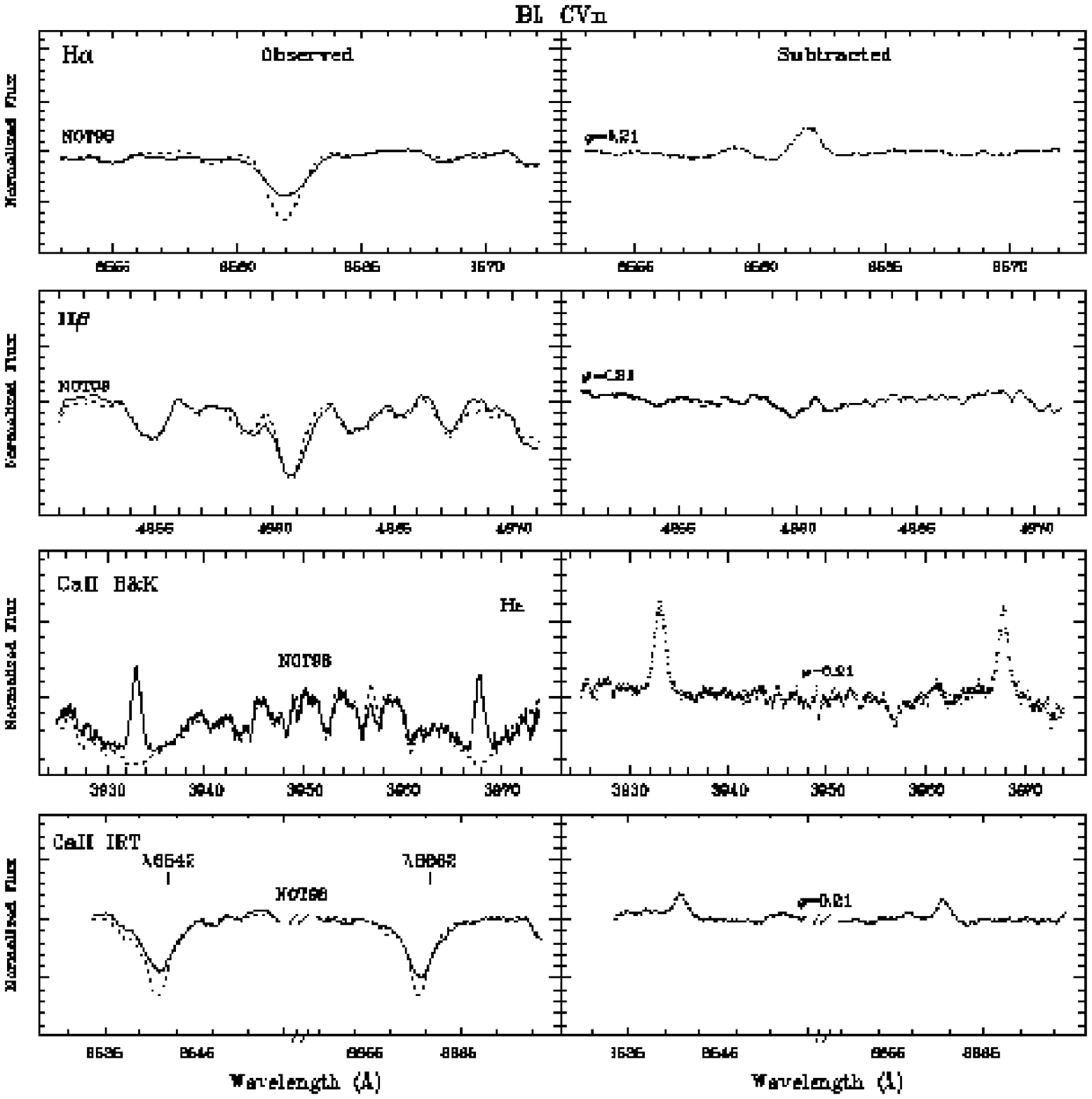,bbllx=36pt,bblly=36pt,bburx=545pt,bbury=544pt,height=15.0cm,width=17.8cm,clip=}}
\caption[ ]{H$\alpha$, H$\beta$, Ca~{\sc ii} H \& K, and Ca~{\sc ii} IRT
spectra of BL CVn.
Observed and synthetic spectra in the left panel and subtracted spectra
in the right panel.
\label{fig:blcvn_all} }
\end{figure*}

\begin{figure*}
{\psfig{figure=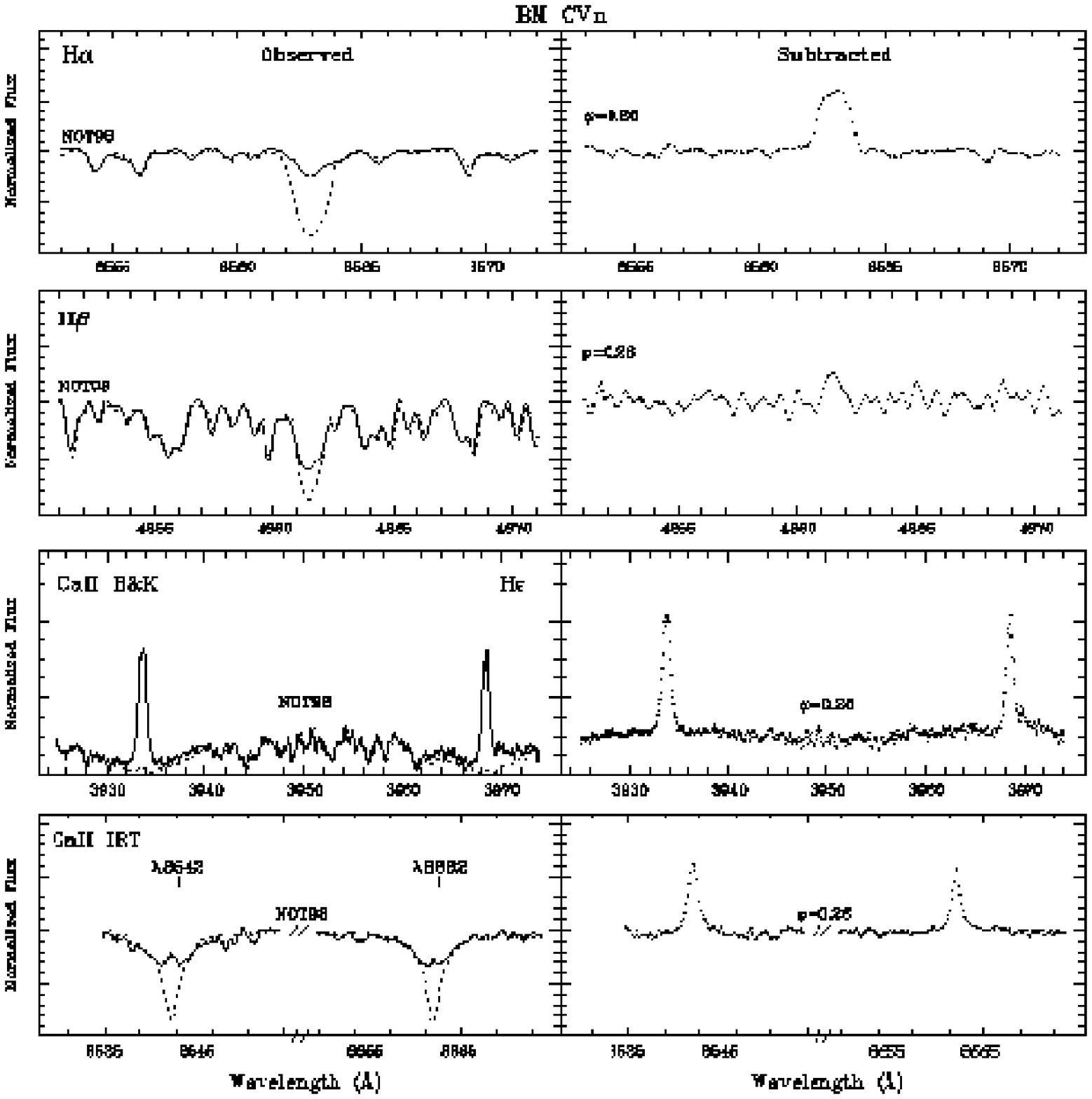,bbllx=36pt,bblly=36pt,bburx=545pt,bbury=544pt,height=15.0cm,width=17.8cm,clip=}}
\caption[ ]{H$\alpha$, H$\beta$, Ca~{\sc ii} H \& K, and Ca~{\sc ii} IRT
spectra of BM CVn.
Observed and synthetic spectra in the left panel and subtracted spectra
in the right panel.
\label{fig:bmcvn_all} }
\end{figure*}


\begin{figure*}
{\psfig{figure=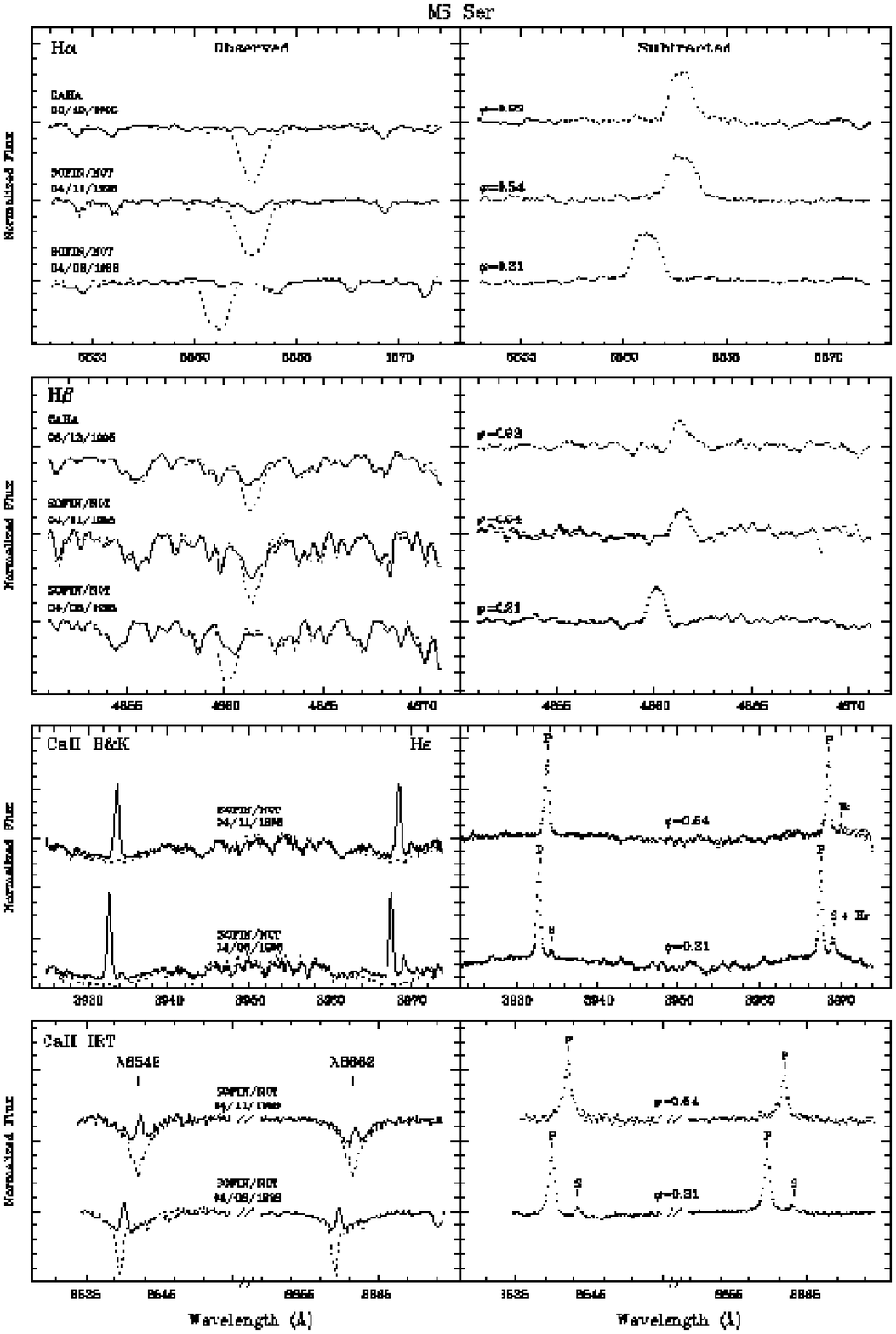,bbllx=36pt,bblly=36pt,bburx=545pt,bbury=790pt,height=22.5cm,width=17.8cm,clip=}}
\caption[ ]{H$\alpha$, H$\beta$, Ca~{\sc ii} H \& K, and Ca~{\sc ii} IRT
spectra of MS Ser.
Observed and synthetic spectra in the left panel and subtracted spectra
in the right panel.
\label{fig:msser_all} }
\end{figure*}

